\documentclass[a4paper,11pt]{article}
\usepackage{jheppub} 
\usepackage[utf8]{inputenc}
\usepackage{physics}
\usepackage{slashed}
\usepackage{caption}
\usepackage{xcolor}
\usepackage{comment}
\usepackage{multirow}
\usepackage{graphics}
\usepackage{float}
\usepackage{cancel}
\usepackage{soul}
\usepackage{cases}
\usepackage{array}
\usepackage{mathtools}  
\usepackage{amsfonts}
\usepackage{hyperref}
\usepackage{amsmath}
\usepackage{amssymb}
\usepackage{tcolorbox}
\usepackage{tikz}

\title{\boldmath\boldmath Four-point functions, Twistors and Supersymmetry in \\
the Symplectic Bi-Grassmannian for CFT$_4$ and AdS$_5$}

\author{Dhruva K.S.}

\affiliation{~}

\emailAdd{dhruvacaesar@gmail.com}

\abstract{We initiate the study of four point functions in the symplectic bi-Grassmannian framework for four dimensional conformal field theories. We derive factorization formulae analogous to those for scattering amplitudes and bootstrap several examples of AdS$_5$ exchange correlation functions involving scalars, photons, fermions, gluons and gravitons. These correlators are rational functions of the minors of the Grassmannian matrices in contrast to their momentum space counterparts which are much more complicated.

We then develop the $GL(2,\mathbb{R})$ twistor space formalism, deriving the Penrose transform and the twistorial Grassmannian where we find remarkably simple expressions. We also find an elegant and natural extension to $\mathcal{N}=1$ supersymmetry, deriving the supersymmetric Penrose transform and the super-twistor Grassmannian. Performing a half-Fourier transform from twistor space, we derive the supersymmetric symplectic bi-Grassmannian corresponding to spinor-helicity variables.

We test our formalism in the context of AdS$_5$ $\mathcal{N}=1$ super-Yang Mills theory where we impose consistent factorization and input the bootstrapped four-fermion correlator to obtain those involving gluons, scalars and fermions. Finally, we discuss the application to supergravity theories.}

\begin{document}
\maketitle

\section{Introduction}
The Grassmannian $\text{Gr}(k,n)$ is the space of $k-$dimensional planes in a $n-$dimensional space. Its role in physics has been exemplified over the past two decades, mainly  in the context of scattering amplitudes \cite{Arkani-Hamed:2012zlh,Elvang:2013cua} and much more recently, in conformal field theory and the description of holographic (A)dS boundary correlators \cite{Arundine:2026fbr, De:2026shn, Bala:2026hdm,Bala:2026bdx,Huang:2026tsh, Bala:2026trw,Arundine:2026myr, Bala:2026lvw}. It provides an efficient way to encode the external kinematic data, makes symmetries manifest, renders recursion relations transparent and paves the way towards a geometric understanding of observables. In recent years, the Grassmannian framework has even been extended to  massive flat space scattering amplitudes \cite{Cachazo:2018hqa,Bering:2022tdr,Cortes:2025xmu}. In the context of three dimensional conformal field theory and (A)dS$_4$ boundary correlators, there has been lots of progress over the years in momentum space, spinor-helicity \cite{Maldacena:2011nz, McFadden:2011kk,  Coriano:2013jba,  Bzowski:2013sza,  Ghosh:2014kba,  Bzowski:2015pba,  Bzowski:2017poo,  Bzowski:2018fql,  Farrow:2018yni,  Isono:2019ihz,  Bautista:2019qxj,  Gillioz:2019lgs,  Baumann:2019oyu,  Baumann:2020dch,  Jain:2020rmw,  Jain:2020puw,  Jain:2021wyn,  Jain:2021qcl,  Jain:2021vrv,  Baumann:2021fxj,  Jain:2021gwa,  Jain:2021whr,  Gillioz:2022yze,  Marotta:2022jrp,  Jain:2023idr,  Bzowski:2023jwt,  S:2024zqp,  Jain:2024bza,  Aharony:2024nqs,  Marotta:2024sce,  Coriano:2024ssu,  Gillioz:2025yfb,  S:2025pmh, De:2026stn}, twistor space \cite{Baumann:2024ttn,  Bala:2025gmz,  Bala:2025jbh,  Bala:2025qxr,  Rost:2025uyj,  S:2025pmh,  Mazumdar:2025egx,  CarrilloGonzalez:2025qjk, Ansari:2025fvi} ultimately culminating in the Grassmannian formulation \cite{Arundine:2026fbr, De:2026shn, Bala:2026hdm,Bala:2026bdx,Huang:2026tsh, Bala:2026trw,Arundine:2026myr, Bala:2026lvw}. Several of the important developments are discussed in the lecture notes \cite{S:2025pmh}.

Given the utility of the Grassmannian formalism for CFT$_3$ and (A)dS$_4$, it is a natural question to test its usefulness in the higher dimensional CFT$_4$, (A)dS$_5$ context. From the conformal field theory perspective, it could potentially open up a geometric picture for off-shell quantities such as correlation functions of gauge invariant local operators, analogous to what we have learned about the structure of on-shell scattering amplitudes especially in $\mathcal{N}=4$ super Yang-Mills theory. From the bulk spacetime perspective, it could potentially be applied to the bootstrap of amplitudes in type IIB string theory on AdS$_5\times S^5$ \cite{Alday:2022lkk,Alday:2023jdk,Alday:2023mvu,Alday:2024ksp, Alday:2025bjp,Alday:2025cxr}, given the simplicity of the Grassmannian framework and its ability to make physical and mathematical properties manifest.

This motivates us to study and develop the Grassmannian construction to the setting of four dimensional conformal field theory and holographic correlators in AdS$_5$ spacetime. This program was first initiated earlier this year in \cite{Bala:2026trw}. 
See also \cite{CarrilloGonzalez:2026eum} where the authors develop a ambitwistor space formalism for CFT$_4$ correlators.  It was shown in \cite{Bala:2026trw} that CFT$_4$ correlators involving symmetric traceless conserved currents admit a representation in terms of a symplectic bi-Grassmannian\footnote{It is interesting to note that massive four point amplitudes on the Coulomb branch of $\mathcal{N}=4$ super Yang-Mills theory are also described by a symplectic Grassmannian \cite{Cachazo:2018hqa,Bering:2022tdr,Cortes:2025xmu}.}. It consists of two $n-$dimensional planes $C$ and $\tilde{C}$ in a $2n$ dimensional space with the constraint that they are symplectically orthogonal to each other. Two and three point functions (more precisely, discontinuities thereof) admit a simple rational representation in contrast to their more complicated momentum space counterparts. Given the central importance of CFT$_4$ especially in the context of AdS$_5$/CFT$_4$ holography, there are many interesting avenues for further applications of the Grassmannian framework such as the study of higher point correlators, extension to supersymmetry, developing the twistor framework, describing generic primary operators and more broadly, to set up a conformal bootstrap program in this geometric language. The aim of this paper is to make progress on several of these fronts. Our main results are summarized below.
\begin{itemize}
    \item We develop the study of four point functions in the symplectic bi-Grassmannian. We derive a \textbf{factorization principle} and bootstrap several \textbf{holographic exchange four point functions} involving $\Delta=2$ scalars, photons, fermions, gluons and gravitons.
    \item We discuss and derive the \textbf{twistor space} realization of the construction which has the advantage of making conformal symmetry completely manifest. We also derive the \textbf{Penrose transform} that converts twistor space quantities to position space and classify conformal invariants.
    \item We extend the twistor formalism to superspace by combining twistors and Grassmann variables to form \textbf{super-twistors}. We derive the \textbf{supersymmetric Penrose transform} that relates super-twistor space to position super-space and classify super-conformal invariants. We obtain the \textbf{super-twistor space conformal Grassmannian} by a simple and elegant extension.
    \item Via the half-Fourier transform, we translate our super-twistor space Grassmannian back to spinor-helicity obtaining the spinor-helicity \textbf{super-conformal symplectic bi-Grassmannian}.
    \item We apply our formalism in the context of \textbf{AdS$_5$ $\mathcal{N}=1$ super Yang-Mills theory}. With a few simple assumptions and inputting the fermion and scalar four point functions, we use supersymmetry to determine other component correlators such as the four gluon correlator. The resulting expressions agree with our bootstrapped results serving as a nice consistency check and highlighting the rigidity of the formalism.
    \item Finally, we discuss the construction of correlators in \textbf{AdS$_5$ $\mathcal{N}=1$ supergravity}. 
\end{itemize}
In terms of the big picture, these results and their simplicity provide further evidence towards the usefulness and naturalness of the Grassmannian framework in the study of conformal field theory. 

A detailed outline of the paper is as follows:
\subsection*{Outline of the paper}
In section \ref{sec:BiGrassmannian}, we review the construction of off-shell spinor-helicity variables and discuss the construction of two and three point functions in the Grassmannian. We also adapt the \textbf{bold} spinor notation to our context. We then classify four point scalar invariants and discuss the construction of covariants.  The subject of section \ref{sec:AdS5bootstrap} is using factorization to bootstrap holographic four point functions using three point data. In section \ref{sec:Twistors}, we obtain the twistorial description of our results using the half-Fourier transform. We also derive the Penrose transform and classify conformal invariants. Section \ref{sec:SuperTwistors} deals with the development of super-twistors in this context. We derive the supersymmetric Penrose transform, classify super-conformal invariants and obtain the super-twistor space Grassmannian. Using super-twistors, we perform an inverse half-Fourier transform to obtain the superspace version of the symplectic bi-Grassmannian in spinor helicity variables in section \ref{sec:SuperBiGrassmannian}. We discuss explicitly, the construction of the supersymmetric delta function for $n=2,3,4$ point functions. Finally, sections \ref{sec:SYM} and \ref{sec:SUGRA} respectively deal with the application of our formalism to $\mathcal{N}=1$ SYM theory and $\mathcal{N}=1$ SUGRA in an AdS$_5$ background. We discuss some potentially interesting future directions in section \ref{sec:Discussion}. 
\subsection*{Outline of the appendices}
We also have a few appendices to complement the material in the main-text. In appendix \ref{app:Factorization}, we explicitly derive the factorization principle in the Grassmannian. We then discuss the details of the half-Fourier transform of the Grassmannian integral in appendix \ref{app:HFT}. In appendix \ref{app:AmbiTwistors}, we discuss the connection between our twistor formalism and ambitwistor space. We present a direct derivation of the super-conformal Grassmannian in appendix \ref{app:SHGrassmann}. We speculate on the extension to higher supersymmetry in appendix \ref{app:ExtendedSUSY}.

\section{The Symplectic Bi-Grassmannian}\label{sec:BiGrassmannian}
The central message of \cite{Bala:2026trw} is that $n-$point correlation functions of symmetric traceless conserved currents in $d=4$ can be represented by a symplectic bi-Grassmannian integral. We begin by stating the pieces that make up the construction, namely, spinor helicity variables and various matrices that form the building blocks of the Grassmannian. We then briefly describe two and three point functions and also set up the \textbf{bold} spinor notation. Finally, we set the stage for the analysis of four point functions by classifying the invariant quantities and discussing the construction of covariants.

\subsection{Off-shell spinor-helicity variables in $d=4$}
The arena in which we work is four dimensional Klein space $\mathbb{R}^{2,2}$. The Klein group $SO(2,2)$ is semi-simple and is homomorphic to $SL(2,\mathbb{R})_L\times SL(2,\mathbb{R})_R$. Given a general momentum vector $p^\mu$, we can thus trade it for real spinors $\lambda_{I\alpha}$ and $\tilde{\lambda}_{I\Dot{\alpha}}$ as follows \cite{Bala:2026trw}:
\begin{align}
    p^{\mu}\to p^\mu(\sigma_\mu)_{\alpha\Dot{\alpha}}=p_{\alpha\Dot{\alpha}}=\epsilon^{IJ}\lambda_{I\alpha}\tilde{\lambda}_{J\Dot{\alpha}}.
\end{align}
$\alpha$ and $\Dot{\alpha}$ are indices of the fundamental representation of $SL(2,\mathbb{R})_L$ and $SL(2,\mathbb{R})_R$ respectively. The additional index $I$ denotes the fundamental index of the $SL(2,\mathbb{R})$ subgroup of the little group that keeps $p_{\alpha\Dot{\alpha}}$ invariant. We also note that there is a $GL(1,\mathbb{R})$ rescaling redundancy $\lambda\to r \lambda,\tilde{\lambda}\to \frac{\tilde{\lambda}}{r}, r\in\mathbb{R}-\{0\}$. Thus the full little group is $SL(2,\mathbb{R})\times GL(1,\mathbb{R})$, bringing down the total number of independent components of the pair $\lambda,\tilde{\lambda}$ down from $8$ to $4$, as is appropriate for a generic four-vector $p^\mu$. 

The observables of interest to us are correlation functions of symmetric traceless conserved currents,
\begin{align}
J_s^{(\alpha_1\cdots\alpha_n)(\Dot{\alpha}_1\cdots\Dot{\alpha}_m)}(p^\mu),m+n=2s.
\end{align}
For $s\in\mathbb{Z}_{>0}$ we have $m=n=s$ whereas for $s\in\mathbb{Z}_{\ge 0}+\frac{1}{2}$ we have $|m-n|=1$.
In the spinor-helicity formalism, we now trade the spacetime spinor indices $\alpha,\Dot{\alpha}$ for the $SL(2,\mathbb{R})$ little group indices. We define the projectors\footnote{Throughout, $I$ denotes the row index of $\lambda$, $\tilde{\lambda}$ whereas the spinor index $\alpha$ or $\Dot{\alpha}$ is a column index. Thus, strictly speaking we should write down $\lambda^I_{~\alpha}$ rather than $\lambda^I_\alpha$ but for brevity we choose to represent it the latter way with the notation understood as explained here.},
\begin{align}\label{polarizationspinors}
    \zeta_{\alpha}^{I}=\frac{\lambda_\alpha^I}{\sqrt{\text{Det}(\lambda)}},\tilde{\zeta}_{\Dot{\alpha}}^{I}=\frac{\tilde{\lambda}_{\Dot{\alpha}}^I}{\sqrt{\text{Det}(\tilde\lambda)}}.
\end{align}
Using these quantities we define,
\begin{align}\label{spinsIJ}
    J_s^{(I_1,\cdots,I_m,J_1,\cdots, J_{n})}(\lambda,\tilde{\lambda})=\zeta_{\alpha_1}^{(I_1}\cdots \zeta_{\alpha_m}^{I_m}\tilde{\zeta}_{\Dot{\alpha}_1}^{J_1}\cdots\tilde{\zeta}_{\Dot{\alpha}_n}^{J_n)}J_s^{\alpha_1,\cdots,\alpha_m,\Dot{\alpha}_1,\cdots,\Dot{\alpha}_n}(p^\mu),~~m+n=2s.
\end{align}
 The total number of independent components is clearly $m+n+1=2s+1$ which for $s=1,2,\cdots$ equals $3,5,\cdots$ which is the correct counting.

Further, we normalize these currents (a momentum dependent rescaling) to ensure a simple action of the special conformal generator \cite{Bala:2026trw}:
\begin{align}\label{Jsrescaled}
    \hat{J}^{(I_1,\cdots,I_m,J_1,\cdots J_{n})}_s(\lambda,\tilde{\lambda})=\frac{J_s^{(I_1,\cdots,I_m,J_1,\cdots, J_{n})}(\lambda,\tilde{\lambda})}{\text{Det}(\lambda)^{\frac{m}{2}}\text{Det}(\tilde{\lambda})^{\frac{n}{2}}},~~m+n=2 s.
\end{align}
This results in the following transformation under the $SL(2,\mathbb{R})\times GL(1,\mathbb{R})$ little group:
\begin{align}\label{littlegroupscaling}
    \hat{J}^{(I_1,\cdots,I_m,J_1,\cdots J_{n})}_s(\lambda,\tilde{\lambda})\to \frac{1}{r^{m-n}}\big(S^{I_1}_{I_1'}\cdots S^{I_m}_{I_m'}\big)\big(S^{J_1}_{J_1'}\cdots S^{J_n}_{J_n'}\big)\hat{J}^{(I_1',\cdots,I_m',J_1',\cdots J_{n}')}_s(\lambda,\tilde{\lambda})
\end{align}
$r\in\mathbb{R}/\{0\}$ is the $GL(1,\mathbb{R})$ scaling parameter and $S_I^{J}$ are the $SL(2,\mathbb{R})$ matrices. For more details on this construction please see \cite{Bala:2026trw}. We denote the correlators of these currents as,
\begin{align}
    \psi_n^{I_1,J_1,\cdots,I_n,J_n,\cdots}=\langle \hat{J}_{s_1}^{I_1 J_1\cdots}\cdots \hat{J}_{s_n}^{I_n J_n\cdots}\rangle.
\end{align} 

\subsection{The Grassmannian construction}
Having defined the spinor-helicity variables, we then repackage them all in the following $2n\times 2$ vectors,
\begin{align}\label{LambdaandLambdatilde}
    \Lambda_{i I}^{\alpha}\equiv\Lambda=\begin{pmatrix}
        (\lambda_{1})_1^1&(\lambda_{1})_1^2\\
        (\lambda_{1})_2^1&(\lambda_{1})_2^2\\
        \vdots&\vdots\\
        (\lambda_{n})_1^1&(\lambda_{n})_1^2\\
        (\lambda_{n})_2^1&(\lambda_{n})_2^2
    \end{pmatrix},\qquad\tilde{\Lambda}_{iI}^{\Dot{\alpha}}\equiv\tilde{\Lambda}=\begin{pmatrix}
        (\tilde{\lambda}_{1})_1^1&(\tilde{\lambda}_{1})_1^2\\
        (\tilde{\lambda}_{1})_2^1&(\tilde{\lambda}_{1})_2^2\\
        \vdots&\vdots\\
        (\tilde{\lambda}_{n})_1^1&(\tilde{\lambda}_{n})_1^2\\
        (\tilde{\lambda}_{n})_2^1&(\tilde{\lambda}_{n})_2^2
    \end{pmatrix}.
\end{align}
The statement of momentum conservation in terms of these quantities is,
\begin{align}\label{momentumcons}
    \Lambda_{iI}^{\alpha}(\Omega^{ij})^{IJ}\tilde{\Lambda}_{jJ}^{\Dot{\alpha}}\equiv\Lambda^T\cdot\Omega\cdot\tilde{\Lambda}=\sum_{i=1}^{n}\lambda_{i I}^{\alpha}\tilde{\lambda}_{i}^{I\dot\alpha}=0,
\end{align}
where the $2n\times 2n$ symplectic form $\Omega$ takes the form,
\begin{align}\label{Omega}(\Omega^{ij})^{IJ} = \delta^{ij}\,\epsilon^{IJ} = \begin{pmatrix} \epsilon & 0 & \cdots & 0 \\ 0 & \epsilon & \cdots & 0 \\ \vdots & \vdots & \ddots & \vdots \\ 0 & 0 & \cdots & \epsilon \end{pmatrix}, \quad \epsilon = \begin{pmatrix} 0 & 1 \\ -1 & 0 \end{pmatrix}.\end{align}
 We also require $n\times 2n$ matrices $\tilde{C}$ and $C$ that are symplectically orthogonal to each other and also to $\Lambda$ and $\tilde{\Lambda}$ respectively. 
\begin{align}\label{CandCtilde}
C_{ij,I} \equiv C =
\left(
\begin{array}{cc|cc|c|cc}
1_1 & 1_2 & 2_1 & 2_2 & \cdots & n_1 & n_2 \\
\hline
(c_{11})_1 & (c_{11})_2 & (c_{12})_1 & (c_{12})_2 & \cdots & (c_{1n})_1 & (c_{1n})_2 \\
(c_{21})_1 & (c_{21})_2 & (c_{22})_1 & (c_{22})_2 & \cdots & (c_{2n})_1 & (c_{2n})_2 \\
\vdots & \vdots & \vdots & \vdots & \ddots & \vdots & \vdots \\
(c_{n1})_1 & (c_{n1})_2 & (c_{n2})_1 & (c_{n2})_2 & \cdots & (c_{nn})_1 & (c_{nn})_2
\end{array}
\right),
\end{align}
and,
\begin{align}
\tilde C_{ij,I} \equiv \tilde C =
\left(
\begin{array}{cc|cc|c|cc}
\tilde{1}_1 & \tilde{1}_2 & \tilde{2}_1 & \tilde{2}_2 & \cdots & \tilde{n}_1 & \tilde{n}_2 \\
\hline
(\tilde c_{11})_1 & (\tilde c_{11})_2 & (\tilde c_{12})_1 & (\tilde c_{12})_2 & \cdots & (\tilde c_{1n})_1 & (\tilde c_{1n})_2 \\
(\tilde c_{21})_1 & (\tilde c_{21})_2 & (\tilde c_{22})_1 & (\tilde c_{22})_2 & \cdots & (\tilde c_{2n})_1 & (\tilde c_{2n})_2 \\
\vdots & \vdots & \vdots & \vdots & \ddots & \vdots & \vdots \\
(\tilde c_{n1})_1 & (\tilde c_{n1})_2 & (\tilde c_{n2})_1 & (\tilde c_{n2})_2 & \cdots & (\tilde c_{nn})_1 & (\tilde c_{nn})_2
\end{array}
\right).
\end{align}
$i=1,\cdots,n$ denotes the row index whereas the pair $(j\in\{1,\cdots,n\},I\in\{1,2\})$ denotes the combined column index. The advantage of this notation is that we can maintain the manifest $SL(2)$ covariance of correlation functions.
The symplectic bi-Grassmannian is then given by,
\begin{align}\label{GrassmannianSH}
\psi_n^{\{I_1,\dots\},\dots,\{I_n,\dots\}}
&=
\int \frac{d^{n\times 2n}C}{\text{Vol}(\text{GL}(n))}
\int\frac{d^{n\times 2n}\tilde C}{\text{Vol}(\text{GL}(n))}\delta^{n\times n}(\tilde C_{ij,I}(\Omega^{jk})^{IJ} C_{kl,J})\notag\\&\times 
\delta^{n\times 2}(\tilde{C}_{ij,I} (\Omega^{jk})^{IJ} \Lambda_{k J}^{\alpha})
\delta^{n\times 2}(C_{ij,I} (\Omega^{jk})^{IJ} \tilde{\Lambda}_{k J}^{\Dot{\alpha}})\,
A_n^{\{I_1,\dots\},..,\{I_n,\dots\}}(C,\tilde C)\notag\\
&=\int DC \int D\tilde{C}\delta^{n\times n}(\tilde{C}\cdot\Omega\cdot C^T)\delta^{n\times 2}(\tilde{C}\cdot\Omega\cdot\Lambda)\delta^{n\times 2}(C\cdot\Omega\cdot\tilde{\Lambda})A_n^{\{I_1,\cdots\},\cdots,\{I_n,\cdots\}},
\end{align}
where $DC=\frac{d^{n\times 2n}C}{\text{Vol}(GL(n))}$ and $D\tilde{C}=\frac{d^{n\times 2n}\tilde{C}}{\text{Vol}(GL(n))}$.
While the spinor-helicity variables already make the physical independent components of the symmetric traceless conserved currents manifest, this representation ensures that the conformal Ward identities are satisfied\cite{Bala:2026trw}. An important point to note is the $GL(n)\times GL(n)$ redundancy of this description that  requires that,
\begin{align}
    A_n(G C,\tilde{G} C)\to \frac{1}{\text{Det}(G)^{n-2}\text{Det}(\tilde{G})^{n-2}}A_n(C,\tilde{C}).
\end{align}

We construct $A_n(C,\tilde{C})$ using the $GL(n)\times GL(n)$ covariant minors of $C$ and $\tilde{C}$. As we see from \eqref{CandCtilde}, the first two columns of $C$ are labeled by $1_I$, the second two by $2_I$ and so on. For $\tilde{C}$, the first two columns are labeled by $\tilde{1}_I$, the second two by $\tilde{2}_I$ etc.. This notation owes itself to the way they transform under the $GL(2)_1\times GL(2)_2\cdots$ little-group transformations of the external spinor-helicity data. This allows us to construct  $A_n^{\{I_1,\cdots\},\cdots,\{I_n,\cdots\}}(C,\tilde{C})$ in a manifestly $GL(2,\mathbb{R})$ covariant manner by constructing it out of minors that themselves carry $SL(2,\mathbb{R})$ indices that ensure correct transformation properties for each operator viz \eqref{littlegroupscaling}. Further, we construct minors such that the correct $GL(1)$ properties of each operator \eqref{littlegroupscaling} are obeyed. Under a little group transformation \eqref{littlegroupscaling} we have,
\begin{align}\label{ianditildetransforms}
    &(\cdots, i^{I},\cdots)\to r_i (S_i)^{I}_{J}(\cdots,i^{J},\cdots),\notag\\
    &(\cdots, \tilde{i}^{I},\cdots)\to \frac{1}{r_i}(S_i)^{I}_{J}(\cdots,i^{J},\cdots),
\end{align}
Finally, we note that after stripping off the momentum conserving delta function, there are a total of $(n-2)^2$ non-trivial integrals that need to be performed.

We now discuss the form of two and three point functions \cite{Bala:2026trw} that will be required in the subsequent sections.  In addition to the notation of \cite{Bala:2026trw}, we also use this as an opportunity to set up a bold spinor notation similar to that of \cite{Arkani-Hamed:2017jhn} in our context which will greatly simplify the notation and ease of viewing the formulae.

\subsection{Two and three point functions}
At the level of two points we have,
\begin{align}
    \psi_2=\int DC\int D\tilde{C}~\delta^{2\times 2}(C\cdot\Omega\cdot\tilde{C}^T)\delta^{2\times 2}(C\cdot\Omega\cdot\tilde{\Lambda})\delta^{2\times 2}(\tilde{C}\cdot\Omega\cdot\Lambda)A_2(C,\tilde{C}),
\end{align}
where we suppress possible little group indices for the external operators. The scalar invariant we can form at two points is $(1_I 2_J)(\tilde{1}^I\tilde{2}^J)$. Two point functions of integer spin$-s$ symmetric traceless conserved currents take the form,
\begin{align}
    A_2^{I_1\cdots I_{2s},J_1\cdots J_{2s}}=\frac{(1^{I_1}2^{J_1})(\tilde{1}^{I_2}\tilde{2}^{J_2})\cdots (1^{I_{2s-1}}2^{J_{2s-1}})(\tilde{1}^{I_{2s}}\tilde{2}^{J_{2s}})}{\big((1_I 2_J)(\tilde{1}^I\tilde{2}^J)\big)^s},
\end{align}
with an implicit symmetrization in $(I_1,\cdots I_{2s})$ and $(J_1\cdots J_{2s})$. We now adapt the bold notation of \cite{Arkani-Hamed:2017jhn} for notational ease.
\begin{align}\label{sstwopoint}
    A_2^{(s,s)}=\bigg(\frac{(\mathbf{1}\mathbf{2})(\mathbf{\tilde{1}}\mathbf{\tilde{2}})}{(\mathbf{1}\mathbf{2})\cdot(\mathbf{\tilde{1}}\mathbf{\tilde{2}})}\bigg)^s
\end{align}
In the numerator we simply have an ordinary product $(\mathbf{1}\mathbf{2})(\mathbf{\tilde{1}}\mathbf{\tilde{2}})$ which indicates that these quantities carry free indices. There are $s$ copies of this product in the numerator so we need to distribute $(I_1,\cdots I_{2s})$ to $\mathbf{1}$ and $\mathbf{\tilde{1}}$ and similarly $(J_1,\cdots J_{2s})$ to $\mathbf{2}$ and $\mathbf{\tilde{2}}$. This process is unambiguous since the sets of indices are completely symmetrized for the symmetric traceless conserved currents. As for the denominator, we have a dot product $(\mathbf{1}\mathbf{2})\cdot(\mathbf{\tilde{1}}\mathbf{\tilde{2}})$. Since there is an independent $SL(2)$ redundancy for each label, it is unambiguous that $\mathbf{1}$ must contract with $\mathbf{\tilde{1}}$ and $\mathbf{2}$ must be contracted with $\mathbf{\tilde{2}}$. We will interchangeably use the bold and  explicit index notation in this paper.
Let us now proceed to $n=3$. \begin{align}
    \psi_3=\int DC\int D\tilde{C}~\delta^{3\times 3}(C\cdot\Omega\cdot\tilde{C}^T)\delta^{3\times 2}(C\cdot\Omega\cdot\tilde{\Lambda})\delta^{3\times 2}(\tilde{C}\cdot\Omega\cdot\Lambda)A_3(C,\tilde{C}),
\end{align}
where we have yet again suppressed possible little group indices for the external operators. The scalar that we form at three points is\footnote{In the index notation this equals $(1^I 1^J 2^K)(\tilde{1}_I \tilde{1}_J\tilde{2}_K)+(2^I 2^J 3^K)(\tilde{2}_I\tilde{2}_J\tilde{3}_K)+(3^I 3^J 1^J)(\tilde{3}_I \tilde{3}_J\tilde{1}_K)$.},
\begin{align}
    \mathcal{K}=(\mathbf{1}\mathbf{1}\mathbf{2})\cdot(\mathbf{\tilde{1}}\mathbf{\tilde{1}}\mathbf{\tilde{2}})+(\mathbf{2}\mathbf{2}\mathbf{3})\cdot(\mathbf{\tilde{2}}\mathbf{\tilde{2}}\mathbf{\tilde{3}})+(\mathbf{3}\mathbf{3}\mathbf{1})\cdot(\mathbf{\tilde{3}}\mathbf{\tilde{3}}\mathbf{\tilde{1}}).
\end{align}
The scalar three point function takes the form\footnote{This and the three point functions to follow are Wightman functions with the middle operator having a space-like momentum. They are obtained from their Euclidean counterparts by taking a discontinuity with respect to $p_1^2$ and then $p_3^2$. The Euclidean versions of many of the CFT$_4$ correlators we consider are generally UV divergent and require renormalization. However, the discontinuity removes the divergent parts and as a consequence, the resulting correlator does not require renormalization. },
\begin{align}\label{A3scalar}
    A_3^{\varphi,\varphi,\varphi}=\frac{1}{|\mathcal{K}|}\to\frac{1}{\mathcal{K}+i\epsilon}.
\end{align}
We put an absolute value to maintain Bose symmetry and the reality property of the correlator. However, in practice, we avoid the absolute value and rather add an $i\epsilon$ prescription while performing integrals to convert the Grassmannian integrands to spinor helicity variables. Let us now list other examples from \cite{Bala:2026trw} while using our new bold spinor notation. Below, $g$ stands for a spin$-1$ current, $G$ stands for the stress tensor and $\chi$ and $\tilde{\chi}$ stand for left handed and right handed Weyl spinors respectively. For a spin$-1$ current and two $\Delta=2$ scalars we obtain,
\begin{align}
    A_3^{g,\varphi,\varphi}&=\frac{(\mathbf{1}\mathbf{2}\mathbf{2})\cdot(\mathbf{\tilde{1}\mathbf{\tilde{2}}\mathbf{\tilde{2}})}}{\mathcal{K}^2},~\text{no contraction between}~\mathbf{1}~\text{and}~\mathbf{\tilde{1}}\notag\\
    &=\frac{(\mathbf{1}\mathbf{2}\cdot\mathbf{2})(\mathbf{\tilde{1}}\mathbf{\tilde{2}}\cdot\mathbf{\tilde{2}})}{2\mathcal{K}^2}.
\end{align}
In going from the first line to the second we introduced the notation $(\mathbf{1}\mathbf{2}\cdot\mathbf{2})$ which stands for $2(\mathbf{1} 2_1 2_2)$. This is how we deal with partially contracted indices in this notation.
Similarly,
\begin{align}
    A_3^{G,\varphi,\varphi}=\frac{\big((\mathbf{1}\mathbf{2}\cdot\mathbf{2})(\mathbf{\tilde{1}\mathbf{\tilde{2}}\cdot\mathbf{\tilde{2}})}\big)^2}{\mathcal{K}^3}.
\end{align}
The generalization to general spin is,
\begin{align}
    A_3^{s,\varphi,\varphi}=\frac{\big((\mathbf{1}\mathbf{2}\cdot\mathbf{2})\cdot(\mathbf{\tilde{1}\mathbf{\tilde{2}}\cdot\mathbf{\tilde{2}})}\big)^s}{\mathcal{K}^{s+1}}.
\end{align}
For two spin$-\frac{1}{2}$ operators and a scalar we find,
\begin{align}
    A_3^{\chi,\tilde{\chi},\varphi}=\frac{(\mathbf{\tilde{1}}\mathbf{\tilde{2}}\mathbf{\tilde{3}})\cdot(\mathbf{2}\cdot\mathbf{2}\mathbf{3})}{\mathcal{K}^2}.
\end{align}
Since there is a contraction $\mathbf{2}\cdot\mathbf{2}$ in the second term in the numerator, the only allowed contraction between it and the first term is between $\mathbf{\tilde{3}}$ and $\mathbf{3}$. A similar notation is needed for a correlator with two spin$-1$ currents and a scalar.
\begin{align}
    A_3^{g,g,\varphi}=\frac{(\mathbf{\tilde{1}}\mathbf{\tilde{2}}\mathbf{\tilde{3}})\cdot(\mathbf{2}\cdot\mathbf{2}\mathbf{3})(\mathbf{1}\mathbf{2}\mathbf{3})\cdot(\mathbf{\tilde{2}}\cdot\mathbf{\tilde{2}}\mathbf{\tilde{3}})}{\mathcal{K}^3}.
\end{align}
For a correlator of three spin$-1$ currents (which must be non-Abelian as a consequence of Furry's theorem) we have two possible correlators. These correspond to in the holographic sense to a bulk AdS$_5$ Yang-Mills and $F^3$ interaction. We are not considering parity odd correlators at the moment but the generalization should not be difficult. The Yang-Mills correlator is given by,
\begin{align}\label{A3YM}
    A_{3,YM}^{g,g,g}=\frac{(\mathbf{1}\mathbf{2}\mathbf{3})(\mathbf{\tilde{1}}\mathbf{\tilde{2}}\mathbf{\tilde{3}})}{\mathcal{K}^2}.
\end{align}
For Einstein gravity we have,
\begin{align}
    A_{3,EG}^{G,G,G}=\frac{\big((\mathbf{1}\mathbf{2}\mathbf{3})(\mathbf{\tilde{1}}\mathbf{\tilde{2}}\mathbf{\tilde{3}})\big)^2}{\mathcal{K}^3}=\mathcal{K}(A_{3,YM}^{g,g,g})^2.
\end{align}
Similarly, we find simple rational expressions for other three point correlators. We now turn to four points.
\subsection{Four point kinematics}
At $n=4$, we have (suppressing the $SL(2,\mathbb{R})$ indices of the external currents),
\begin{align}
    \psi_4=\int DC\int D\tilde{C}~\delta^{4\times 4}(C\cdot \Omega\cdot\tilde{C}^T)\delta^{4\times 2}(C.\Omega\cdot\tilde{\Lambda}^T)\delta^{4\times 2}(\tilde{C}\cdot \Omega\cdot\Lambda^T)A_4.
\end{align}
We define the following set of the natural six (external $SL(2,\mathbb{R})_i$) invariants:
\begin{align}\label{mandelstam1}
    &S=(1_I 1^I 2_J 2^J),~~\tilde{S}=(\tilde{1}_I\tilde{1}^{I}\tilde{2}_J \tilde{2}^J),\notag\\
     &T=(1_I 1^I 4_J 4^J),~~\tilde{T}=(\tilde{1}_I\tilde{1}^{I}\tilde{4}_J \tilde{4}^J),\notag\\ &U=(1_I 1^I 3_J 3^J),~~\tilde{U}=(\tilde{1}_I\tilde{1}^{I}\tilde{3}_J \tilde{3}^J).
\end{align}
In the bold notation these quantities are given by,
\begin{align}\label{mandelstam1bold}
    &S=(\mathbf{1}\cdot\mathbf{1}\mathbf{2}\cdot\mathbf{2}),~~\tilde{S}=(\mathbf{\tilde{1}}\cdot\mathbf{\tilde{1}}\mathbf{\tilde{2}}\cdot\mathbf{\tilde{2}}),\notag\\
    &T=(\mathbf{1}\cdot\mathbf{1}\mathbf{4}\cdot\mathbf{4}),~~\tilde{T}=(\mathbf{\tilde{1}}\cdot\mathbf{\tilde{1}}\mathbf{\tilde{4}}\cdot\mathbf{\tilde{4}}),\notag\\
    &U=(\mathbf{1}\cdot\mathbf{1}\mathbf{3}\cdot\mathbf{3}),~~\tilde{U}=(\mathbf{\tilde{1}}\cdot\mathbf{\tilde{1}}\mathbf{\tilde{3}}\cdot\mathbf{\tilde{3}}).
\end{align}
To ensure the external $GL(1,\mathbb{R})_i$ invariance is met we define the quadratics\footnote{We used the elementary identities $(\mathbf{1}\cdot\mathbf{1}\mathbf{2}\cdot\mathbf{2})(\mathbf{\tilde{1}}\cdot\mathbf{\tilde{1}}\mathbf{\tilde{2}}\cdot\mathbf{\tilde{2}})=2^4(1_1 1_2 2_1 2_2)(\tilde{1}_1\tilde{1}_2\tilde{2}_1\tilde{2}_2)$ equals $2^2(1_I 1_J 2_K 2_L)(\tilde{1}^I\tilde{1}^J\tilde{2}^K\tilde{2}^L)=4(\mathbf{1}\mathbf{1}\mathbf{2}\mathbf{2})\cdot(\mathbf{\tilde{1}}\mathbf{\tilde{1}}\mathbf{\tilde{2}}\mathbf{\tilde{2}})$},
\begin{align}\label{mandelstammathcal}
    &\mathcal{S}=S \tilde{S}=4(\mathbf{1}\mathbf{1}\mathbf{2}\mathbf{2})\cdot(\mathbf{\tilde{1}}\mathbf{\tilde{1}}\mathbf{\tilde{2}}\mathbf{\tilde{2}}),\notag\\ &\mathcal{T}=T\tilde{T}=4(\mathbf{1}\mathbf{1}\mathbf{4}\mathbf{4})\cdot(\mathbf{\tilde{1}}\mathbf{\tilde{1}}\mathbf{\tilde{4}}\mathbf{\tilde{4}}),\notag\\&\mathcal{U}=U\tilde{U}=4(\mathbf{1}\mathbf{1}\mathbf{3}\mathbf{3})\cdot(\mathbf{\tilde{1}}\mathbf{\tilde{1}}\mathbf{\tilde{3}}\mathbf{\tilde{3}}).
\end{align}
These quantities are $GL(2,\mathbb{R})_i,i=1,2,3,4$ invariant and covariant under the internal $GL(4)\times GL(4)$ gauge symmetry of the Grassmannian. In particular, they transform with a factor of $\text{Det}(G)\text{Det}(\tilde{G})$ for $G,\tilde{G}\in \text{GL}(4,\mathbb{R})\times GL(4,\mathbb{R})$ as is clear from their definition. 

These quantities, defined similarly to the flat space Mandelstam variables, obey the analog of the Mandelstam relation,
\begin{align}
    \mathcal{S}+\mathcal{T}+\mathcal{U}=2(\mathbf{1}\mathbf{2}\mathbf{3}\mathbf{4})\cdot(\mathbf{\tilde{1}}\mathbf{\tilde{2}}\mathbf{\tilde{3}}\mathbf{\tilde{4}}).
\end{align}
Other invariants can be written in terms of these quantities. For example,
\begin{align}
    (\mathbf{1}\mathbf{2}\mathbf{3}\cdot\mathbf{3})\cdot(\mathbf{\tilde{1}}\mathbf{\tilde{2}}\mathbf{\tilde{3}}\cdot\mathbf{\tilde{3}})=(1_I 2_J 3_K 3^K)(\tilde{1}^I \tilde{2}^J \tilde{3}_L \tilde{3}^L)=16\bigg(\mathcal{S}-\mathcal{T}-\mathcal{U}\bigg).
\end{align}
As for covariants, we can  appropriately include minors with $SL(2,\mathbb{R})$ indices. For example, if we were dealing with a spin$-1$ four point function a natural choice for a numerator is,
\begin{align}
    (\mathbf{1}\mathbf{2}\mathbf{3}\mathbf{4})(\mathbf{\tilde{1}}\mathbf{\tilde{2}}\mathbf{\tilde{3}}\mathbf{\tilde{4}})=(1^{I_1}2^{I_2}3^{I_3}4^{I_4})(\tilde{1}^{J_1}\tilde{2}^{J_2}\tilde{3}^{J_3}\tilde{4}^{J_4}),
\end{align}
with the symmetrization implicit.
 On the other hand, an ansatz for a correlator involving two spin$-1$ currents and two scalars is,
\begin{align}
    (\mathbf{1}\mathbf{2}\mathbf{3}\cdot\mathbf{3})(\mathbf{\tilde{1}}\mathbf{\tilde{2}}\mathbf{\tilde{3}}\cdot\mathbf{\tilde{3}})=(1^{I_1}2^{J_1}3_1 3_2)(\tilde{1}^{I_2}\tilde{2}^{J_2}\tilde{3}_1\tilde{3}_2).
\end{align}
Essentially, the manifest $SL(2,\mathbb{R})$ index notation makes it extremely easy and efficient to classify invariants as well as covariants, while the bold notation makes dealing with a multitude of indices manageable. With this, we now turn to the study of explicit examples of four point correlators in the context of AdS$_5$/CFT$_4$.
\section{Factorization and Bootstrap for AdS$_5$ boundary correlators}\label{sec:AdS5bootstrap}
The aim of this section is to bootstrap several examples of holographic correlators in the Grassmannian language. We derive a factorization formula for four point functions in terms of a sum of products of three point functions. We then set up the bootstrap principle that allows us to construct holographic four point correlators up to local contact terms, just from the three point data. We discuss examples of scalars exchanging scalars, spin$-J$ particles, scalar Compton scattering and non-abelian gauge theory. We also discuss the singularity structure of the bootstrapped correlators and discuss its potentially interesting geometric interpretation. We conclude with a discussion of the graviton four point function and a suggestive double copy relation between four point functions in Yang-Mills theory and Einstein gravity.
\subsection{$A_4\sim A_3\times A_3$: The bootstrap principle}
Consider a four point holographic correlator\footnote{More precisely, a discontinuity thereof with respect to some of the external momenta such that it obeys a \textit{homogeneous} current conservation Ward-Takahashi identity. This is the same philosophy taken in \cite{Arundine:2026fbr} for the analogous observables bootstrapped in (A)dS$_4$.} $\psi_4$ that receives contributions from exchanges of massless gauge bosons dual to operators with $\Delta=s+2$ saturating the unitarity bound. We suppress the little group indices of the external operators for notational simplicity. It is easy to show that when we take a discontinuity with respect to $s^2=|p_1+p_2|^2$, the momentum space correlator factorizes\footnote{Essentially, this operation converts the bulk to bulk Feynman propagator into its factorized Wightman counterpart, see for instance \cite{Ansari:2025fvi}.}.
\begin{align}
    \text{Disc}_{s^2}(\psi_4)=\sum_{s}a_s\int\frac{d^4 q}{(2\pi)^4}\psi_3^{\mu_1\cdots\mu_s}(p_1,p_2,q)\Pi_{\mu_1\cdots\mu_s;\nu_1\cdots\nu_s}(q)\psi_3^{\nu_1\cdots\nu_s}(-q,p_3,p_4).
\end{align}
$\Pi$ is a transverse traceless projector and $a_s$ weighs the contribution of a particular spin exchange. We can recast this formula in spinor-helicity variables where the factorization takes the form,
\begin{align}\label{SHDisc}
    \text{Disc}_{s^2}(\psi_4)=\sum_{s}a_s\int\frac{d^4 q}{(2\pi)^4}\psi_3^{I_1\cdots I_{2s}}(p_1,p_2,q)\epsilon_{I_1J_1}\cdots\epsilon_{I_{2s}J_{2s}}\psi_3^{J_1\cdots J_{2s}}(-q,p_3,p_4).
\end{align}
This representation also includes the generalization to half-integer spin exchanges. Given this formula, our strategy is to express $\psi_4$ as well as the $\psi_3$ correlators in the Grassmannian language. We then use this to derive the analogous factorization statement for the Grassmannian integrand. The derivation is quite lengthy and technical and thus we relegate it to appendix \ref{app:Factorization} and discuss only the final result here. Let us however, motivate the construction. The RHS of \eqref{SHDisc} (after solving for $q$ using momentum conservation) is a product of two stripped three point functions and the momentum conserving delta function. Each of these stripped three point functions can be written as an integral over a single parameter in the Grassmannian representation, thus resulting in a double integral required to obtain $\text{Disc}_{s^2}\psi_4$. Following our general counting of $(n-2)^2$ non-trivial integrals for $n$ point functions, we expect to need to perform $4$ for $n=4$. Since the discontinuity of the four point function only requires $2$ integrals, it must localize two integrals. In other words, it must localize two minors to zero. Since we are considering the $s$ channel, the most natural candidates for this are $S$ and $\tilde{S}$ given in \eqref{mandelstam1} and is indeed what the explicit calculation reveals.

After all the dust settles, we obtain the following result for the right hand side of \eqref{SHDisc}:
\begin{align}\label{SHDiscRHS}
    \int DC\int D\tilde{C}~\delta(C\cdot\Omega\cdot\tilde{C}^T)\delta(C\cdot\Omega\cdot\tilde{\Lambda}^T)\delta(\tilde{C}\cdot\Omega\cdot \Lambda^T)\delta(S)\delta(\tilde{S})A_{3L}^{I_1\cdots I_{2s}}A_{3R,I_1\cdots I_{2s}}.
\end{align}
As we derive in the appendix, one can also show (assuming that $A_4$ only has simple poles in $S$ and $\tilde{S}$) that the left hand side of \eqref{SHDisc} equals,
\begin{align}\label{SHDiscLHS}
    \int DC\int D\tilde{C}~\delta(C\cdot\Omega\cdot\tilde{C}^T)\delta(C\cdot\Omega\cdot\tilde{\Lambda}^T)\delta(\tilde{C}\cdot\Omega\cdot \Lambda^T)\delta(S)\delta(\tilde{S})\text{Res}_{\tilde{S}=0}\text{Res}_{S=0}A_4.
\end{align}
Equating \eqref{SHDiscLHS} and \eqref{SHDiscRHS} and accounting for the possibility of multiple currents being exchanged, we find,
\begin{align}\label{A4inA3A3}
    \text{Res}_{\tilde{S}=0}\text{Res}_{S=0}A_4=\sum_{s}a_s A_{3L}^{I_1\cdots I_{2s}}A_{3R,I_1\cdots I_{2s}}.
\end{align}
Given \eqref{A4inA3A3}, and the assumption of simple poles only, we can bootstrap holographic correlators receiving contributions from particles with a given spin, up to quantities that have a vanishing residue at $S=0$ or $\tilde{S}=0$. AdS$_5$ contact diagrams are an example of such quantities.  Contact diagrams do not depend on the exchanged momenta or are polynomial in them and thus their discontinuity with respect to $s^2$ is zero. Therefore, we can recover the exchange correlators up to these  contributions. Thus if we write\footnote{In principle, $B_4$ can contain terms of the form $\frac{a_4(C,\tilde{C})}{S}+\frac{\tilde{a}_4(C,\tilde{C})}{\tilde{S}}+\text{Regular in}~S,\tilde{S}$.},
\begin{align}
    A_4(C,\tilde{C})=\frac{R_4(C,\tilde{C})}{S\tilde{S}}+B_4(C,\tilde{C}),
\end{align}
The factorization formula \eqref{A4inA3A3} allows us to determine the most singular term $R_4$, which is our goal in this paper. 
\subsection{Examples involving scalars and spin$-1$}
We now proceed to discuss various examples of correlators that we construct using \eqref{A4inA3A3}. We begin with the simplest example of scalars exchanging scalars with a $\phi^3$ type AdS$_5$ interaction vertex.
\subsubsection{$\langle \varphi\varphi\varphi\varphi\rangle$ with a scalar exchange}
We have,
\begin{align}
    \text{Res}_{\tilde{S}=0}\text{Res}_{S=0}A_{4,s}^{(\varphi,\varphi|\varphi|\varphi,\varphi)}=A_{3L}^{(\varphi,\varphi,\varphi)}A_{3R}^{(\varphi,\varphi,\varphi)}.
\end{align}
The scalar three point functions are given by\footnote{We suppress the absolute values on the denominators since we supplement these quantities with an $i\epsilon$ prescription while performing the Grassmannian integrals.},
\begin{align}
    A_{3L}^{(\varphi,\varphi,\varphi)}=\frac{1}{\mathcal{K}_L}, A_{3R}^{(\varphi,\varphi,\varphi)}=\frac{1}{\mathcal{K}_R}.
\end{align}
Working in the factorization kinematics where $S=\tilde{S}=0$ and hence $\mathcal{S}=0$, we can show that the following identity holds:
\begin{align}
    \mathcal{K}_L\mathcal{K}_R&=4(1_1 1_2 2_I)(\tilde{1}_1\tilde{1}_2 2^I)(3_1 3_2 4_J)(\tilde{3}_1\tilde{3}_2 4^J)=-\frac{1}{4}(T\tilde{T}-U\tilde{U})=-\frac{1}{4}(\mathcal{T}-\mathcal{U}).
\end{align}
Thus, we have,
\begin{align}
    \text{Res}_{\tilde{S}=0}\text{Res}_{S=0}A_4^{(\varphi\varphi|\varphi|\varphi\varphi)}\propto \frac{1}{\mathcal{T}-\mathcal{U}}.
\end{align}
With the assumption of a simple pole in $\mathcal{S}$, the most natural uplift of this residue to the full expression is,
\begin{align}\label{scalar4pointscalarex}
    A_4^{(\varphi\varphi|\varphi|\varphi\varphi)}=\frac{1}{\mathcal{S}(\mathcal{S}+\mathcal{T}-\mathcal{U})}=\frac{1}{S\tilde{S}(S\tilde{S}+T\tilde{T}-U\tilde{U})}.
\end{align}

\subsubsection{$\langle \varphi\varphi\varphi\varphi\rangle$ with a spinning exchange}
For a spin$-1$ exchange we have,
\begin{align}
    \text{Res}_{\tilde{S}=0}\text{Res}_{S=0}A_4^{(\varphi\varphi|g|\varphi\varphi)}=A_{3L}^{(\varphi\varphi g){IJ}}A^{(g\varphi\varphi)}_{3R,IJ}.
\end{align}
In this case, the three point functions are given by,
\begin{align}
    A_{3L}^{(\varphi\varphi g)IJ}=\frac{(s_L^{(I} 1_1 1_2)(\tilde{s}_L^{J)} \tilde{1}_1\tilde{1}_2)}{\mathcal{K}_L^2},~~A^{(g \varphi\varphi)}_{3R,IJ}=\frac{(s_{R(I}3_1 3_2)(\tilde{s}_{RJ)}\tilde{3}_1\tilde{3}_2)}{\mathcal{K}_R^2}.
\end{align}
One can show that the product of the numerators obeys the following identity:
\begin{align}
    (s_L^{(I} 1_1 1_2)(\tilde{s}_L^{J)} \tilde{1}_1\tilde{1}_2)(s_{R(I}3_1 3_2)(\tilde{s}_{RJ)}\tilde{3}_1\tilde{3}_2)=\frac{1}{32}(T\tilde{T}+U\tilde{U})=\frac{1}{32}\mathcal{T}+\mathcal{U}.
\end{align}
Thus we find,
\begin{align}
    \text{Res}_{S=0}\text{Res}_{\tilde{S}=0}A_4^{(\varphi\varphi|g|\varphi\varphi)}\propto \frac{(\mathcal{T}+\mathcal{U})}{(\mathcal{T}-\mathcal{U})^2}.
\end{align}
The simplest uplift is,
\begin{align}
    A_4^{(\varphi\varphi|g|\varphi\varphi)}=\frac{(\mathcal{T}+\mathcal{U})}{\mathcal{S}(\mathcal{S}+\mathcal{T}-\mathcal{U})^2}=\frac{1}{2(\mathcal{S}+\mathcal{T}-\mathcal{U})^2}C_1^{(1)}\bigg(\frac{\mathcal{T}+\mathcal{U}}{\mathcal{S}}\bigg),
\end{align}
where $\mathcal{C}_n^{(m)}(x)$ denotes the Gegenbauer polynomial. In this form, an uplift to a spin$-J$ exchange is straightforward and reads,
\begin{align}\label{scalarfourpointJexchange}
    A_4^{(\varphi\varphi|J|\varphi\varphi)}=\frac{1}{2^J}\frac{\mathcal{S}^{J-1}}{(\mathcal{S}+\mathcal{T}-\mathcal{U})^{J+1}}C_J^{(1)}\bigg(\frac{\mathcal{T}+\mathcal{U}}{\mathcal{S}}\bigg).
\end{align}
The double residue at $S=0,\tilde{S}=0$ is,
\begin{align}
    \text{Res}_{\tilde{S}=0}\text{Res}_{S=0}A_4^{(J)}=\frac{(\mathcal{T}+\mathcal{U})^J}{(\mathcal{T}-\mathcal{U})^{J+1}}=\frac{\bigg((s_L^{(I} 1_1 1_2)(\tilde{s}_L^{J)} \tilde{1}_1\tilde{1}_2)(s_{R(I}3_1 3_2)(\tilde{s}_{RJ)}\tilde{3}_1\tilde{3}_2)\bigg)^{J}}{(\mathcal{K}_L\mathcal{K}_R)^{J+1}}=A_{3L}^{I_1\cdots I_{2s}}A_{3R,I_1\cdots I_{2s}},
\end{align}
which are indeed the correct expressions for the three point functions with two scalars and a spin $s$ current. Any other solution should be related to this one by AdS$_5$ contact diagrams. For example, choosing Legendre polynomials instead of the Gegenbauer polynomial (and ensuring the normalization such that they have equal residues at $\mathcal{S}=0$) results in a difference (say $J=2$) of $\frac{\mathcal{S}}{\mathcal{S}+\mathcal{T}-\mathcal{U})^3}$ which has no pole in $\mathcal{S}=S\tilde{S}$.
\subsubsection{$\langle \varphi\varphi gg\rangle$ with a scalar exchange}
We now discuss an example where two of the external particles have spin. We first discuss the $s-$channel exchange of a scalar particle. The three point seeds are,
\begin{align}
    &A_3^{\varphi,\varphi,\varphi}=\frac{1}{\mathcal{K}_L},\notag\\
    &A_{3}^{\varphi,g,g}=\frac{(\tilde{3}^{I_3}\tilde{4}^{I_4}\tilde{s}_R^I)(4_1 4_2 s_{R I})(3^{J_3}4^{J_4}s_{RJ})(\tilde{4}_1\tilde{4}_2\tilde{s}_R^I)}
{\mathcal{K}_R^3}
\end{align}
After using various identities to re-express the three point product in terms of four point minors we find,
\begin{align}
    A_3^{\varphi,\varphi,\varphi}A_{3}^{\varphi,g,g}=\frac{(3^{I_3}1_1 1_2 2_I)(\tilde{3}^{J_3}\tilde{1}_1\tilde{1}_2 2^I)(4^{I_4}1_1 1_2 2_I)(\tilde{4}^{J_4}\tilde{1}_1\tilde{1}_2 2^I)}{(\mathcal{T}-\mathcal{U})^3}.
\end{align}
Uplifting this to the full exchange contribution we obtain,
\begin{align}\label{phiphiggwithphiexchange}
    A_4^{(\varphi\varphi|\varphi|gg)}=\frac{(3^{I_3}1_1 1_2 2_I)(\tilde{3}^{J_3}\tilde{1}_1\tilde{1}_2 2^I)(4^{I_4}1_1 1_2 2_I)(\tilde{4}^{J_4}\tilde{1}_1\tilde{1}_2 2^I)}{\mathcal{S}(\mathcal{S}+\mathcal{T}-\mathcal{U})^3}
\end{align}
\subsubsection{$\langle \varphi\varphi gg\rangle$ with a spinning exchange}
We now consider a gluon exchange. For the gluon three point function we consider the Yang-Mills three point interaction. We have,
\begin{align}
    A_{3L}^{\varphi\varphi,g}=\frac{(s_L^{(I}1_1 1_2)(\tilde{s}_{L}^{J)}\tilde{1}_1\tilde{1}_2)}{\mathcal{K}_L^2}, A_{3R}^{ggg}=\frac{(3^{I_3}4^{I_4}s_{R(I})(\tilde{3}^{J_3}\tilde{4}^{J_4}\tilde{s}_{RJ)})}{\mathcal{K}_R^2}.
\end{align}
Following the same method as for the previous examples, we find the below expression that has the product of the above two three point functions as its residue.
\begin{align}\label{phiphiggwithgexchange}
    A_{4s}^{\varphi\varphi|g|gg}=\frac{(3^{I_3}4^{I_4}1_1 1_2)(\tilde{3}^{J_3}\tilde{4}^{J_4}\tilde{1}_1\tilde{1}_2)}{\mathcal{S}(\mathcal{S}+\mathcal{T}-\mathcal{U})^2}.
\end{align}
We now proceed to a correlator of four gluons.
\subsubsection{$\langle gggg\rangle$: AdS$_5$ $SU(N)$ Yang-Mills gluon correlator}
For the Yang-Mills theory gluon four point function, the relevant three point data is,
\begin{align}
    A_{3L}^{I_1 J_1,I_2 J_2,I J}=\frac{(1^{I_1}2^{I_2}s_L^I)(\tilde{1}^{J_1}\tilde{2}^{J_2}\tilde{s}_L^{J})}{\mathcal{K}_L^2},A_{3R,IJ}^{I_3J_3,I_4J_4}=\frac{(3^{I_3}4^{I_4}s_{RI})(\tilde{3}^{J_3}\tilde{4}^{J_4}\tilde{s}_{RJ})}{\mathcal{K}_R^2}.
\end{align}
We find the following elegant identity that relates the product of the $3\times 3$ minors to their $4\times 4$ counterparts:
\begin{align}
    (1^{I_1}2^{I_2}s_L^I)(\tilde{1}^{J_1}\tilde{2}^{J_2}\tilde{s}_L^{J})(3^{I_3}4^{I_4}s_{RI})(\tilde{3}^{J_3}\tilde{4}^{J_4}\tilde{s}_{RJ})=(1^{I_1}2^{I_2}3^{I_3}4^{I_4})(\tilde{1}^{J_1}\tilde{2}^{J_2}\tilde{3}^{J_3}\tilde{4}^{J_4}).
\end{align}
Thus we have,
\begin{align}
    \text{Res}_{\tilde{S}=0}\text{Res}_{S=0}A_4^{(gg|g|gg)}=\frac{(1^{I_1}2^{I_2}3^{I_3}4^{I_4})(\tilde{1}^{J_1}\tilde{2}^{J_2}\tilde{3}^{J_3}\tilde{4}^{J_4})}{(\mathcal{T}-\mathcal{U})^2}.
\end{align}
The simplest and most natural uplift to the full $s-$channel correlator is,
\begin{align}\label{ggggYMexchangeschannel}
    A_{4,s}^{(gg|g|gg)}=\frac{(1^{I_1}2^{I_2}3^{I_3}4^{I_4})(\tilde{1}^{J_1}\tilde{2}^{J_2}\tilde{3}^{J_3}\tilde{4}^{J_4})}{\mathcal{S}(\mathcal{S}+\mathcal{T}-\mathcal{U})^2}=\frac{(\mathbf{1}\mathbf{2}\mathbf{3}\mathbf{4})(\mathbf{\tilde{1}}\mathbf{\tilde{2}}\mathbf{\tilde{3}}\mathbf{\tilde{4}})}{\mathcal{S}(\mathcal{S}+\mathcal{T}-\mathcal{U})^2}.
\end{align}
For the colour ordered gluon correlator we need to include the contribution of the $t-$channel thus yielding,
\begin{align}\label{ggggYMexchange}
    A_{4}^{(gg|g|gg)}(1234)=\frac{(\mathbf{1}\mathbf{2}\mathbf{3}\mathbf{4})(\mathbf{\tilde{1}}\mathbf{\tilde{2}}\mathbf{\tilde{3}}\mathbf{\tilde{4}})}{(\mathcal{S}+\mathcal{T}-\mathcal{U})^2}\bigg(\frac{1}{\mathcal{S}}+\frac{1}{\mathcal{T}}\bigg).
\end{align}
We will later find that this expression actually follows from the super-conformal Ward identities that relate this correlator to a four-fermion and a four-scalar correlator, thus serving as an additional check on its consistency.
\subsection{The singularity structure}
All our results have the denominator $\mathcal{S}+\mathcal{T}-\mathcal{U}$. Note that the minus sign in $\mathcal{U}$ singles out $(1,3)$ and indicates that this correlator potentially corresponds to a discontinuity with respect to $p_1^2,p_3^2$ (or equivalently a discontinuity with respect to $p_2^2,p_4^2$). Let us add some more weight to this argument focusing on the scalar four point function due to a scalar exchange \eqref{scalar4pointscalarex}. The three point Grassmann data $A_{3L}^{\varphi\varphi\varphi}, A_{3R}^{\varphi\varphi\varphi}$ we used corresponds to Wightman functions with the middle operator having space-like momenta \cite{Bala:2026trw}. This corresponds to a discontinuity of the Euclidean correlator with respect to momenta magnitude squares of the first and third operator.
\begin{align}
    &\psi_{3L}^{\varphi,\varphi,\varphi}(p_1,p_2,s)=\int_0^\infty dz~z~J_0(p_1 z)K_0(p_2 z)J_0(s z)=\frac{1}{\sqrt{-\mathcal{J}^2(p_1,p_2,s)}},\notag\\
    &\psi_{3R}^{\varphi,\varphi,\varphi}(p_3,p_4,s)=\int_0^\infty dz~z~J_0(p_3 z)K_0(p_4 z)J_0(s z)=\frac{1}{\sqrt{-\mathcal{J}^2(p_3,p_4,s)}},
\end{align}
where,
\begin{align}
    \mathcal{J}^2(a,b,c)=(a+b+c)(a+b-c)(a-b+c)(-a+b+c).
\end{align}
The stitching operation in momentum space corresponds to the operation,
\begin{align}
    &\psi_{4s}^{\varphi,\varphi|\varphi|\varphi,\varphi}=-i\int_0^\infty \frac{k~dk}{k^2+s^2-i\epsilon}\psi_{3L}^{\varphi,\varphi,\varphi}(p_1,p_2,k)\psi_{3R}^{\varphi,\varphi,\varphi}(p_3,p_4,k)\notag\\
    &=\int_0^\infty dz~\sqrt{z}\int_0^\infty dz'~\sqrt{z'}J_0(p_1 z)K_0(p_2 z)\mathcal{G}_{F}(z,z',s)J_0(p_3 z')K_0(p_4 z'),
\end{align}
where we identified the spectral representation of the Feynman propagator,
\begin{align}
    \mathcal{G}_F(z,z',s)=-i\int_0^\infty \frac{k dk}{k^2+s^2-i\epsilon}\sqrt{z}\sqrt{z'}J_0(k z)J_0(k z').
\end{align}
The operation of reconstructing the double residue at $S=0,\tilde{S}=0$ in the Grassmannian corresponds to the above operation in momentum space. The Euclidean AdS$_5$ four point function is given by,
\begin{align}
    \psi_{4s,\text{Euclid}}^{\varphi,\varphi|\varphi|\varphi,\varphi}=\int_0^\infty dz\sqrt{z}\int_0^\infty dz'\sqrt{z'}K_0(p_1 z)K_0(p_2 z)\mathcal{G}_F(z,z',s)K_0(p_3 z)K_0(p_4 z).
\end{align}
Therefore\footnote{Note that we also analytically continue the momenta $p_1^\mu,p_3^\mu$ to be time-like in this operation implicitly.},
\begin{align}
    \psi_{4s}^{\varphi,\varphi|\varphi|\varphi,\varphi}=\text{Disc}_{p_1^2}\text{Disc}_{p_3^2}\psi_{4s,\text{Euclid}}^{\varphi,\varphi|\varphi|\varphi,\varphi}.
\end{align}
We used the formula,
\begin{align}
    \text{Disc}_{p^2}K_0(\sqrt{p^2}z)=K_0(p z)-K_0(-p z)=i\pi I_0(\sqrt{p^2} z)=i \pi J_0(|p|z), |p|=\sqrt{-p^2}.
\end{align}
Thus, what we compute in the Grassmannian is this particular discontinuity of the correlator. Given the fact that there are four non-trivial integrals to perform to obtain the spinor-helicity result, we have many prescriptions and possible contour choices to evaluate the integral. A detailed study of the integration procedure would be very interesting and is left to future work.
\subsection{The Geometry of $\mathcal{S}+\mathcal{T}-\mathcal{U}$: A Klein quadric in kinematic space}
There is yet another way to understand why the particular combination $\mathcal{S}+\mathcal{T}-\mathcal{U}$ appears in all our formulae. First of all, consider the Mandelstam like invariants $S,\tilde{S},T,\tilde{T},U,\tilde{U}$ given in \eqref{mandelstam1}. On the support of the symplectic orthogonality condition, one can show that the following identities hold (we also write down the values of these quantities in a particular chart for concreteness).
\begin{align}
    &S=(1_I 1^I 2_J 2^J)=(\tilde{3}_I\tilde{3}^J\tilde{4}_I\tilde{4}^J)=4\big(b_{32}b_{41}-b_{31}b_{42}\big)\notag\\
    &\tilde{S}=(\tilde{1}_I \tilde{1}^I \tilde{2}_J \tilde{2}^J)=(3_I 3^I 4_J 4^J)=4\big(b_{23}b_{14}-b_{13}b_{24}\big),\notag\\& T=(1_I 1^I 4_J 4^J)=(\tilde{2}_I\tilde{2}^J\tilde{3}_I\tilde{3}^J)=4\big(b_{24}b_{31}-b_{21}b_{34}\big)\notag\\
    &\tilde{T}=(\tilde{1}_I \tilde{1}^I \tilde{4}_J \tilde{4}^J)=(2_I 2^I 3_J 3^J)=4\big(b_{13}b_{42}-b_{12}b_{43}\big),\notag\\
    &U=(1_I 1^I 3_J 3^J)=(\tilde{2}_I\tilde{2}^J\tilde{4}_I\tilde{4}^J)=4\big(b_{21}b_{43}-b_{23}b_{41}\big)\notag\\
    &\tilde{U}=(\tilde{1}_I \tilde{1}^I \tilde{3}_J \tilde{3}^J)=(2_I 2^I 4_J 4^J)=4\big(b_{12}b_{34}-b_{14}b_{32}\big).
\end{align}
Note in particular that these invariants do not depend on the diagonal components $b_{11},b_{22},b_{33},b_{44}$. We now define the following anti-symmetric $4\times 4$ matrix,
\begin{align}
    \mathbb{P}=\begin{pmatrix}
        0&S&U&T\\
        -S&0&\tilde{T}&\tilde{U}\\
        -U&-\tilde{T}&0&\tilde{S}\\
        -T&-\tilde{U}&-\tilde{S}&0
    \end{pmatrix}.
\end{align}
The Pfaffian of this matrix is,
\begin{align}
    \text{Pf}(\mathbb{P})=S\tilde{S}+T \tilde{T}-U\tilde{U}=\mathcal{S}+\mathcal{T}-\mathcal{U},
\end{align}
which is the pole structure present in all the correlators we have bootstrapped. What is interesting is that the six entries of an antisymmetric matrix are homogeneous coordinates on $\mathbb{CP}^5$. The most natural metric on this space is the AdS$_5$ metric \cite{Adamo:2016rtr}. The condition $\text{Pf}(\mathbb{P})=0$ then restricts to a hypersurface on $\mathbb{CP}^{5}$. This surface is the Klein quadric and represents the boundary of the AdS$_5$ spacetime\cite{Adamo:2016rtr}. Whether this interpretation is just a simple mathematical observation or indicates something more fundamental remains to be seen and we leave it to a future work.

\subsection{Extension to graviton correlators: Double copy?}
We can play the same game with gravity and try to bootstrap the exchange graviton four point correlator using the factorization principle. The Einstein gravity three point function is a double copy of its Yang-Mills counterpart. The three point functions are,
\begin{align}
    &A_{3L}^{I_1 J_1 K_1 L_1,I_2 J_2 K_2 L_2,I J KL}=\mathcal{K}_L\bigg(\frac{(1^{I_1}2^{I_2}s_L^I)(\tilde{1}^{J_1}\tilde{2}^{J_2}\tilde{s}_L^{J})}{\mathcal{K}_L^2}\bigg)\bigg(\frac{(1^{K_1}2^{K_2}s_L^K)(\tilde{1}^{L_1}\tilde{2}^{L_2}\tilde{s}_L^{L})}{\mathcal{K}_L^2}\bigg),\notag\\&A_{3R,IJKL}^{I_3J_3K_3L_3,I_4J_4K_4L_4}=\mathcal{K}_R\bigg(\frac{(3^{I_3}4^{I_4}s_{RI})(\tilde{3}^{J_3}\tilde{4}^{I_4}\tilde{s}_{RJ})}{\mathcal{K}_R^2}\bigg)\bigg(\frac{(3^{K_3}4^{K_4}s_{RK})(\tilde{3}^{L_3}\tilde{4}^{L_4}\tilde{s}_{RL})}{\mathcal{K}_R^2}\bigg).
\end{align}
Using the same identities as for the Yang-Mills correlator we find,
\begin{align}
    \text{Res}_{\tilde{S}=0}\text{Res}_{S=0}A_{4,s}^{(GG|G|GG)}=\frac{(\mathbf{1}\mathbf{2}\mathbf{3}\mathbf{4})^2(\mathbf{\tilde{1}}\mathbf{\tilde{2}}\mathbf{\tilde{3}}\mathbf{\tilde{4}})^2}{(\mathcal{T}-\mathcal{U})^3}.
\end{align}
We uplift this double residue to,
\begin{align}\label{GGGGwithGexchange}
    A_{4,s}^{(GG|G|GG)}=\frac{(\mathbf{1}\mathbf{2}\mathbf{3}\mathbf{4})^2(\mathbf{\tilde{1}}\mathbf{\tilde{2}}\mathbf{\tilde{3}}\mathbf{\tilde{4}})^2}{\mathcal{S}(\mathcal{S}+\mathcal{T}-\mathcal{U})^3}=\frac{1}{(\mathcal{S}+\mathcal{T}-\mathcal{U})}\big(A_{4,s}^{(gg|g|gg)}\big)^2,
\end{align}
which is an extremely simple squaring double copy relation! 

One can proceed this way and attempt to construct higher spin correlators. Another interesting direction to pursue is the supersymmetrization of our results so far. This would allow an additional consistency check on the bootstrapped answers since supersymmetry relates various correlators to each other. However, rather than focussing on developing a super-space spinor-helicity formalism, we take a brief detour to the world of twistors which will allow an extremely straightforward generalization to superspace. In any case, a twistor reformulation of the symplectic Bi-Grassmannian is of interest independently and we now turn to it.

\section{The Grassmannian in Twistor Space}\label{sec:Twistors}
So far, our discussion of the conformal Grassmannian has been related to the spinor helicity formalism. However, given the fact that we are dealing with conformal correlators and that  \textit{twistors} make conformal symmetry completely manifest, it is a natural question to ask for a twist in the Grassmannian. We do exactly that in this section and show that it is an extremely simple exercise to obtain any twistorial correlator given its spinor helicity Grassmannian representation. Along the way, we discuss key formulae in the twistor framework, namely the Half-Fourier transform and the Penrose transform. The latter is especially useful as it connects twistor space to position space.
\subsection{The half-Fourier transform}
The half-Fourier transform was first introduced by Witten in the context of spinor-helicity variables for massless particles \cite{Witten:2003nn}. In our case, we have off-shell spinor helicity variables but the idea of the half-Fourier transform remains the same, that being, obtaining a set of variables in which the action of the conformal generators is linear. The momentum and special conformal generators in the off-shell spinor-helicity formalism are given by \cite{Bala:2026trw},
\begin{align}
    P_{\alpha\Dot{\alpha}}=\lambda_{I}^{\alpha}\tilde{\lambda}^{I\Dot{\alpha}},~~K_{\alpha\Dot{\alpha}}=\frac{\partial^2}{\partial\lambda_I^\alpha\partial\tilde{\lambda}^{I\Dot{\alpha}}}.
\end{align}
One is multiplicative whereas the other is second order. Clearly, this representation obscures special conformal invariance and thus a better choice of variables is desirable. We perform a half-Fourier transform for symmetric traceless conserved currents:
\begin{align}\label{halfFourier}
    J_s^{I_1\cdots I_m J_1\cdots J_n}(Z)= J_s^{I_1\cdots I_m J_1\cdots J_n}(\lambda,\mu)=\int \frac{d^4\tilde{\lambda}}{(2\pi)^4}e^{i\tilde{\lambda}\cdot\tilde{\mu}}\frac{J_s^{I_1\cdots I_{m}J_1\cdots J_n}(\lambda,\tilde{\lambda})}{(\text{Det}(\lambda))^{\frac{m}{2}}(\text{Det}(\tilde{\lambda}))^{\frac{n}{2}}},~n+m=2s, |n-m|=1.
\end{align}
In these variables, the action of the conformal generators are linearized and are all encapsulated within,
\begin{align}\label{TABtwistor}
    T^A_B=Z^{IA}\frac{\partial}{\partial Z^{IB}}-\frac{\delta^A_B}{4}Z^{IC}\frac{\partial}{\partial Z^{IC}}.
\end{align}
Note that these are identical in form to the on-shell twistor space generators \cite{Adamo:2017qyl}, just with an additional sum over the $I$ index.
These matrices obey the $SL(4,\mathbb{R})$ conformal algebra (appropriate for Klein signature),
\begin{align}
    [T^A_B,T^C_D]=\delta^A_D T^C_B-\delta^C_B T^A_D.
\end{align}

Thus, we have obtained a simple representation of the conformal group. We started from spinor-helicity/momentum space and performed a four dimensional integral to obtain twistors. However, a slightly more familiar four dimensional integral takes us from momentum to position space: The Fourier transform. In the next subsection, we clarify the relation between twistor space and position space.
\subsection{The Penrose transform}
The cornerstone of twistor theory is the Penrose transform \cite{Penrose:1967wn, Wolf:2010av,Adamo:2017qyl}. It relates functions (or distributions, members of a cohomology class depending on the spacetime signature and the reality condition of the spinors \cite{Adamo:2017qyl,Wolf:2010av}) in twistor space to operators/fields in position space. We adapt the derivation of the Penrose transform from the half-Fourier and Fourier transforms in \cite{Bala:2025qxr} which was done in CFT$_3$, to the CFT$_4$ setting.

We begin with the spinor-helicity formula for the rescaled currents:
\begin{align}
    \hat{J}_s^{I_1\cdots I_m J_1\cdots J_n}(\lambda,\tilde{\lambda})=\frac{\lambda^{I_1\alpha_1}\cdots\lambda^{I_m \alpha_m}\tilde{\lambda}^{J_1\Dot{\alpha}_1}\cdots\tilde{\lambda}^{J_n \Dot{\alpha}_n}}{\text{Det}(\lambda)^m\text{Det}(\tilde{\lambda})^n}\hat{J}_{s\alpha_1\cdots\alpha_m,\Dot{\alpha}_1\cdots\Dot{\alpha}_n}(p)
\end{align}
One can invert this formula using the identity,
\begin{align}
    \lambda^{I\alpha}\lambda^J_\alpha=-\text{Det}(\lambda)\epsilon^{IJ},
\end{align}
and similarly for $\tilde{\lambda}$. We thus find,
\begin{align}
    J_{s}^{\alpha_1\cdots\alpha_m,\Dot{\alpha}_1\cdots\Dot{\alpha}_n}(p)=\frac{\lambda^{I_1\alpha_1}\cdots\lambda^{I_m \alpha_m}\tilde{\lambda}^{J_1\Dot{\alpha}_1}\cdots\tilde{\lambda}^{J_n \Dot{\alpha}_n}}{\text{Det}(\lambda)^{\frac{m}{2}}\text{Det}(\tilde{\lambda})^{\frac{n}{2}}}J_{sI_1\cdots I_mJ_1\cdots J_n}(\lambda,\tilde{\lambda}).
\end{align}
The right hand side of the above equation is precisely what appears in the integrand of the half-Fourier transform \eqref{halfFourier}. Therefore we can invert it resulting in,
\begin{align}
    J_{s}^{\alpha_1\cdots\alpha_m,\Dot{\alpha}_1\cdots\Dot{\alpha}_n}(p)=i^n \int d^4 \tilde{\mu}\lambda^{I_1\alpha_1}\cdots\lambda^{I_m \alpha_m}\frac{\partial}{\partial \mu_{J_1\Dot{\alpha}_1}}\cdots\frac{\partial}{\partial\mu_{J_n\Dot{\alpha}_n}}e^{-i\tilde{\lambda}\cdot\tilde{\mu}}J_{sI_1\cdots I_mJ_1\cdots J_n}(\lambda,\mu).
\end{align}
Performing the Fourier transform on both sides of the equation and using the fact that,
\begin{align}
    \int \frac{d^4 p}{(2\pi)^4}e^{ip\cdot x}=\int \frac{d^4\lambda d^4\tilde{\lambda}}{(2\pi)^4\text{Vol}(GL(2))}e^{i\lambda_{I\alpha}\tilde{\lambda}^I_{\Dot{\alpha}}x^{\alpha\Dot{\alpha}}},
\end{align}
we can perform the $\tilde{\lambda}$ integral trivially resulting in,
\begin{align}\label{PenroseTransform}
    J_s^{\alpha_1\cdots \alpha_m,\Dot{\alpha}_1\cdots\Dot{\alpha}_n}(x)&=i^n\int \frac{d^4\lambda d^4\mu}{\text{Vol}(GL(2))}\delta^4(\mu^{I\Dot{\alpha}}-x^{\alpha\Dot{\alpha}}\lambda^{I}_{\alpha})\lambda^{I_1\alpha_1}\cdots\lambda^{I_m \alpha_m}\frac{\partial}{\partial\mu_{J_1\Dot{\alpha}_1}}\cdots\frac{\partial}{\partial\mu_{J_n\Dot{\alpha}_n}}J_{sI_1\cdots I_m J_1\cdots J_n}(\lambda,\mu)\notag\\
    &=i^n\int d^4\lambda \bigg(\prod_{j=1}^{m}\lambda^{I_j \alpha_j}\prod_{k=1}^{n}\frac{\partial}{\partial \mu_{J_k \Dot{\alpha}_k}}\bigg)J_{s,I_1,\cdots ,I_m,J_1,\cdots,J_n}(Z)|_{X}
\end{align}
This is the generalization of the Penrose transform for free massless particles to symmetric traceless conserved currents\footnote{It would be interesting to compare this with the ambitwistor Penrose transform used in \cite{CarrilloGonzalez:2026eum}}. Note that we obtain the incidence relation,
\begin{align}
    X=\{\mu^{I\Dot{\alpha}}=x^{\alpha\Dot{\alpha}}\lambda^I_\alpha\}.
\end{align}
This represents two copies of the usual incidence relation $(I=1,2)$ and thus the usual geometry of twistor space \cite{Adamo:2017qyl} applies to each copy. Note the projective rescaling property of the integrand. Under $Z\to r Z\equiv(\lambda,\mu)\to r(\lambda,\mu)$ we find that,
\begin{align}\label{twistorrescaling}
    J_{sI_1\cdots I_mJ_1\cdots J_n}(rZ)=\frac{1}{r^{4+m-n}}J_{sI_1\cdots I_mJ_1\cdots J_n},
\end{align}
For symmetric traceless integer spin conserved currents which have $m=n=s$ we obtain the uniform $\frac{1}{r^4}$ scaling. For half-integer spin conserved currents which have $|m-n|=1$, we either obtain a $\frac{1}{r^3}$ or $\frac{1}{r^5}$ scaling. Thus, using the Penrose transform, one can convert any twistor space expression into position space if desired.
\subsection{Classification of conformal invariants}
Next, we classify conformal invariants. This amounts to classifying $SL(4,\mathbb{R})$ invariants with the currents in the correlator transforming as given earlier in \eqref{TABtwistor}. The conformal Ward identities that constrain their correlators read,
\begin{align}
    \sum_{i=1}^{n}(T_{i})^{A}_{B}\langle J_{s_1}\cdots J_{s_n}\rangle=0,
\end{align}
where we suppressed possible little group $SL(2)$ indices. These equations also have to be supplemented by the rescaling property \eqref{twistorrescaling}.
The most natural invariant we can form are delta functions of the form,
\begin{align}\label{twistorinvariantansatz}
    \int [d\tilde{C}]\delta^{n\times 4}(\tilde{C}_{ij,I}Z_j^{IA})f(\tilde{C}),
\end{align}
which is indeed what we will find in our twistor space Grassmannian expressions. This object is clearly $SL(4)$ invariant and one can arrange for the rescaling property \eqref{twistorrescaling} by choosing $f(\tilde{C})$ appropriately. The representation \eqref{twistorinvariantansatz} is actually the twistor space Grassmannian for an appropriate choice of the measure $[d\tilde{C}]$! The most natural guess leads us to the Grassmannian. Before we prove this statement, let us note that there exist other invariants formed using the invariant Levi-Civita symbol $\epsilon^{ABCD}$. For example,
\begin{align}
    \epsilon^{ABCD}Z_{iA}^{I}Z_{jB}^{J}Z_{kC}^{K}Z_{lD}^{L}.
\end{align}
These quantities will not feature in what follows but it would be interesting to understand what class of correlators utilize them such as perhaps, those that are odd under parity.
\subsection{The Grassmannian construction}
Given the Grassmannian in spinor-helicity variables, we perform the direct half-Fourier transform to determine its twistor space counterpart. First, we perform the integral over the $C$ matrix in the symplectic Bi-Grassmannian which yields,
\begin{align}
    \psi_n(\Lambda,\tilde{\Lambda})=\int \frac{d^{n\times 2n}\tilde{C}}{\text{Vol}(GL(n))}\delta^{n\times 2}(\tilde{C}_{\perp}\cdot\Omega\cdot\tilde{\Lambda})\delta^{n\times 2}(\tilde{C}\cdot\Omega\cdot\Lambda)A_n(C=\tilde{C}_{\perp},\tilde{C}),
\end{align}
where $\tilde{C}_{\perp}$ satisfies $\tilde{C}_{\perp}\cdot\Omega\cdot\tilde{C}=0$ is a solution to the symplectic orthogonality constraint. In this form, performing the half-Fourier transform is easy. We use the identity,
\begin{align}\label{hftid}
    \int d^{2n\times 2}\tilde{\Lambda}e^{i\Tr{\tilde{\Lambda}^T\cdot M}}\delta^{n\times 2}(\tilde{C}_{\perp}\cdot\Omega\cdot\tilde{\Lambda})=\delta^{n\times 2}(\tilde{C}\cdot\Omega\cdot M).
\end{align}
where we have defined the $2n\times 2$ vector,
\begin{align}
    M_{iI}^{\Dot{\alpha}}\equiv M=\begin{pmatrix}
        (\mu_{1})_1^1&(\mu_{1})_1^2\\
        (\mu_{1})_2^1&(\mu_{1})_2^2\\
        \vdots&\vdots\\
        (\mu_{n})_1^1&(\mu_{n})_1^2\\
        (\mu_{n})_2^1&(\mu_{n})_2^2
    \end{pmatrix}.
\end{align}
We prove the identity \eqref{hftid} in appendix \ref{app:HFT}. Thus we find,
\begin{align}
    &\psi_n(\Lambda,M)=\int \frac{d^{n\times 2n}\tilde{C}}{\text{Vol}(GL(n))}\delta^{n\times 2}(\tilde{C}\cdot\Omega\cdot\Lambda)\delta^{n\times 2}(\tilde{C}\cdot\Omega\cdot M)A_n(C=\tilde{C}_{\perp},\tilde{C}),
\end{align}
where we have defined the collective twistor variable,
\begin{align}
    Z_{i}^{IA}=\begin{pmatrix}
        \Lambda_{i}^{I\alpha}\\
        M_{i}^{I\Dot{\alpha}}
    \end{pmatrix}.
\end{align}
The twistor space conformal Grassmannian takes the form,
\begin{align}\label{twistorGrassmannian}
    \psi_n(Z)&=\int \frac{d^{n\times 2n}\tilde{C}}{\text{Vol}(GL(n))}\delta^{n\times 4}(\tilde{C}\cdot\Omega\cdot Z)A_n(C=\tilde{C}_{\perp},\tilde{C})\notag\\
    &=\int \frac{d^{n\times 2n}\tilde{C}}{\text{Vol}(GL(n))}\delta^{n\times 4}(\tilde{C}\cdot\Omega\cdot Z)\tilde{A}_n(\tilde{C}),
\end{align}
which is extremely simple and is exactly the kind of solution we guessed when solving the conformal Ward identities. Under a $GL(n)$ transformation of the $\tilde{C}$ matrix we require,
\begin{align}
    \tilde{A}_n(\tilde{G}\tilde{C})=\frac{1}{\text{Det}(\tilde{G})^{2(n-2)}}\tilde{A}_n(\tilde{C}).
\end{align}
Covariance is taken care by the explicit $SL(2)$ indices of the minors that we appropriately choose for $\tilde{A}_n$ just like in the previous sections.
 As for the examples of correlators we have considered so far, it is a simple matter to use them in the above formula appropriate for twistor space. The only thing we require are the expressions for the minors of $C$ in terms of the minors of  $\tilde{C}$.
\subsection{Correlation functions}
 We now discuss several examples for $n=2,3,4$ point correlators, re-writing the minors of $C$ in terms of those of $\tilde{C}$.
\subsubsection{$n=2$}
Let's start with $s=1$ for simplicity. The two point function is given by \eqref{sstwopoint},
\begin{align}
    A^{g,g}_2(C,\tilde{C})=\frac{(1^{I_1} 2^{I_2})(\tilde{1}^{J_1}\tilde{2}^{J_2})}{(1_I 2_J)(1^I 2^J)}.
\end{align}
Let us now set $\tilde{C}=C_{\perp}$. An explicit gauge choice is,
\begin{align}
    &\tilde{C}=\begin{pmatrix}
        a_{11}&1&a_{12}&0\\
        a_{21}&0&a_{22}&1
    \end{pmatrix},\notag\\
    &C=\begin{pmatrix}
        a_{11}&1&a_{21}&0\\
        a_{12}&0&a_{22}&1
    \end{pmatrix}.
\end{align}
It is easy to show in this particular parametrization that $(\tilde{1}^{J_1}\tilde{2}^{J_2})=(1^{J_1}2^{J_2})$. In a generic chart that spans a portion of the Grassmannian integral, the more general relation is\footnote{To be more precise, we choose to work in parametrizations where there is no non-trivial Jacobian factor picked up when solving for the components of $C$ in terms of those of $\tilde{C}$. In parameterizations where non-trivial Jacobian factors arise, one has to do more work, taking care of the Jacobian factors and reorganizing expressions. In any case, it is easiest to work in a particular chart and then uplift the result to the general case and check for the consistency of the resulting expression.},
\begin{align}
    (\tilde{1}^{J_1}\tilde{2}^{J_2})|_{C=\tilde{C}_{\perp}}=\pm (1^{J_1}2^{J_2}).
\end{align}
The two point function is thus given by,
\begin{align}
    &\tilde{A}_2^{g,g}=\frac{(\tilde{1}^{I_1}\tilde{2}^{I_2})(\tilde{1}^{J_1}\tilde{2}^{J_2})}{(\tilde{1}_I\tilde{2}_J)(\tilde{1}^I\tilde{2}^{J})}=\frac{(\mathbf{\tilde{1}}\mathbf{\tilde{2})^2}}{(\mathbf{\tilde{1}\mathbf{\tilde{2}})\cdot(\mathbf{\tilde{1}}\mathbf{\tilde{2}})}}.
\end{align}
The generalization to spin$-s$ is simple and takes the form,
\begin{align}
    A_2^{s,s}=\bigg(\frac{(\mathbf{\tilde{1}}\mathbf{\tilde{2})^2}}{(\mathbf{\tilde{1}\mathbf{\tilde{2}})\cdot(\mathbf{\tilde{1}}\mathbf{\tilde{2}})}}\bigg)^s.
\end{align}
\subsubsection{$n=3$}
We consider the scalar three point function \eqref{A3scalar} as the first example.
\begin{align}
    A_3^{\varphi\varphi\varphi}=\frac{1}{\mathcal{K}}.
\end{align}
Choosing the parameterization,
\begin{align}
    &C=\begin{pmatrix}
        c_{11}&1&c_{12}&0&c_{13}&0\\
        c_{21}&0&c_{22}&1&c_{23}&0\\
        c_{31}&0&c_{32}&0&c_{33}&1
    \end{pmatrix},
    \tilde{C}=\begin{pmatrix}
        c_{11}&1&c_{21}&0&c_{31}&0\\
        c_{12}&0&c_{22}&1&c_{32}&0\\
        c_{13}&0&c_{23}&0&c_{33}&1
        \end{pmatrix},
\end{align}
one can show that the three point invariant $\mathcal{K}$ satisfies\footnote{This formula can differ by a minus sign for a different choice of parametrization.},
\begin{align}
    \mathcal{K}=\frac{1}{2}(\tilde{1}_I\tilde{2}_J\tilde{2}^J)(\tilde{1}^I \tilde{3}_K\tilde{3}^K).
\end{align}
Thus we find,
\begin{align}
    A_3^{\varphi\varphi\varphi}=\frac{2}{(\tilde{1}_I\tilde{2}_J\tilde{2}^J)(\tilde{1}^I \tilde{3}_K\tilde{3}^K)}.
\end{align}
Similarly, using the formula,
\begin{align}
    (1^I 2^J 3^K)=(\tilde{1}^I\tilde{2}^J\tilde{3}^K),
\end{align}
we find the Yang-Mills three point function \eqref{A3YM} to be,
\begin{align}
    A_3^{g,g,g}=4\frac{(\mathbf{\tilde{1}}\mathbf{\tilde{2}}\mathbf{\tilde{3}})^2}{\big((\mathbf{\tilde{1}\mathbf{\tilde{2}\cdot\mathbf{\tilde{2}})\cdot(\mathbf{\tilde{1}}\mathbf{\tilde{3}}\cdot\mathbf{\tilde{3}})}}\big)^2}.
\end{align}
A similar analysis can be carried out for the other three point functions we have considered.
\subsubsection{$n=4$}
At $n=4$, we constructed three minors of $C$ and three of $\tilde{C}$ namely $S,T,U$ and $\tilde{S},\tilde{T},\tilde{U}$ respectively. Consider the parameterization,
\begin{align}
    &C=\begin{pmatrix}
        c_{11}&1&c_{12}&0&c_{13}&0&c_{14}&0\\
        c_{21}&0&c_{22}&1&c_{23}&0&c_{24}&0\\
        c_{31}&0&c_{32}&0&c_{33}&1&c_{34}&0\\
        c_{41}&0&c_{42}&0&c_{43}&0&c_{44}&1
    \end{pmatrix},
    \notag\\
    &\tilde{C}=\begin{pmatrix}
        c_{11}&1&c_{21}&0&c_{31}&0&c_{41}&0\\
        c_{12}&0&c_{22}&1&c_{32}&0&c_{42}&0\\
        c_{13}&0&c_{23}&0&c_{33}&1&c_{43}&0\\
        c_{14}&0&c_{24}&0&c_{34}&0&c_{44}&1
    \end{pmatrix}.
\end{align}
We find,
\begin{align}\label{4ptctoctilde}
    &S=4(1_1 1_2 2_1 2_2)=4(\tilde{3}_1\tilde{3}_2\tilde{4}_1\tilde{4}_2),\notag\\
    &T=4(1_1 1_2 4_1 4_2)=4(\tilde{2}_1\tilde{2}_2\tilde{3}_1\tilde{3}_2),\notag\\
    &U=4(1_1 1_2 3_1 3_2)=4(\tilde{2}_1\tilde{2}_2\tilde{4}_1\tilde{4}_2).
\end{align}
Thus, we find for the $GL$ invariant Mandelstams \eqref{mandelstammathcal},
\begin{align}
    &\mathcal{S}=16(\tilde{1}_1\tilde{1}_2\tilde{2}_1\tilde{2}_2)(\tilde{3}_1\tilde{3}_2\tilde{4}_1\tilde{4}_2),\notag\\
    &\mathcal{T}=16(\tilde{1}_1\tilde{1}_2\tilde{4}_1\tilde{4}_2)(\tilde{2}_1\tilde{2}_2\tilde{3}_1\tilde{3}_2),\notag\\
    &\mathcal{U}=16(\tilde{1}_1\tilde{1}_2\tilde{3}_1\tilde{3}_2)(\tilde{2}_1\tilde{2}_2\tilde{4}_1\tilde{4}_2).
\end{align}
In a general parametrization (one in which we do not pick up a non-trivial Jacobian when solving for $C$ in terms of $\tilde{C}$), the formulae \eqref{4ptctoctilde} could potentially pick up a sign. Finally, we can recast covariants too using formulae such as,
\begin{align}
    (1^I 2^J 3^K 4^L)=(\tilde{1}^I\tilde{2}^J\tilde{3}^K\tilde{4}^L).
\end{align}
The Yang-Mills $s-$channel  contribution \eqref{ggggYMexchangeschannel} for instance becomes,
\begin{align}
    A_{4,s}^{gg|g|gg}=\frac{(\mathbf{\tilde{1}}\mathbf{\tilde{2}}\mathbf{\tilde{3}}\mathbf{\tilde{4}})^2}{\mathcal{S}(\mathcal{S}+\mathcal{T}-\mathcal{U})^2},
\end{align}
It is easy to carry out a similar analysis for other examples and thus we do not present the results here.
\subsection{Dual Twistors}
Before we conclude this section, recall that twistors can be traded for dual-twistors using the twistor Fourier transform. In our case, we have,
\begin{align}
    f(W)=\int \frac{d^{4\times 2}Z}{(2\pi)^4}e^{iW_{IA}Z^{IA}}f(Z).
\end{align}
$Z^{IA}$ is in the fundamental representation of $SL(4)$ whereas $W^I_A$ is in the anti-fundamental representation (and both also carry a fundamental index of the $SL(2)$ little group of course). Using this formula, it is a simple matter to derive the conformal generators, the Penrose transform and the dual-twistor space Grassmannian. The latter is given by,
\begin{align}
    \psi_n(W)=\int \frac{d^{n\times 2n}C}{\text{Vol}(GL(n))}\delta^{n\times 4}(C\cdot\Omega\cdot W)A_n(C,\tilde{C}=C_{\perp}),
\end{align}
where $W$ equals,
\begin{align}
    W_i^{IA}=\begin{pmatrix}
        W_{1}^{IA}\\
        \vdots\\
        W_n^{IA}
    \end{pmatrix}=\begin{pmatrix}
        \tilde{\mu}_{1}^{I\alpha}\\
        \tilde{\lambda}_{1}^{I\dot{\alpha}}\\
        \vdots\\
        \tilde{\mu}_n^{I\alpha}\\
        \tilde{\lambda}_n^{I\Dot{\alpha}}
    \end{pmatrix}.
\end{align}
One can now play the same game, recasting minors of the $\tilde{C}$ matrix in terms of those of $C$ to obtain $A_n(C,\tilde{C}=C_{\perp})$. With this, we conclude our section on the $SL(2)$ twistor space construction.

One can also represent each operator as a function of a twistor and a dual-twistor which is known as the ambitwistor formalism \cite{Adamo:2017qyl, Adamo:2016rtr,CarrilloGonzalez:2026eum}. We discuss such a construction in appendix \ref{app:AmbiTwistors}. We now proceed to the super-twistor space formalism.
 \section{The  Grassmannian in Super-Twistor Space}\label{sec:SuperTwistors}
Twistors, in addition to making conformal invariance manifest also admit a natural generalization to super-twistors that one can use to describe super-conformal correlators \cite{Adamo:2017qyl,Adamo:2011pv}. We discuss and derive the super-Penrose transform that relates twistor space to position space. We derive the super-conformal Grassmannian for $\mathcal{N}=1$ SCFT$_4$ in twistor space. This also serves as a stepping stone to obtain the symplectic Bi-Grassmannian for super-correlators in super spinor-helicity variables in the next section.
\subsection{From twistors to super-twistors}
Given the twistor space conformal generators \eqref{TABtwistor}, the generalization to $\mathcal{N}=1$ superspace is straightforward. We introduce an auxiliary Grassmann coordinate $\eta^{I}$ and define the super-twistor,
\begin{align}
    \mathcal{Z}^{I\mathcal{A}}=(Z^{IA},\eta^{I}).
\end{align}
The super-conformal generators of the super-group $SL(4,\mathbb{R}|1,\mathbb{R})$ acting on the primaries with $\Delta=s+2$ take the form,
\begin{align}
    \mathcal{T}^{\mathcal{A}}_{\mathcal{B}}=\mathcal{Z}^{I\mathcal{A}}\frac{\partial}{\partial\mathcal{Z}^{I\mathcal{B}}}-\frac{\delta^{\mathcal{A}}_{\mathcal{B}}}{3}\text{STr}\big(\mathcal{Z}^{I\mathcal{C}}\frac{\partial}{\partial\mathcal{Z}^{I\mathcal{C}}}\big),
\end{align}
where $\text{STr}\big(\mathcal{Z}^{I\mathcal{C}}\frac{\partial}{\partial\mathcal{Z}^{I\mathcal{C}}}\big)=\sum_{\mathcal{C}}(-1)^{|\mathcal{C}|}\mathcal{Z}^{I\mathcal{C}}\frac{\partial}{\partial\mathcal{Z}^{I\mathcal{C}}}$
These generators are defined such that they have zero super-trace \cite{Frappat:1996pb}.
The super-generators obey the graded Lie super-algebra,
\begin{align}
    [\mathcal{T}^{\mathcal{A}}_{\mathcal{B}},\mathcal{T}^{\mathcal{C}}_{\mathcal{D}}\}=\delta^{\mathcal{C}}_{\mathcal{B}}\mathcal{T}^{\mathcal{A}}_{\mathcal{D}}-(-1)^{(|\mathcal{A}|+|\mathcal{B}|)(|\mathcal{C}|+|\mathcal{D}|)}\delta^{\mathcal{A}}_{\mathcal{D}}\mathcal{T}^{\mathcal{C}}_{\mathcal{B}},
\end{align}
where $|\mathcal{A}|$ denotes the grade. If $\mathcal{A}$ is a bosonic index it equals $0$ whereas if it is fermionic, it equals $1$. We now turn to the super-fields of interest.
\subsection{The super-fields of interest}
We work with half-integer spin super-currents with $s\in\mathbb{Z}_{\ge 0}+\frac{1}{2}$ in $\mathcal{N}=1$ theories. The super-field expansion is \cite{Herderschee:2019ofc},
\begin{align}\label{Neq1supercurrent}
    \mathbf{J}^{I_1\cdots I_mJ_1\cdots J_n}_s(\mathcal{Z})=\tilde{J}^{I_1\cdots I_mJ_1\cdots J_n}_{s}(Z)+\eta^{I_1}J_{s-\frac{1}{2}}^{I_2\cdots I_m J_1\cdots J_n}(Z)+\eta_J J_{s+\frac{1}{2}}^{JI_1\cdots I_mJ_1\cdots J_n}(Z)-\frac{\eta^2}{2}J_s^{I_1\cdots I_m J_1\cdots J_n}(Z).
\end{align}
Under a projective rescaling $\mathcal{Z}=(Z,\eta)\to (r Z,r \eta)$ $\tilde{J}_s$ transforms with a factor of $\frac{1}{r^3}$, $J_s$ with a factor of $\frac{1}{r^5}$ while the integer spin $J_{s-\frac{1}{2}}$ and $J_{s+\frac{1}{2}}$ transform with a factor of $\frac{1}{r^4}$. The Grassmann variables $\eta$ precisely take into account this scaling ensuring that the super-current transforms with a factor of $\frac{1}{r^3}$.

\subsection{The super-Penrose transform}
Given the Penrose transform of each component current, we can derive a supersymmetric version of the same. Starting with \eqref{Neq1supercurrent}, we perform the $GL(2)$ projective integral over $\lambda$ while imposing the incidence relations. For notational brevity we first work with $s=\frac{1}{2}$ and generalize to general spin later. The super-current \eqref{Neq1supercurrent} is thus,
\begin{align}\label{Jhalf}
    \mathbf{J}^I_{\frac{1}{2}}(\mathcal{Z})=\tilde{O}_{\frac{1}{2}}^{I}(Z)+\eta^I O_2(Z)+\eta_J J^{IJ}(Z)-\frac{\eta^2}{2}O_{\frac{1}{2}}(Z).
\end{align}
We express the position space super-current via,
\begin{align}\label{superPenroseansatz}
    \mathbf{J}_{\frac{1}{2},\Dot{\alpha}}(x,\theta,\tilde{\theta})=\int \frac{d^4 \lambda}{\text{Vol}(GL(2))}\frac{\partial}{\partial\mu^{I\Dot{\alpha}}}\mathbf{J}_{\frac{1}{2}}^{I}(Z,\eta)|_{\mathcal{X}},
\end{align}
where $\mathcal{X}$ are super-incidence relations that we need to determine. \eqref{superPenroseansatz} is consistent with the $GL(1)$ weight $\frac{1}{r^3}$  of $\mathbf{J}_{\frac{1}{2}}$ ensuring the integral remains invariant under $\lambda\to r \lambda,\mu\to r \mu,\eta\to r \eta$.
To proceed, we write down the following ansatz for the super-incidence relations:
\begin{align}
    \mathcal{X}=\{\mu^{I\Dot{\alpha}}=x^{\alpha\Dot{\alpha}}\lambda^I_\alpha+c_1\theta^\alpha\tilde{\theta}^{\Dot{\alpha}}\lambda_\alpha^I,~~\eta^I=c_2 \theta^{\alpha}\lambda_\alpha^I\}.
\end{align}
$c_1$ and $c_2$ are arbitrary coefficients for the moment. Let us begin by expanding $\mathbf{J}_{\frac{1}{2}}^{I}(Z,\eta)$ into its components using \eqref{Jhalf} and performing a Taylor expansion in $\mu^{I\Dot{\alpha}}$. We define,
\begin{align}
    \mu_0^{I\Dot{\alpha}}=x^{\alpha\Dot{\alpha}}\lambda_{\alpha}^I.
\end{align}
Then, 
\begin{align}
    \tilde{O}_{\frac{1}{2}}^{I}(\lambda,\mu_0+c_1 \theta^\beta\tilde{\theta}^{\Dot{\beta}}\lambda_\beta^I)&=\tilde{O}_{\frac{1}{2}}^I(\lambda,\mu_0)+c_1\theta^\beta\tilde{\theta}^{\Dot{\beta}}\lambda_\beta^J\frac{\partial}{\partial \mu_0^{J\Dot{\beta}}}\tilde{O}_{\frac{1}{2}}^I(\lambda,\mu_0)+\frac{c_1^2}{2}\theta^\beta\tilde{\theta}^{\Dot{\beta}}\theta^\gamma\tilde{\theta}^{\Dot{\gamma}}\lambda_{\beta}^J\lambda_{\gamma}^K\frac{\partial^2}{\partial \mu_0^{J\Dot{\beta}}\partial\mu_0^{K\Dot{\gamma}}}\tilde{O}_{\frac{1}{2}}^I(\lambda,\mu_0)\notag\\
    &=\tilde{O}_{\frac{1}{2}}^{I}(\lambda,\mu_0)+c_1\theta^\beta\tilde{\theta}^{\Dot{\beta}}\frac{\partial}{\partial x^{\beta\Dot{\beta}}}\tilde{O}_{\frac{1}{2}}^I(\lambda,\mu_0)-\frac{c_1^2\theta^2\tilde{\theta}^2}{8}\epsilon^{\beta\gamma}\epsilon^{\Dot{\beta}\Dot{\gamma}}\frac{\partial^2}{\partial x^{\beta\Dot{\beta}}\partial x^{\gamma\Dot{\gamma}}}\tilde{O}_{\frac{1}{2}}^I(\lambda,\mu_0).
\end{align}
In the second term, we used the formula,
\begin{align}
    \frac{\partial}{\partial x^{\alpha\Dot{\alpha}}}f(\lambda,\mu_0^{I\Dot{\alpha}}=x^{\alpha\Dot{\alpha}}\lambda^I_\alpha)=\lambda^J_\alpha\frac{\partial}{\partial \mu_0^{J\Dot{\alpha}}}f(\lambda,\mu_0^{I\Dot{\alpha}}=x^{\alpha\Dot{\alpha}}\lambda^I_\alpha).
\end{align}
In the third term we used,
\begin{align}
    \theta^\alpha\theta^\beta=\frac{\epsilon^{\alpha\beta}}{2}\theta^2,
\end{align}
and a similar formula for $\tilde{\theta}$. Finally, we use the identity,
\begin{align}
    \epsilon^{\beta\gamma}\epsilon^{\Dot{\beta}\Dot{\gamma}}\partial_{\beta\Dot{\beta}}\partial_{\gamma\Dot{\gamma}}=-2\Box.
\end{align}
This results in,
\begin{align}\label{Ohtildeformula1}
     \tilde{O}_{\frac{1}{2}}^{I}(\lambda,\mu_0+c_1 \theta^\beta\tilde{\theta}^{\Dot{\beta}}\lambda_\beta^I)&=\tilde{O}_{\frac{1}{2}}^{I}(\lambda,\mu_0)+c_1\theta^\beta\tilde{\theta}^{\Dot{\beta}}\frac{\partial}{\partial x^{\beta\Dot{\beta}}}\tilde{O}_{\frac{1}{2}}^I(\lambda,\mu_0)+\frac{c_1^2\theta^2\tilde{\theta}^2}{4}\Box\tilde{O}_{\frac{1}{2}}^I(\lambda,\mu_0)\notag\\
     &=\big(1+c_1\theta^\beta\tilde{\theta}^{\Dot{\beta}}\frac{\partial}{\partial x^{\beta\Dot{\beta}}}+\frac{c_1^2\theta^2\tilde{\theta}^2}{4}\Box\big)\tilde{O}_{\frac{1}{2}}^I(\lambda,\mu_0).
\end{align}

For the second term in the super-field expansion \eqref{Jhalf} we have,
\begin{align}\label{O2formula}
    &\eta^I O_2(\lambda,\mu)=c_2\theta^\alpha\lambda^I_\alpha\big(O_2(\lambda,\mu_0)+c_1\theta^\beta\tilde{\theta}^{\Dot{\beta}}\lambda^J_{\beta}\frac{\partial}{\partial \mu_0^{J\Dot{\beta}}}O_2(\lambda,\mu_0)\big)\notag\\
    &=c_2\theta^\alpha\lambda_\alpha^I\big(1+c_1\theta^\beta\tilde{\theta}^{\Dot{\beta}}\frac{\partial}{\partial x^{\beta\Dot{\beta}}}\big)O_2(\lambda,\mu_0).
\end{align}
For the spin$-1$ component of \eqref{Jhalf}, we obtain,
\begin{align}\label{JIJformula}
    \eta_J J^{IJ}(\lambda,\mu)=c_2\theta^\alpha\lambda_{J\alpha}\bigg(1+c_1\theta^\beta\tilde{\theta}^{\Dot{\beta}}\frac{\partial}{\partial x^{\beta\Dot{\beta}}}\big)J^{IJ}(\lambda,\mu_0).
\end{align}
Finally, the last term of \eqref{Jhalf} is given by,
\begin{align}\label{Ohformula}
    -\frac{\eta^2}{2}O_{\frac{1}{2}}^I(\lambda,\mu)&=-c_2^2\frac{\epsilon_{JK}}{2}\theta^\alpha \lambda^J_\alpha\theta^\beta\lambda^K_\beta O_{\frac{1}{2}}^I(\lambda,\mu_0)=-c_2^2\frac{\theta^2}{4}\epsilon_{JK}\epsilon^{\alpha\beta}\lambda_\alpha^J\lambda_\beta^KO_{\frac{1}{2}}^{I}(\lambda,\mu_0)\notag\\
    &=-c_2^2\frac{\theta^2}{2}\text{Det}(\lambda)O_{\frac{1}{2}}^{I}(\lambda,\mu_0).
\end{align}
Putting together \eqref{Ohtildeformula1}, \eqref{O2formula}, \eqref{JIJformula} and \eqref{Ohformula} and substituting it in \eqref{superPenroseansatz} results in\footnote{We use the fact that the partial derivative with respect to $\mu$ equals the partial derivative with respect to $\mu_0$ for fixed $\lambda$.}
\begin{align}
   \mathbf{J}_{\frac{1}{2},\Dot{\alpha}}(x,\theta,\tilde{\theta})&=\int \frac{d^4 \lambda}{\text{Vol}(GL(2))}\frac{\partial}{\partial\mu^{I\Dot{\alpha}}}\mathbf{J}_{\frac{1}{2}}^{I}(Z,\eta)|_{\mathcal{X}}\notag\\
   &=\int \frac{d^4 \lambda}{\text{Vol}(GL(2))}\frac{\partial}{\partial\mu_0^{I\Dot{\alpha}}}\Bigg[\big(1+c_1\theta^\beta\tilde{\theta}^{\Dot{\beta}}\frac{\partial}{\partial x^{\beta\Dot{\beta}}}+\frac{c_1^2\theta^2\tilde{\theta}^2}{4}\Box\big)\tilde{O}_{\frac{1}{2}}^I(\lambda,\mu_0)\notag\\
   &+\big(c_2\theta^\alpha\lambda_\alpha^I\big(1+c_1\theta^\beta\tilde{\theta}^{\Dot{\beta}}\frac{\partial}{\partial x^{\beta\Dot{\beta}}}\big)O_2(\lambda,\mu_0)\big)+c_2\theta^\alpha\lambda_{J\alpha}\bigg(1+c_1\theta^\beta\tilde{\theta}^{\Dot{\beta}}\frac{\partial}{\partial x^{\beta\Dot{\beta}}}\big)J^{IJ}(\lambda,\mu_0)\notag\\
   &-c_2^2\frac{\theta^2}{2}\text{Det}(\lambda)O_{\frac{1}{2}}^{I}(\lambda,\mu_0)\Bigg]\notag\\&=\big(1+c_1\theta^\beta\tilde{\theta}^{\Dot{\beta}}\frac{\partial}{\partial x^{\beta\Dot{\beta}}}+\frac{c_1^2\theta^2\tilde{\theta}^2}{4}\Box\big)\tilde{O}_{\frac{1}{2},\Dot{\alpha}}(x)+c_2\theta^\alpha \frac{\partial}{\partial  x^{\alpha\Dot{\alpha}}}O_2(x)+\frac{c_1 c_2}{2}\epsilon^{\alpha\beta}\theta^2\tilde{\theta}^{\Dot{\beta}}\frac{\partial^2}{\partial x^{\alpha\Dot{\alpha}}\partial x^{\beta\Dot{\beta}}}O_2(x)\notag\\
   &+c_2\theta^\alpha J_{\alpha\Dot{\alpha}}(x)+\frac{c_1 c_2}{2}\epsilon^{\alpha\beta}\theta^2\tilde{\theta}^{\Dot{\beta}}\frac{\partial}{\partial x^{\beta\Dot{\beta}}}J_{\alpha \Dot{\alpha}}(x)-\frac{c_2^2\theta^2}{2}\frac{\partial}{\partial x^{\alpha \Dot{\alpha}}}O_{\frac{1}{2}}^\alpha(x).
\end{align}
For the final term we used the formula\footnote{Note that $I$ labels the row index and $\alpha$ labels the column index for $\lambda$ whereas for $\lambda^{-1}$, it is $\alpha$ that labels the row and $I$, the column.},
\begin{align}
    \frac{\partial}{\partial \mu_{0}^{I\Dot{\alpha}}}=(\lambda^{-1})_{~I}^{\alpha}\frac{\partial}{\partial x^{\alpha\Dot{\alpha}}}~~, (\lambda^{-1})^{\alpha}_{~I}=\frac{\lambda^{~\alpha}_I}{\text{Det}(\lambda)}
\end{align}
Thus, the position space super-field takes the form,
\begin{align}
    &\mathbf{J}_{\frac{1}{2},\Dot{\alpha}}(x,\theta,\tilde{\theta})=\big(1+c_1\theta^\beta\tilde{\theta}^{\Dot{\beta}}\frac{\partial}{\partial x^{\beta\Dot{\beta}}}+\frac{c_1^2\theta^2\tilde{\theta}^2}{4}\Box\big)\tilde{O}_{\frac{1}{2},\Dot{\alpha}}(x)+\bigg(c_2\theta^\alpha \frac{\partial}{\partial  x^{\alpha \Dot{\alpha}}}+\frac{c_1 c_2}{2}\epsilon^{\alpha\beta}\theta^2\tilde{\theta}^{\Dot{\beta}}\frac{\partial^2}{\partial x^{\alpha \Dot{\alpha}}\partial x^{\beta\Dot{\beta}}}\bigg)O_2(x)\notag\\
   &+\bigg(c_2\theta^\alpha +\frac{c_1 c_2}{2}\epsilon^{\alpha\beta}\theta^2\tilde{\theta}^{\Dot{\beta}}\frac{\partial}{\partial x^{\beta\Dot{\beta}}}\bigg)J_{\alpha \Dot{\alpha}}(x)-\frac{c_2^2\theta^2}{2}\frac{\partial}{\partial x^{\alpha \Dot{\alpha}}}O_{\frac{1}{2}}^\alpha(x).
\end{align}
$c_2$ can be absorbed into the normalization of the $O_2$, $J^{\alpha\Dot{\alpha}}$ and $O_{\frac{1}{2}}$ and thus we can set it equal to $1$. However,  $c_1$ has to take a specific value. To see why, we first define some important quantities.
\begin{align}
\tilde{D}_{\Dot{\alpha}}=\frac{\partial}{\partial \tilde{\theta}^{\Dot{\alpha}}}+\frac{i}{2}\theta^\alpha\frac{\partial}{\partial x^{\alpha\Dot{\alpha}}},~~D_{\alpha}=\frac{\partial}{\partial\theta^\alpha}+\frac{i}{2}\tilde{\theta}^{\Dot{\alpha}}\frac{\partial}{\partial x^{\alpha\Dot{\alpha}}}.
\end{align}
The anti-commutator of these two quantities satisfies,
\begin{align}
    \{D_{\alpha},\tilde{D}_{\Dot{\alpha}}\}=i\frac{\partial}{\partial x^{\alpha\Dot{\alpha}}}=-p_{\alpha\Dot{\alpha}}.
\end{align}
The supersymmetry generators on the other hand take the form,
\begin{align}
\tilde{Q}_{\Dot{\alpha}}=\frac{\partial}{\partial \tilde{\theta}^{\Dot{\alpha}}}-\frac{i}{2}\theta^\alpha\frac{\partial}{\partial x^{\alpha\Dot{\alpha}}},~~Q_{\alpha}=\frac{\partial}{\partial\theta^\alpha}-\frac{i}{2}\tilde{\theta}^{\Dot{\alpha}}\frac{\partial}{\partial x^{\alpha\Dot{\alpha}}}.
\end{align}
The anti-commutator of the supersymmetry generators is of course, the momentum matrix.
\begin{align}
    \{Q_{\alpha},\tilde{Q}_{\Dot{\alpha}}\}=-i\frac{\partial}{\partial x^{\alpha\Dot{\alpha}}}=p_{\alpha\Dot{\alpha}}.
\end{align}
Our conventions for the Fourier transform are,
\begin{align}
    f(x)=\int \frac{d^4p}{(2\pi)^4}e^{i\lambda_{I\alpha}\tilde{\lambda}^I_{\Dot{\alpha}}x^{\alpha\Dot{\alpha}}}f(p).
\end{align}
Returning to the problem at hand, we apply $\tilde{D}_{\Dot{\alpha}}$ on the super-current which should yield zero due to the shortening condition (conservation). 
\begin{align}
    \tilde{D}_{\Dot{\alpha}}\mathbf{J}_{\frac{1}{2}}^{\Dot{\alpha}}(x,\theta,\tilde{\theta})=0.
\end{align}
This results in a set of equations all of which uniquely fix the value of $c_1=\frac{i}{2}$. As for $c_2$, we set it equal to $1$ as we discussed earlier since it can be reabsorbed into the normalization of the operators. Essentially, the bottom component is $\order{c_2^0}$, the middle components $O_2$ and $J_{\alpha\Dot{\alpha}}$ are $\order{c_2}$ whereas the top component $O_{\frac{1}{2}}^{\alpha}$ is  $\order{c_2^2}$ so a natural choice is simply $c_2=1$. If we keep $c_2$ arbitrary, then it comes out as an overall coefficient in correlators involving this super-field. For example, two point super-correlators go like $c_2^2$, three point super-correlators like $c_2^3$ and so on. Therefore we find,
\begin{align}
    &\mathbf{J}_{\frac{1}{2},\Dot{\alpha}}(x,\theta,\tilde{\theta})=\big(1+\frac{i}{2}\theta^\beta\tilde{\theta}^{\Dot{\beta}}\frac{\partial}{\partial x^{\beta\Dot{\beta}}}-\frac{\theta^2\tilde{\theta}^2}{16}\Box\big)\tilde{O}_{\frac{1}{2},\Dot{\alpha}}(x)+\bigg(\theta^\alpha \frac{\partial}{\partial  x^{\alpha \Dot{\alpha}}}+\frac{i}{4}\epsilon^{\alpha\beta}\theta^2\tilde{\theta}^{\Dot{\beta}}\frac{\partial^2}{\partial x^{\alpha \Dot{\alpha}}\partial x^{\beta\Dot{\beta}}}\bigg)O_2(x)\notag\\
   &+\bigg(\theta^\alpha +\frac{i}{4}\epsilon^{\alpha\beta}\theta^2\tilde{\theta}^{\Dot{\beta}}\frac{\partial}{\partial x^{\beta\Dot{\beta}}}\bigg)J_{\alpha \Dot{\alpha}}(x)-\frac{\theta^2}{2}\frac{\partial}{\partial x^{\alpha \Dot{\alpha}}}O_{\frac{1}{2}}^\alpha(x).
\end{align}
Generalizing to arbitrary spin, we find that the super-Penrose transform takes the form\footnote{The position space super-current is given by,
\begin{align}
&\mathbf{J}_{s,\alpha_1\cdots\alpha_m,\Dot{\alpha}_1\cdots\Dot{\alpha}_n}(x,\theta,\tilde{\theta})=\big(1+\frac{i}{2}\theta^\beta\tilde{\theta}^{\Dot{\beta}}\frac{\partial}{\partial x^{\beta\Dot{\beta}}}-\frac{\theta^2\tilde{\theta}^2}{16}\Box\big)\tilde{J}_{s,\alpha_1,\cdots\alpha_m,\Dot{\alpha}_1\cdots\Dot{\alpha}_n}(x)\notag\\
    &+\bigg(\theta^\alpha \frac{\partial}{\partial  x^{\alpha \Dot{\alpha}_1}}+\frac{i}{4}\epsilon^{\alpha\beta}\theta^2\tilde{\theta}^{\Dot{\beta}}\frac{\partial^2}{\partial x^{\alpha \Dot{\alpha}_1}\partial x^{\beta\Dot{\beta}}}\bigg)J_{s-\frac{1}{2},\alpha_1\cdots\alpha_m,\Dot{\alpha}_2\cdots\Dot{\alpha}_n}(x)\notag\\
   &+\bigg(\theta^\alpha +\frac{i}{4}\epsilon^{\alpha\beta}\theta^2\tilde{\theta}^{\Dot{\beta}}\frac{\partial}{\partial x^{\beta\Dot{\beta}}}\bigg)J_{s+\frac{1}{2},\alpha\alpha_1\cdots\alpha_m,\Dot{\alpha}_1\cdots\Dot{\alpha}_m}(x)-\frac{\theta^2}{2}\frac{\partial}{\partial x^{\alpha \Dot{\alpha}_1}}J_{s,\alpha_1\cdots\alpha_m,\Dot{\alpha}_2\cdots\Dot{\alpha}_n}^\alpha(x),\notag
\end{align}
where $s\in\frac{2\mathbb{Z}_{\ge 0}+1}{2}$ and $n=s+\frac{1}{2}$, $m=s-\frac{1}{2}$ with $n-m=1$.

}

\begin{align}\label{superPenrose}
    \mathbf{J}_{s,\alpha_1\cdots \alpha_m,\Dot{\alpha}_1\cdots\Dot{\alpha}_n}(x,\theta,\tilde{\theta})=\int \frac{d^4 \lambda}{\text{Vol}(GL(2))}\lambda_{I_1\alpha_1}\cdots\lambda_{I_m\alpha_m}\frac{\partial}{\partial\mu^{J_1}_{\Dot{\alpha}_1}}\cdots\frac{\partial}{\partial \mu^{J_n}_{\Dot{\alpha}_n}}\mathbf{J}_{s}^{I_1\cdots I_m J_1\cdots J_n}(\mathcal{Z})|_{\mathcal{X}},
\end{align}
where the super-incidence relations $\mathcal{X}$ are given by,
\begin{align}
    \mathcal{X}=\{\mu^{I\Dot{\alpha}}=x^{\alpha\Dot{\alpha}}\lambda^I_\alpha+\frac{i}{2}\theta^\alpha\tilde{\theta}^{\Dot{\alpha}}\lambda_\alpha^I,~~\eta^I= \theta^{\alpha}\lambda_\alpha^I\}.
\end{align}
This is extremely similar in form to the result for on-shell massless super-fields \cite{Ferber:1977qx}, with the main difference being the presence of the $SL(2)$ index.
This concludes our discussion of the supersymmetric Penrose transform that reconstructs the position super-space data from super-twistor space. We now turn to the classification of super-conformal invariant quantities in twistor space.
\subsection{The classification of super-conformal invariants}
The super-conformal Ward identities are given by,
\begin{align}
    \sum_{i=1}^{n}\mathcal{T}_{i\mathcal{B}}^{\mathcal{A}}\mathbf{\Psi}_n(\mathcal{Z}_1,\cdots,\mathcal{Z}_n)=0.
\end{align}
Given its identical form to its non supersymmetric counterpart, the solution space takes the same form too with twistors replaced by super-twistors. In particular, the most natural solution is the projective super delta function,
\begin{align}
    \int [d\tilde{C}]\delta(\tilde{C}_{ij,I}\mathcal{Z}_j^{I\mathcal{A}})f(\tilde{C})=\int [d\tilde{C}]\delta(\tilde{C}_{ij,I}Z_j^I)\delta(\tilde{C}_{ij,I}\eta_j^I)f(\tilde{C}).
\end{align}
This is exactly the form we will find in the super-Grassmannian construction. There are also solutions to the Ward identities involving the Levi-Civita symbol $\epsilon^{\mathcal{A}\mathcal{B}\mathcal{C}\mathcal{D}}$ but they will not feature in our analysis.

\subsection{The super-Grassmannian construction}
Lets start with the twistor space Grassmannian \eqref{twistorGrassmannian} which we repeat here for convenience:
\begin{align}
    \psi_n(Z)&=\int \frac{d^{n\times 2n}\tilde{C}}{\text{Vol}(GL(n))}\delta^{n\times 4}(\tilde{C}\cdot\Omega\cdot Z)A_n(C=\tilde{C}_{\perp},\tilde{C})\notag\\
    &=\int \frac{d^{n\times 2n}\tilde{C}}{\text{Vol}(GL(n))}\delta^{n\times 4}(\tilde{C}\cdot\Omega\cdot Z)\tilde{A}_n(\tilde{C}).
\end{align}
To obtain its super-twistor space counterpart we replace $Z\to \mathcal{Z}$ finding,
\begin{align}\label{supertwistorGrassmannian}
    \mathbf{\Psi}_n(\mathcal{Z})&=\int \frac{d^{n\times 2n}\tilde{C}}{\text{Vol}(GL(n))}\delta^{n\times 4|n\times 1}(\tilde{C}\cdot\Omega\cdot \mathcal{Z})A_n(C=\tilde{C}_{\perp},\tilde{C})\notag\\
    &=\int \frac{d^{n\times 2n}\tilde{C}}{\text{Vol}(GL(n))}\delta^{n\times 4|n\times 1}(\tilde{C}\cdot\Omega\cdot \mathcal{Z})\tilde{A}_n(\tilde{C}),
\end{align}
where we have,
\begin{align}
    \delta^{n\times 4|n\times 1}(\tilde{C}\cdot \Omega\cdot\mathcal{Z})=\delta^{n\times 4}(\tilde{C}\cdot\Omega\cdot Z)\delta^{n\times 1}(\tilde{C}\cdot\Omega\cdot \eta),
\end{align}
and the $2n\times 1$ Grassmann vector is given by,
\begin{align}
    \eta^I\equiv \eta=\begin{pmatrix}
        \eta_1^I\\
        \vdots\\
        \eta_n^I
    \end{pmatrix}.
\end{align}
Under a $GL(n)$ transformation of the $\tilde{C}$ matrix, we require,
\begin{align}
    \tilde{A}_n(\tilde{G}\tilde{C})=\frac{1}{\text{Det}(\tilde{G})^{2n-3}}\tilde{A}_n(\tilde{C}).
\end{align}
The easiest way to determine $\tilde{A}_n$ is to match with one of the known component correlators in the Grassmannan expansion and use it to read off the remaining component correlators. Before doing this however, we will first use the super-twistor space Grassmannian to derive its super spinor-helicity counterpart which was initially our main motivation for the twistor space detour. We will then discuss explicit examples of $2,3$ and $4$ point super-correlators.
\section{The Supersymmetric Symplectic Bi-Grassmannian}\label{sec:SuperBiGrassmannian}
We have used the natural generalization of twistors to super-twistors to derive the Grassmannian in super-twistor space. Using the inverse half-Fourier transform and undoing the gauge fixing of the $C$ matrix, we can now obtain the super-Grassmannian in spinor helicity variables. This is particularly useful as it allows for a direct comparison with our earlier bootstrapped results which have elegant expressions in terms of the minors of both the $C$ and $\tilde{C}$ matrices. Of course, one can continue to use minors of just $\tilde{C}$ to represent the Grassmannian $A_n$ by virtue of the symplectic orthogonality constraint. This is perhaps, an aesthetic choice, but we choose to work with both minors of $C$ and $\tilde{C}$ as it makes more manifest the $GL(1)$ rescaling property of the external operators.
\subsection{The construction}
Let us start with the supertwistor space Grassmannian \eqref{supertwistorGrassmannian} representation and write it in component notation.
\begin{align}
        \mathbf{\Psi}_n(\mathcal{Z})=\int \frac{d^{n\times 2n}\tilde{C}}{\text{Vol}(GL(n))}\delta^{n\times 2}(\tilde{C}\cdot\Omega\cdot\Lambda)\delta^{n\times 2}(\tilde{C}\cdot\Omega\cdot M)\delta^{n\times 1}(\tilde{C}\cdot\Omega\cdot\eta)\mathbf{A}_n(C=\tilde{C}_{\perp},\tilde{C}).
\end{align}
Performing an inverse half-Fourier transform from $\mu$ to $\tilde{\lambda}$ for each operator results in\footnote{One can follow analogous steps to the derivation in appendix \ref{app:HFT} to arrive at this result.},
\begin{align}
        \mathbf{\Psi}_n(\Lambda,\tilde{\Lambda},\eta)=\int \frac{d^{n\times 2n}\tilde{C}}{\text{Vol}(GL(n))}\delta^{n\times 2}(\tilde{C}\cdot\Omega\cdot\Lambda)\delta^{n\times 2}(\tilde{C}_{\perp}\cdot\Omega\cdot \tilde{\Lambda})\delta^{n\times 1}(\tilde{C}\cdot\Omega\cdot\eta)\mathbf{A}_n(C=\tilde{C}_{\perp},\tilde{C}).
\end{align}
Restoring the $C$ integral with its $\text{GL}(n)$ redundancy and the symplectic orthogonality constraint yields,
\begin{align}\label{SHSUSYGrassmannian}
    &\mathbf{\Psi}_n(\Lambda,\tilde{\Lambda},\eta)\notag\\&=\int \frac{d^{n\times 2n}C}{\text{Vol}(GL(n))}\int \frac{d^{n\times 2n}\tilde{C}}{\text{Vol}(GL(n))}\delta^{n\times n}(C\cdot\Omega\cdot\tilde{C}^T)\delta^{n\times 2}(C\cdot\Omega\cdot\tilde{\Lambda})\delta^{n\times 2}(\tilde{C}\cdot\Omega\cdot\Lambda)\delta^{n\times 1}(\tilde{C}\cdot\Omega\cdot\eta)\mathbf{A}_n(C,\tilde{C}).
\end{align}
For invariance under $GL(n)\times GL(n)$, we require,
\begin{align}\label{mathbfAnGGtildescaling}
    \mathbf{A}_n(G C,\tilde{G}\tilde{C})=\frac{1}{\text{Det}(G)^{n-2}}\frac{1}{\text{Det}(\tilde{G})^{n-1}}\mathbf{A}_n(C,\tilde{C}).
\end{align}
We see that the super-twistor construction has resulted in the quickest and simplest route to obtain the super-conformal symplectic bi-Grassmannian in spinor-helicity variables. A direct construction in spinor-helicity variables would first require finding the right Grassmann parameters, solving the supersymmetry and special super-conformal symmetry Ward identities and then trying to repackage this in the Grassmannian integral representation. In appendix \ref{app:SHGrassmann}, we perform this brute force analysis which ultimately results in \eqref{SHSUSYGrassmannian}. The simplicity of the twistor construction has allowed us to circumvent this tedious process and also automatically prove that our answer is super-conformal invariant.

We now evaluate the supersymmetric delta function explicitly for $n=2,3,4$ which will set the stage for our applications to theories of interest in the following sections. As mentioned earlier, we can either work with both the minors of $C$ and $\tilde{C}$ or just those of the latter. We make the former choice in what follows. Given the super-current expansion \eqref{Neq1supercurrent} in twistor space, we obtain the spinor-helicity counterpart,
\begin{align}\label{Neq1supercurrentSH}
    \mathbf{J}^{I_1\cdots I_mJ_1\cdots J_n}_s(\lambda,\tilde{\lambda},\eta)&=\tilde{J}^{I_1\cdots I_mJ_1\cdots J_n}_{s}(\lambda,\tilde{\lambda})+\eta^{I_1}J_{s-\frac{1}{2}}^{I_2\cdots I_m J_1\cdots J_n}(\lambda,\tilde{\lambda})+\eta_J J_{s+\frac{1}{2}}^{JI_1\cdots I_mJ_1\cdots J_n}(\lambda,\tilde{\lambda})\notag\\&-\frac{\eta^2}{2}J_s^{I_1\cdots I_m J_1\cdots J_n}(\lambda,\tilde{\lambda}).
\end{align}
We take $m=s-\frac{1}{2}$ and $n=s+\frac{1}{2}$ for our currents that have $s\in \mathbb{Z}_{\ge 0}+\frac{1}{2}$. Below, we first discuss the construction of two point functions of this super-current. We then calculate the explicit form of the supersymmetric delta function for $n=3,4$ which will be used in the subsequent sections for application to particular theories.

\subsection{Two point correlators}

For $n=2$, the supersymmetric delta function evaluates to,
\begin{align}\label{deltaCeta2pt}
    \delta^{2\times 1}(\tilde{C}\cdot\Omega\cdot\eta)=\frac{\eta_1^2}{2}(\tilde{1}_1\tilde{1}_2)+\frac{\eta_2^2}{2}(\tilde{2}_1\tilde{2}_2)+(\tilde{1}^{I_1}\tilde{2}^{I_2})\eta_{1I_1}\eta_{2I_2}.
\end{align}
For $s\in\mathbb{Z}_{\ge 0}+\frac{1}{2}$, we make the ansatz,
\begin{align}
    \mathbf{A}_2^{s,s}(C,\tilde{C})&=\bigg(\frac{(\mathbf{1}\mathbf{2})}{(\mathbf{1}\mathbf{2})\cdot(\mathbf{\tilde{1}}\mathbf{\tilde{2})}}\bigg)\bigg(\frac{(\mathbf{1}\mathbf{2})(\mathbf{\tilde{1}}\mathbf{\tilde{2}})}{(\mathbf{1}\mathbf{2})\cdot(\mathbf{\tilde{1}}\mathbf{\tilde{2})}}\bigg)^{s-\frac{1}{2}}.
\end{align}
For example we take $s=\frac{1}{2}$. We find,
\begin{align}
    \mathbf{A}_2^{\frac{1}{2},\frac{1}{2}}(C,\tilde{C})=\bigg(\frac{(\mathbf{1}\mathbf{2})}{(\mathbf{1}\mathbf{2})\cdot(\mathbf{\tilde{1}}\mathbf{\tilde{2})}}\bigg)=\bigg(\frac{(1^{I_1}2^{I_2})}{(1_I 2_J)(\tilde{1}^I\tilde{2}^J)}\bigg).
\end{align}
Let us now read off the component correlators using the super-current expansion and \eqref{deltaCeta2pt}. First we obtain,
\begin{align}\label{twopointsusy}
    \delta^{2\times 1}(\tilde{C}\cdot\Omega\cdot \eta)\mathbf{A}_2^{\frac{1}{2},\frac{1}{2}}=\bigg(\frac{\eta_1^2}{2}(\tilde{1}_1 \tilde{1}_2)+\frac{\eta_2^2}{2}(\tilde{2}_1\tilde{2}_2)\bigg)\bigg(\frac{(1^{I_1}2^{I_2})}{(1_I 2_J)(\tilde{1}^I\tilde{2}^J)}\bigg)+\eta_{1J_1}\eta_{2J_2}\bigg(\frac{(1^{I_1}2^{I_2})(\tilde{1}^{J_1}\tilde{2}^{J_2})}{(1_I 2_J)(\tilde{1}^I\tilde{2}^J)}\bigg).
\end{align}
The $O_2$ scalar correlator is given by the trace part of the coefficient of the $\order{\eta_1\eta_2}$ term. This is clearly a constant, consistent with the scalar two point result. The symmetric part on the other hand, corresponds to the spin$-1$ two point function which we find to be,
\begin{align}
    A_2^{g,g}=\bigg(\frac{(1^{(I_1}2^{(I_2})(\tilde{1}^{J_1)}\tilde{2}^{J_2)})}{(1_I 2_J)(\tilde{1}^I\tilde{2}^J)}\bigg)=\frac{(\mathbf{1}\mathbf{2})(\mathbf{\tilde{1}}\mathbf{\tilde{2}})}{(\mathbf{1}\mathbf{2})\cdot(\mathbf{\tilde{1}}\mathbf{\tilde{2}})}.
\end{align}
The coefficient of $\eta_1^2$ and $\eta_2^2$ denote the two orderings of the spin$-\frac{1}{2}$ two point function (anti-chiral chiral and chiral anti-chiral). We find,
\begin{align}
    A_{2}^{\chi,\tilde{\chi}}=(\tilde{1}_1\tilde{1}_2)\frac{(1^{I_1}2^{I_2})}{(1_I 2_J)(\tilde{1}^I\tilde{2}^J)},
\end{align}
which is indeed the correct answer \cite{Bala:2026trw}. Similarly, we find the correct $A_2^{\chi,\tilde{\chi}}$ component two point function. 
\subsection{Three point correlators}
We find, setting $n=3$ in the general formula, the following expression for the supersymmetric delta function:
\begin{align}\label{susydelta3pt}
    \delta^{3\times 1}(\tilde{C}\cdot\Omega\cdot\eta)&=\frac{\eta_1^2}{2}\bigg((\tilde{1}_1\tilde{1}_2 \tilde{2}^{I_2})\eta_{2I_2}+(\tilde{1}_1\tilde{1}_2 \tilde{3}^{I_3})\eta_{3I_3}\bigg)+\frac{\eta_2^2}{2}\bigg((\tilde{2}_1\tilde{2}_2 1^{I_1})\eta_{1I_1}+(\tilde{2}_1\tilde{2}_2\tilde{3}^{I_3})\eta_{3I_3}\bigg)\notag\\
    &+\frac{\eta_3^2}{2}\bigg((\tilde{3}_1\tilde{3}_2\tilde{1}^{I_1})\eta_{1I_1}+(\tilde{3}_1\tilde{3}_2\tilde{2}^{I_2})\eta_{2I_2})\bigg)+(\tilde{1}^{I_1}\tilde{2}^{I_2}\tilde{3}^{I_3})\eta_{1I_1}\eta_{2I_2}\eta_{3I_3}.
\end{align}
The full super-correlator is thus given by,
\begin{align}\label{Psi3SUSYgen}
    \mathbf{\Psi}_3=\int DC D\tilde{C}~\delta^{3\times 3}(C\cdot\Omega\cdot\tilde{C}^T)\delta^{3\times 2}(C\cdot \Omega\cdot\tilde{\Lambda})\delta^{3\times 2}(\tilde{C}\cdot\Omega\cdot\Lambda)\delta^{3\times 1}(\tilde{C}\cdot\Omega\cdot\eta)\mathbf{A}_3.
\end{align}
In the sections to follow, we will discuss examples of three point functions in two examples of $\mathcal{N}=1$ supersymmetric theories.
\subsection{Four point correlators}
Evaluating the supersymmetric delta function for $n=4$ results in,
\begin{align}\label{deltaCdotXi4pt}
    &\delta^{4\times 1}(\tilde{C}\cdot\Omega\cdot \eta)=(\tilde{C}_{1j,I}\eta_j^^I)(\tilde{C}_{2j,I}\eta_j^^I)(\tilde{C}_{3j,I}\eta_j^^I)(\tilde{C}_{4j,I}\eta_j^^I)\notag\\
    &=(\tilde{1}^{J_1}\tilde{2}^{J_2}\tilde{3}^{J_3}\tilde{4}^{J_4})\eta_{1J_1}\eta_{2J_2}\eta_{3J_3}\eta_{4J_4}\notag\\
    &+\frac{\eta_1^2}{2}\bigg((\tilde{1}_1\tilde{1}_2 \tilde{2}^{J_2}\tilde{3}^{J_3})\eta_{2J_2}\eta_{3J_3}+(\tilde{1}_1\tilde{1}_2 \tilde{2}^{J_2}\tilde{4}^{J_4})\eta_{2J_2}\eta_{4J_4}+(\tilde{1}_1\tilde{1}_2 \tilde{3}^{J_3}\tilde{4}^{J_4})\eta_{3J_3}\eta_{4J_4}\bigg)\notag\\
    &+\frac{\eta_2^2}{2}\bigg((\tilde{2}_1\tilde{2}_2\tilde{1}^{J_1}\tilde{3}^{J_3})\eta_{1J_1}\eta_{3J_3}+(\tilde{2}_1\tilde{2}_2\tilde{1}^{J_1}\tilde{4}^{J_4})\eta_{1J_1}\eta_{4J_4}+(\tilde{2}_1\tilde{2}_2\tilde{3}^{J_3}\tilde{4}^{J_4})\eta_{3J_3}\eta_{4J_4}\bigg)\notag\\
    &+\frac{\eta_3^2}{2}\bigg((\tilde{3}_1\tilde{3}_2\tilde{1}^{J_1}\tilde{2}^{J_2})\eta_{1J_1}\eta_{2J_2}+(\tilde{3}_1\tilde{3}_2\tilde{1}^{J_1}\tilde{4}^{J_4})\eta_{1J_1}\eta_{4J_4}+(\tilde{3}_1\tilde{3}_2\tilde{2}^{J_2}\tilde{4}^{J_4})\eta_{2J_2}\eta_{4J_4}\bigg)\notag\\
    &+\frac{\eta_4^2}{2}\bigg((\tilde{4}_1\tilde{4}_2\tilde{1}^{J_1}\tilde{2}^{J_2})\eta_{1J_1}\eta_{2J_2}+(\tilde{4}_1\tilde{4}_2\tilde{1}^{J_1}\tilde{3}^{J_3})\eta_{1J_1}\eta_{3J_3}+(\tilde{4}_1\tilde{4}_2\tilde{2}^{J_2}\tilde{3}^{J_3})\eta_{2J_2}\eta_{3J_3}\bigg)\notag\\
    &+\eta_1^2\eta_2^2 \tilde{S}+\eta_1^2\eta_3^2\tilde{U}+\eta_1^2\eta_4^2 \tilde{T}+\eta_3^2\eta_4^2 S+\eta_2^2\eta_4^2 U+\eta_2^2\eta_3^2 T,
\end{align}
where $\tilde{S}=4(\tilde{1}_1\tilde{1}_2\tilde{2}_1\tilde{2}_2)$, $\tilde{T}$ and $\tilde{U}$ are $(2\leftrightarrow 4)$ and $(2\leftrightarrow 3)$ exchanges of $\tilde{S}$, $S=4(\tilde{3}_1\tilde{3}_2\tilde{4}_1\tilde{4}_2)$ and $T$ and $U$ are $(2\leftrightarrow 4)$ and $(2\leftrightarrow 3)$ exchanges of $S$.

The full super-correlator is thus given by,
\begin{align}\label{Psi4SUSYgen}
    \mathbf{\Psi}_4=\int DC D\tilde{C}~\delta^{4\times 4}(C\cdot\Omega\cdot\tilde{C}^T)\delta^{4\times 2}(C\cdot \Omega\cdot\tilde{\Lambda})\delta^{4\times 2}(\tilde{C}\cdot\Omega\cdot\Lambda)\delta^{4\times 1}(\tilde{C}\cdot\Omega\cdot\eta)\mathbf{A}_4.
\end{align}
We will discuss particular examples of four point functions in the next two sections.
\section{$\mathcal{N}=1$ AdS$_5$ SYM Theory}\label{sec:SYM}
Having dealt with the general formalism for many a section now, we specialize to the interesting specific theory of $\mathcal{N}=1$ super Yang-Mills theory on a rigid AdS$_5$ spacetime. After discussing the spectrum of particles and their interactions, we then set up the super-conformal Grassmannian, evaluate the supersymmetric delta function and input one correlator, using it to obtain the remaining ones. Although we use inputs from the bulk Lagrangian to bootstrap the correlators (namely, the allowed field interactions), our analysis is entirely from the CFT$_4$ perspective and we do not explicitly perform any bulk Witten diagram computations.
\subsection{The spectrum and interactions}
Our main references for this subsection are \cite{Hosomichi:2012ek,Qiu:2013pta} which discuss $\mathcal{N}=1$ SYM in $\mathbb{R}^5$ and $S^5$.
  The Lagrangian density in five dimensional flat space features the Yang-Mills term, kinetic terms for the minimally coupled scalar and fermion fields as well as a Yukawa interaction.
The theory in flat space has the $R-$symmetry group $SU(2)$ (which is $SL(2)$ in the $(3,2)$ signature that we are working in). The gauge field and the scalar are $R-$symmetry singlets whereas the gluinos are in the doublet representation. Putting the theory on a AdS$_5$ background and demanding the closure of the supersymmetry algebra introduces masses for the scalar and fermionic fields. It also breaks the $R-$symmetry group from $SU(2)$ to $U(1)$ (which is $SL(2)$ to $GL(1)$ in our spacetime signature). For the masses we find,
\begin{align}
    m_{\varphi}^2=-4\implies \Delta=2,
\end{align}
where we used the AdS$_5$/CFT$_4$ relation $m_{\varphi}^2=\Delta(\Delta-4)$. This field $\varphi$ is the bulk dual of the $\Delta=2$ scalar which is one we have discussed in many prior examples. Similarly, we find that the gluinos have scaling dimension $\frac{5}{2}$. The gauge-field is massless and has scaling dimension $3$. The spectrum of particles in the theory include a gluon $g$, fermions $\chi$ and $\tilde{\chi}$ and a scalar $\varphi$. All these fields are in the adjoint representation of the gauge group $SU(N)$. Thus, for every colour, there are $4$ bosonic degrees of freedom ($3$ for the massless vector field in five dimensional spacetime and $1$ for the scalar) and $4$ fermionic degrees of freedom ($2$ for $\chi$ and $2$ for $\tilde{\chi}$). The conformal theory description of these bulk fields should thus consist of a conserved spin$-1$ current $J$, a $\Delta=2$ scalar $O_2$ and fermionic $\Delta=\frac{5}{2}$ spinors $O_{\frac{1}{2}}$ and $\tilde{O}_{\frac{1}{2}}$. We can encapsulate these physical degrees of freedom of this theory into the following on-shell super-field (on-shell from the AdS$_5$ bulk perspective):
\begin{align}\label{supercurrentexpansionSYM}
    \mathbf{J}^{I}_{\frac{1}{2}}=\tilde{O}_{\frac{1}{2}}^{I}+\eta^{I}O_2+\eta_J J^{IJ}-\frac{\eta_1^2}{2}O_{\frac{1}{2}}^{I}.
\end{align}
As we discussed, $\tilde{O}_{\frac{1}{2}}$ and $O_{\frac{1}{2}}$ represent left handed and right handed Weyl fermions respectively. $O_2$ is a $\Delta=2$ scalar and $J^{IJ}$ is a spin$-1$ conserved current dual to the bulk gluon with all the operators in the adjoint representation of $SU(N)$.
\subsection{Three point function}
For $s_1=s_2=s_3=\frac{1}{2}$ in the general three point function \eqref{Psi3SUSYgen} we write down the ansatz,
\begin{align}\label{A3susy3point}
    \mathbf{A}_3^{I_1 I_2 I_3}=\frac{(1^{I_1}2^{I_2}3^{I_3})}{\mathcal{K}^2},
\end{align}
that is consistent with covariance property \eqref{mathbfAnGGtildescaling}. Since the bottom component of the super-field is $\tilde{O}_{\frac{1}{2}}$, this ansatz is also consistent with the external $GL(1)$ rescaling. Multiplying this with the supersymmetric delta function \eqref{susydelta3pt} results in,
\begin{align}\label{susy3point}
     &\mathbf{F}_3^{I_1 I_2 I_3}(C,\tilde{C},\eta)\notag\\&=\delta^{3\times 1}(\tilde{C}\cdot\Omega\cdot\eta)\mathbf{A}_3^{I_1 I_2 I_3}=\Bigg[\frac{\eta_1^2}{2}\bigg((\tilde{1}_1\tilde{1}_2 \tilde{2}^{J_2})\eta_{2J_2}+(\tilde{1}_1\tilde{1}_2 \tilde{3}^{J_3})\eta_{3J_3}\bigg)+\frac{\eta_2^2}{2}\bigg((\tilde{2}_1\tilde{2}_2 1^{J_1})\eta_{1J_1}+(\tilde{2}_1\tilde{2}_2\tilde{3}^{J_3})\eta_{3J_3}\bigg)\notag\\
    &+\frac{\eta_3^2}{2}\bigg((\tilde{3}_1\tilde{3}_2\tilde{1}^{J_1})\eta_{1J_1}+(\tilde{3}_1\tilde{3}_2\tilde{2}^{J_2})\eta_{2J_2})\bigg)+(\tilde{1}^{J_1}\tilde{2}^{J_2}\tilde{3}^{J_3})\eta_{1J_1}\eta_{2J_2}\eta_{3J_3}\Bigg]\frac{(1^{I_1}2^{I_2}3^{I_3})}{\mathcal{K}^2}.
\end{align}
Using the super-field expansion \eqref{supercurrentexpansionSYM}, we find,
\begin{align}
    A_3^{\tilde{\chi}^{I_1},\chi^{I_2},\varphi}=\frac{1}{4}\frac{\partial}{\partial \eta_3^{I_3}}\epsilon^{IJ}\frac{\partial^2}{\partial \eta_2^I\partial \eta_2^J}\mathbf{F}^{I_1I_2I_3}_3|_{\eta_1=\eta_3=0}=\frac{(1^{I_1}2^{I_2}3^{I_3})(\tilde{2}_1\tilde{2}_2\tilde{3}_{I_3})}{2\mathcal{K}^2}.
\end{align}
Similarly,
\begin{align}
    A_3^{\chi^{I_1},\tilde{\chi}^{I_2},\varphi}=\frac{1}{4}\frac{\partial}{\partial \eta_3^{I_3}}\epsilon^{IJ}\frac{\partial^2}{\partial\eta_1^I\partial \eta_1^J}\mathbf{F}^{I_1I_2I_3}_3|_{\eta_2=\eta_3=0}=\frac{(1^{I_1}2^{I_2}3^{I_3})(\tilde{1}_1\tilde{1}_2\tilde{3}_{I_3})}{2\mathcal{K}^2}=-\frac{(\tilde{1}^{I_1}\tilde{2}^{I_2}\tilde{3}^{I_3})(2_1 2_2 3_{I_3})}{2\mathcal{K}^2},
\end{align}
which matches with the result in \cite{Bala:2026trw}. For the scalar three point function we consider,
\begin{align}
    A_3^{\varphi,\varphi,\varphi}=\frac{1}{8}\frac{\partial^3}{\partial \eta_3^{I_3}\partial\eta_2^{I_2}\partial\eta_1^{I_1}}\mathbf{F}_3^{I_1I_2I_3}=\frac{(1^{I_1}2^{I_2}3^{I_3})(\tilde{1}_{I_1}\tilde{2}_{I_2}\tilde{3}_{I_3})}{\mathcal{K}^2}=0,
\end{align}
by virtue of the three point relation,
\begin{align}
    (1^{I_1}2^{I_2}3^{I_3})(\tilde{1}_{I_1}\tilde{2}_{I_2}\tilde{3}_{I_3})=0.
\end{align}
This result is perfectly consistent with the fact that $\mathcal{N}=1$ SYM in five dimensions has no cubic scalar self-coupling term. Proceeding this way we can determine the remaining component correlators. The gluon three point function for example is,
\begin{align}
    A_3^{g,g,g}=\frac{(1^{(I_1}2^{(I_2}3^{(I_3})(\tilde{1}^{J_1)}\tilde{2}^{J_2)}\tilde{3}^{J_3)})}{\mathcal{K}^2}=\frac{(\mathbf{1}\mathbf{2}\mathbf{3})(\mathbf{\tilde{1}}\mathbf{\tilde{2}}\mathbf{\tilde{3}})}{\mathcal{K}^2},
\end{align}
which is the correct result for the Yang-Mills three point function. Note that the supersymmetry does not yield a $F^3$ interaction contribution which is again a consistent well known fact. We also find,
\begin{align}
    A_3^{\chi,\tilde{\chi},g}=\frac{(1^{I_1}2^{I_2}3^{(I_3})(\tilde{1}_1\tilde{1}_2\tilde{3}^{J_3)})}{\mathcal{K}^2},
\end{align}
and so on. Let us now proceed to the four point case.
\subsection{The four point gluon super-correlator}
We return to the expanded supersymmetric delta function \eqref{deltaCdotXi4pt} and the super-correlator \eqref{Psi4SUSYgen}. We focus on the colour ordered correlator where the $s$ and $t$ channels contribute. Using the super-field expansion, we see that the coefficient of $\eta_1^2\eta_3^2$ is $4A_4^{\chi^{I_1}\tilde{\chi}^{I_2}\chi^{I_3}\tilde{\chi}^{I_4}}$. Thus,
\begin{align}
    \mathbf{A}_4^{I_1I_2I_3I_4}=\frac{1}{(\tilde{1}_1\tilde{1}_2\tilde{3}_1\tilde{3}_2)}A_4^{\chi^{I_1}\tilde{\chi}^{I_2}\chi^{I_3}\tilde{\chi}^{I_4}}.
\end{align}
$A_4^{\chi^{I_1}\tilde{\chi}^{I_2}\chi^{I_3}\tilde{\chi}^{I_4}}$ receives contributions from gluon and scalar exchanges in both the $s$ and $t$ channels. Thus,
\begin{align}
    \text{Res}_{S=0}\text{Res}_{\tilde{S}=0}\mathbf{A}^{I_1I_2I_3I_4}&=\frac{1}{(\tilde{1}_1\tilde{1}_2\tilde{3}_1\tilde{3}_2)}\bigg(a_{\varphi}A_{3L}^{\chi^{I_1}\tilde{\chi}^{I_2}\varphi}A_{3R}^{\chi^{I_3}\tilde{\chi}^{I_4}\varphi}+a_{g}A_{3L}^{\chi^{I_1}\tilde{\chi}^{I_2} g_{IJ}}A_{3R}^{\chi^{I_3}\tilde{\chi}^{I_4} g^{IJ}}\bigg)\notag\\
    &=\frac{1}{(\tilde{1}_1\tilde{1}_2\tilde{3}_1\tilde{3}_2)\mathcal{K}_L^2\mathcal{K}_R^2}\bigg(a_{\varphi}(\tilde{1}^{I_1}\tilde{2}^{I_2}\tilde{s}_L^I)(2_1 2_2 s_{L,I})(\tilde{3}^{I_3}\tilde{4}^{I_4}\tilde{s}_R^J)(4_1 4_2 s_{R,J})\notag\\
    &+a_{g}(\tilde{1}^{I_1}\tilde{2}^{I_2}\tilde{s}_L^I)(2_1 2_2 s_{L}^{J})(\tilde{3}^{I_3}\tilde{4}^{I_4}\tilde{s}_{R,I})(4_1 4_2 s_{R,J})\bigg)\notag\\
    &=\frac{16}{(\tilde{1}_1\tilde{1}_2\tilde{3}_1\tilde{3}_2)}\bigg(\frac{-a_{\varphi}}{(\mathcal{T}-\mathcal{U})^2}(\tilde{1}^{I_1}\tilde{4}^{I_4}\tilde{3}_1\tilde{3}_2)(2^{I_2}3^{I_3}4_1 4_2)\notag\\&+\frac{a_g}{(\mathcal{T}-\mathcal{U})^2}(1^{I_1}2^{I_2}3^{I_3}4^{I_4})(\tilde{1}_1\tilde{1}_2\tilde{3}_1\tilde{3}_2)+\frac{a_g}{2}(1^{I_1}4^{I_4}2_1 2_2)(\tilde{2}^{I_2}\tilde{3}^{I_3}\tilde{1}_1\tilde{1}_2)\bigg).
\end{align}
A natural uplift (and also including the $t-$channel contribution taking into account the fermionic statistics) is,
\begin{align}\label{susybootstrap}
    \mathbf{A}^{I_1 I_2 I_3 I_4}=&\frac{16}{(\tilde{1}_1\tilde{1}_2\tilde{3}_1\tilde{3}_2)(\mathcal{S}+\mathcal{T}-\mathcal{U})^2}\bigg(-\frac{a_{\varphi}}{\mathcal{S}}(\tilde{1}^{I_1}\tilde{4}^{I_4}\tilde{3}_1\tilde{3}_2)(2^{I_2}3^{I_3}4_1 4_2)+\frac{a_g}{2\mathcal{S}}(1^{I_1}4^{I_4}2_1 2_2)(\tilde{2}^{I_2}\tilde{3}^{I_3}\tilde{1}_1\tilde{1}_2)\notag\\
    &-\big(-\frac{a_{\varphi}}{\mathcal{T}}(\tilde{1}^{I_1}\tilde{2}^{I_2}\tilde{3}_1\tilde{3}_2)(4^{I_4}3^{I_3}2_1 2_2)+\frac{a_g}{2\mathcal{T}}(1^{I_1}2^{I_2}4_1 4_2)(\tilde{4}^{I_4}\tilde{3}^{I_3}\tilde{1}_1\tilde{1}_2)\big)\bigg)\notag\\&+\frac{16 a_g(1^{I_1}2^{I_2}3^{I_3}4^{I_4})}{(\mathcal{S}+\mathcal{T}-\mathcal{U})^2}\bigg(\frac{1}{\mathcal{S}}+\frac{1}{\mathcal{T}}\bigg)+\mathbf{B}^{I_1 I_2 I_3 I_4},
\end{align}
where $\mathbf{B}^{I_1 I_2 I_3 I_4}$ has zero double-
residue at $S=0,\tilde{S}=0$ and $T=0,\tilde{T}=0$ (although it could potentially have poles at just $S=0$ or $\tilde{S}=0$ and similarly in the t and u channels).
The scalar four point function is obtained using,
\begin{align}
    &A^{\varphi\varphi\varphi\varphi}=(\tilde{1}_{I_1}\tilde{2}_{I_2}\tilde{3}_{I_3}\tilde{4}_{I_4})\mathbf{A}^{I_1 I_2 I_3 I_4}\notag\\
    &=\frac{\mathcal{S}^2 \left(5 \text{$a_g$}-2 a_{\varphi}\right)+2 \mathcal{S} \mathcal{T} \left(\text{$a_g$}+6 a_{\varphi}\right)+\mathcal{S} \mathcal{U} \left(3 \text{$a_g $}+2 a_{\varphi}\right)+\mathcal{T} \left(5 \text{$a_g $}\mathcal{T}-2 a_{\varphi} \mathcal{T}+3 a_g \mathcal{U}+2 a_{\varphi}\mathcal{U}\right)}{2 \mathcal{S} \mathcal{T} (\mathcal{S}+\mathcal{T}-\mathcal{U})^2}\notag\\
    &+(\tilde{1}_{I_1}\tilde{2}_{I_2}\tilde{3}_{I_3}\tilde{4}_{I_4})\mathbf{B}^{I_1 I_2 I_3 I_4}.
\end{align}
The double residue at $S=0,\tilde{S}=0$ is,
\begin{align}
    \text{Res}_{S=0}\text{Res}_{\tilde{S}=0}A^{\varphi\varphi\varphi\varphi}=\frac{(5\mathcal{T}+3\mathcal{U})a_g+2(-\mathcal{T}+\mathcal{U})a_{\varphi}}{2(\mathcal{T}-\mathcal{U})^2}.
\end{align}
Independently, we know that this equals,
\begin{align}
    \frac{(\mathcal{T}+\mathcal{U})}{(\mathcal{T}-\mathcal{U})^2},
\end{align}
which implies $a_g=\frac{1}{2}$, $a_\varphi=\frac{a_g}{2}$. This choice also results in the correct $t-$channel residue. Ultimately, what this does is set the super-correlator equal to,
\begin{align}\label{A4susy}
\mathbf{A}^{I_1 I_2 I_3 I_4}=\frac{8(1^{I_1}2^{I_2}3^{I_3}4^{I_4})}{(\mathcal{S}+\mathcal{T}-\mathcal{U})^2}\bigg(\frac{1}{\mathcal{S}}+\frac{1}{\mathcal{T}}\bigg)+\mathbf{B}^{I_1 I_2 I_3 I_4}.
\end{align}
At $S=0,\tilde{S}=0$, this quantity has a double-residue,
\begin{align}
    \frac{8(1^{I_1}2^{I_2}3^{I_3}4^{I_4})}{(\mathcal{T}-\mathcal{U})^2}|_{S=0,\tilde{S}=0}=\frac{(1^{I_1}2^{I_2}s_{L I})(3^{I_3}4^{I_4}s_{R}^{I})}{\mathcal{K}_L^2\mathcal{K}_R^2},
\end{align}
which represents a consistent factorization into $\mathbf{A}_{3L}\times\mathbf{A}_{3R}$ \eqref{A3susy3point}.
The scalar correlator then comes out to be,
\begin{align}
    A_4^{\varphi\varphi\varphi\varphi}=\frac{1}{(\mathcal{S}+\mathcal{T}-\mathcal{U})^2}\bigg(\frac{\mathcal{T}+\mathcal{U}}{\mathcal{S}}+\frac{\mathcal{S}+\mathcal{U}}{\mathcal{T}}+2\bigg)+(\tilde{1}_{I_1}\tilde{2}_{I_2}\tilde{3}_{I_3}\tilde{4}_{I_4})\mathbf{B}^{I_1 I_2 I_3 I_4}.
\end{align}
The gluon four point correlator takes the form,
\begin{align}
    A_4^{gggg}=\frac{8(1^{I_1}2^{I_2}3^{I_3}4^{I_4})(\tilde{1}^{J_1}\tilde{2}^{J_2}\tilde{3}^{J_3}\tilde{4}^{J_4})}{(\mathcal{S}+\mathcal{T}-\mathcal{U})^2}\bigg(\frac{1}{\mathcal{S}}+\frac{1}{\mathcal{T}}\bigg)+\mathbf{B}^{I_1 I_2 I_3 I_4}(\tilde{1}^{J_1}\tilde{2}^{J_2}\tilde{3}^{J_3}\tilde{4}^{J_4}),
\end{align}
which also clearly has the correct double-residues at $S=0,\tilde{S}=0$ and $T=0,\tilde{T}=0$ and is the answer we guessed while bootstrapping the gluon four point exchange contribution \eqref{ggggYMexchange}.

 Thus, we find that our super-correlator has the correct residues and thus fully represents the exchange contributions to the super-correlator. Therefore, up to possible AdS$_5$ contact diagram ambiguities, this represents the correct answer. A correct choice of $\mathbf{B}^{I_1 I_2 I_3 I_4}$ should fix all these ambiguities, enabling a full reconstruction of the correlator. We leave a more detailed analysis of the full correlator to a future work.

Returning to the remaining components, one can extract other component correlators where we find perfect agreement with the bootstrap yielding
consistent residues at the poles. This serves as a dual check on both our bootstrap algorithm as well as our supersymmetry construction to reproduce consistently factorizing component correlators.
\section{Extension to 
Supergravity}\label{sec:SUGRA}
Before we conclude, we now briefly discuss the extension of our formalism to supergravity theories in AdS$_5$. The graviton is dual to the stress tensor in AdS/CFT. In $\mathcal{N}=1$ supersymmetry, the graviton, gravi-photon and gravitino form a super-multiplet. From the conformal field theory perspective we consider the stress tensor multiplet:
\begin{align}
    \mathbf{J}_{\frac{3}{2}}^{IJK}=\tilde{J}_{\frac{3}{2}}^{IJK}+\eta^{I}J^{JK}+\eta_L T^{IJKL}-\frac{\eta^2}{2}\tilde{J}_{\frac{3}{2}}^{IJK}.
\end{align}
 Consider the following ansatz for the three point function (there is an implicit symmetrization):
 \begin{align}
     \mathbf{A}_3^{I_1 J_1 K_1,I_2 J_2 K_2,I_3 J_3 K_3}=\frac{(1^{I_1}2^{I_2}3^{I_3})(1^{J_1}2^{J_2}3^{J_3})(\tilde{1}^{K_1}\tilde{2}^{K_2}\tilde{3}^{K_3})}{\mathcal{K}^3}.
 \end{align}
 Using the super-field expansion and the supersymmetric delta function \eqref{susydelta3pt}, we find for example, the following graviton three point function 
 \begin{align}
     \frac{(\mathbf{1}\mathbf{2}\mathbf{3})^2(\mathbf{\tilde{1}}\mathbf{\tilde{2}}\mathbf{\tilde{3}})^2}{\mathcal{K}^3},
 \end{align}
 which is the result for Einstein gravity \cite{Bala:2026trw}. We also find that the spin$-1$ three point function evaluates to zero as a consequence of the identity $(1_I 2_J 3_K)(\tilde{1}^I\tilde{2}^J\tilde{3}^K)=0$ which is consistent with Furry's theorem for Abelian spin$-1$ currents. 

 Proceeding to four points, we are led to the ansatz,
 \begin{align}
     \mathbf{A}_4=\frac{(\mathbf{1}\mathbf{2}\mathbf{3}\mathbf{4})^2(\mathbf{\tilde{1}}\mathbf{\tilde{2}}\mathbf{\tilde{3}}\mathbf{\tilde{4}})}{(\mathcal{S}+\mathcal{T}-\mathcal{U})^3}\bigg(\frac{1}{\mathcal{S}}+\frac{1}{\mathcal{T}}\bigg).
 \end{align}
 This gives a graviton four point function,
 \begin{align}
     A_4^{GGGG}=\frac{(\mathbf{1}\mathbf{2}\mathbf{3}\mathbf{4})^2(\mathbf{\tilde{1}}\mathbf{\tilde{2}}\mathbf{\tilde{3}}\mathbf{\tilde{4}})^2}{(\mathcal{S}+\mathcal{T}-\mathcal{U})^3}\bigg(\frac{1}{\mathcal{S}}+\frac{1}{\mathcal{T}}\bigg),
 \end{align}
 which very much looks like its Yang-Mills counterpart. It consistently factorizes at $\mathcal{S}=0$ and $\mathcal{T}=0$ and thus represents the correct correlator up to possible contact diagram contributions. We leave a more detailed study of supergravity to the future.

\section{Discussion}\label{sec:Discussion}
We have developed various aspects of the Grassmannian formalism for four dimensional conformal field theories and five dimensional AdS boundary correlators in this paper. We developed the bootstrap principle for four point functions by deriving a factorization formula in terms of a product of three point functions. Using this we bootstrapped several examples of holographic exchange correlators demanding that they have the correct residues at the poles, very similar to the scattering amplitudes bootstrap principle. We also derived the twistor space Grassmannian via a half-Fourier transform. Further, we derived the Penrose transform that converts twistor space data to position space. Using the natural extension of twistors to super-twistors, we derived the supersymmetric Penrose transform for conserved super-currents and the super Grassmannian in super-twistor space. Via an inverse half-Fourier transform we then obtained the supersymmetric symplectic bi-Grassmannian in spinor-helicity variables. We discussed the construction of $n=2,3,4$ point super-correlators. We specialized to $\mathcal{N}=1$ super Yang-Mills theory on a rigid AdS$_5$ background and bootstrapped the four point super-correlator up to possible contact diagram ambiguities. We also discussed the extension to supergravity. We summarize our main results in table \ref{tab:main-results}. 

There are many interesting avenues for future work. The first is to extend the formalism to five and higher point functions by developing a bootstrap principle analogous to what we have done here at four points. Even within the study of four point functions, the answers we have bootstrapped correctly account for the exchange contributions (from the holographic perspective). Explicitly figuring out the required integration cycles to perform the four non-trivial Grassmannian integrals and obtaining the results in spinor-helicity variables is an important task to pursue. As we discussed in section \ref{sec:AdS5bootstrap}, the correlators we have constructed should correspond to certain discontinuities. Understanding the interplay between the contour choices and the discontinuities would be interesting similar to what was discussed in \cite{Arundine:2026fbr}. Classifying contact diagrams is also another important direction to pursue since our bootstrap correctly accounts for exchange contributions with contact contributions undetermined. Another fascinating problem would be to understand the flat-space limit of our results which should correspond to five dimensional flat space scattering amplitudes while also serving as another bootstrap condition to determine the full four point function. Developing recursion relations analogous to BCFW recursion \cite{Britto:2005fq}, such as what was done recently in CFT$_3$/AdS$_4$ \cite{Bala:2026lvw} is also an interesting area of research. In terms of our supersymmetric construction, a natural direction to pursue is to further study AdS$_5$ supergravity theories and derive the graviton four point function. Understanding the double copy and the colour-kinematics duality \cite{Bern:2019prr} in this framework would also be fascinating. Overall, it would be interesting to obtain a more geometric understanding of CFT$_4$/AdS$_5$ correlators similar to what has been achieved for scattering amplitudes \cite{Arkani-Hamed:2012zlh}.
\begin{table}[h!]
\centering
\renewcommand{\arraystretch}{1.5} 
\begin{tabular}{|l|l|}
\hline
\textbf{Main Result} & \textbf{Equation(s)} \\ \hline
Factorization formula in the bi-Grassmannian & \eqref{A4inA3A3}\\
\hline
 Bootstrapped four point functions &  \eqref{scalarfourpointJexchange}, \eqref{phiphiggwithphiexchange}, \eqref{phiphiggwithgexchange}, \eqref{ggggYMexchange}, \eqref{GGGGwithGexchange}, \eqref{susybootstrap}\\ \hline
Penrose transform for $\Delta=s+2$, $s\ge 0,s\in \frac{\mathbb{Z}_{\ge 0}}{2}$ & \eqref{PenroseTransform}\\ \hline
Twistor space Grassmannian & \eqref{twistorGrassmannian}\\ \hline
$\mathcal{N}=1$ Penrose transform for conserved super-currents & \eqref{superPenrose}\\ \hline
$\mathcal{N}=1$ super-Twistor space Grassmannian & \eqref{supertwistorGrassmannian}\\ \hline 
Supersymmetric symplectic bi-Grassmannian & \eqref{SHSUSYGrassmannian}\\ \hline 
$n=2,3,4$ point super-correlators & \eqref{twopointsusy}, \eqref{susy3point}, \eqref{A4susy}\\ \hline
\end{tabular}
\caption{Summary of Results}
\label{tab:main-results}
\end{table}


\appendix

\section{The Factorization Principle}\label{app:Factorization}
In this appendix, we derive one of our main technical results: The factorization formula in the language of the symplectic bi-Grassmannian. Our starting point is \eqref{SHDisc} which we repeat here for convenience. This formula is exactly the same as the starting point for the analogous derivation in (A)dS$_4$/CFT$_3$ performed in \cite{Arundine:2026fbr}.
\begin{align}\label{SHDisc1}
    \text{Disc}_{s^2}(\psi_4&)=\sum_{s}a_s\int\frac{d^4 q}{(2\pi)^4}\psi_3^{I_1\cdots I_{2s}}(p_1,p_2,q)\epsilon_{I_1J_1}\cdots\epsilon_{I_{2s}J_{2s}}\psi_3^{J_1\cdots J_{2s}}(-q,p_3,p_4)\notag\\
    &=\sum_{s}a_s\int\frac{d^4 \lambda d^4\tilde{\lambda}}{(2\pi)^4\text{Vol}(GL(2))}\psi_3^{I_1\cdots I_{2s}}(p_1,p_2,\lambda^{I\alpha},\tilde{\lambda}^{I\Dot{\alpha}})\epsilon_{I_1J_1}\cdots\epsilon_{I_{2s}J_{2s}}\psi_3^{J_1\cdots J_{2s}}(\lambda^{I\alpha},-\tilde{\lambda}^{I\Dot{\alpha}},p_3,p_4).
\end{align}
Our aim now is to cast both the left hand side and right hand side of this equation as four point Grassmannian integrals and compare the resulting integrands. This will be our four point factorization formula.
\subsection{The product of three point functions}
For the left three point function, we gauge-fix and work with the following matrices:
\begin{align}
    C_{3L}=\begin{pmatrix}
        c_{11}&1&c_{12}&0&c_{1s_L}&0\\
        c_{21}&0&c_{22}&1&c_{2s_L}&0\\
        c_{s_L 1}&0&c_{s_L 2}&0&c_{s_L s_L}&1
    \end{pmatrix},~\tilde{C}_{3L}=\begin{pmatrix}
        c_{11}&1&c_{21}&0&c_{s_L 1}&0\\
        c_{12}&0&c_{22}&1&c_{s_L 2}&0\\
        c_{1 s_L }&0&c_{2 s_L }&0&c_{s_L s_L}&1
    \end{pmatrix}.
\end{align}
The spinor-helicity variables are encapsulated into,
\begin{align}
    \Lambda_{3L}=\begin{pmatrix}
        \lambda_1^1&\lambda_1^2\\
        \rho_1^1&\rho_1^2\\
        \lambda_2^1&\lambda_2^2\\
        \rho_2^1&\rho_2^2\\
        \lambda^1&\lambda^2\\
        \rho^1&\rho^2
    \end{pmatrix},~\tilde{\Lambda}_{3L}=\begin{pmatrix}
        \tilde{\lambda}_1^1&\tilde{\lambda}_1^2\\
        \tilde{\rho}_1^1&\tilde{\rho}_1^2\\
        \tilde{\lambda}_2^1&\tilde{\lambda}_2^2\\
        \tilde{\rho}_2^1&\tilde{\rho}_2^2\\
        \tilde{\lambda}^1&\tilde{\lambda}^2\\
        \tilde{\rho}^1&\tilde{\rho}^2
    \end{pmatrix}
\end{align}
As for the right three point function, we choose a similar gauge fixing resulting in,
\begin{align}
    C_{3R}=\begin{pmatrix}
        c_{33}&1&c_{34}&0&c_{3s_R}&0\\
        c_{43}&0&c_{44}&1&c_{4s_R}&0\\
        c_{s_R 3}&0&c_{s_R 4}&0&c_{s_R s_R}&1
    \end{pmatrix},~\tilde{C}_{3R}=\begin{pmatrix}
        c_{33}&1&c_{43}&0&c_{s_R 3}&0\\
        c_{34}&0&c_{44}&1&c_{s_R 4}&0\\
        c_{3 s_R }&0&c_{4 s_R }&0&c_{s_R s_R}&1
    \end{pmatrix}.
\end{align}
The spinor-helicity variables are encapsulated into,
\begin{align}
    \Lambda_{3R}=\begin{pmatrix}
        \lambda_3^1&\lambda_3^2\\
        \rho_3^1&\rho_3^2\\
        \lambda_4^1&\lambda_4^2\\
        \rho_4^1&\rho_4^2\\
        \lambda^1&\lambda^2\\
        \rho^1&\rho^2
    \end{pmatrix},~\tilde{\Lambda}_{3R}=\begin{pmatrix}
        \tilde{\lambda}_3^1&\tilde{\lambda}_3^2\\
        \tilde{\rho}_3^1&\tilde{\rho}_3^2\\
        \tilde{\lambda}_4^1&\tilde{\lambda}_4^2\\
        \tilde{\rho}_4^1&\tilde{\rho}_4^2\\
        -\tilde{\lambda}^1&-\tilde{\lambda}^2\\
        -\tilde{\rho}^1&-\tilde{\rho}^2
    \end{pmatrix}
\end{align}
Thus, the RHS of \eqref{SHDisc1} (suppressing the sum $\sum_{s} a_s$) takes the form,
\begin{align}\label{3x3step1}
    &\int\frac{d^2 \lambda d^2\rho d^2\tilde{\lambda}d^2\tilde{\rho}}{(2\pi)^4\text{Vol}(GL(2))}\bigg(\int \prod_{i,j=1,2,s_L}dc_{ij}~\delta^{3\times 2}(C_{3L,ij,I}\cdot\tilde{\Lambda}_{3L,j}^{I\Dot{\alpha}})\delta^{3\times 2}(\tilde{C}_{3L,ij,I}\Lambda_{3L,j}^{I\alpha})A_{3L}(C_{3L},\tilde{C}_{3L})\bigg)\notag\\&\times\bigg(\int\prod_{i,j=3,4,s_R}dc_{ij}~\delta^{3\times 2}(C_{3R,ij,I}\cdot\tilde{\Lambda}_{3R,j}^{I\Dot{\alpha}})\delta^{3\times 2}(\tilde{C}_{3R,ij,I}\Lambda_{3R,j}^{I\alpha})A_{3R}(C_{3R},\tilde{C}_{3R})\bigg).
\end{align}
We have $9+9=18$ integrals over the $c_{ij}$ and $8$ integrals over the intermediate momenta for a total of $26$ integrals to solve. There are a total of $12+12=24$ delta functions and also the volume factor that we need to take into account which can be used to perform $4$ integrals. $4$ of the delta functions should result in the momentum conserving delta function and thus this leaves $2$ non-trivial integrals. This is consistent with the fact that a general four point function has $4$ non-trivial integrals to perform, thus indicating that $2$ additional minors are localized when taking a discontinuity.

Returning to \eqref{3x3step1}, we use $8$ of the delta functions to determine $\lambda,\rho,\tilde{\lambda}$ and $\tilde{\rho}$. These are,
\begin{align}
    &\delta^2(c_{s_L 1}\tilde{\lambda}_1^{\Dot{\alpha}}+c_{s_L 2}\tilde{\lambda}_2^{\Dot{\alpha}}+c_{s_L s_L}\tilde{\lambda}^{\Dot{\alpha}}+\tilde{\rho}^{\Dot{\alpha}})\delta^2(c_{s_R 3}\tilde{\lambda}_3^{\Dot{\alpha}}+c_{s_R 4}\tilde{\lambda}_4^{\Dot{\alpha}}-c_{s_R s_R}\tilde{\lambda}^{\Dot{\alpha}}-\tilde{\rho}^{\Dot{\alpha}})\notag\\
    &\times \delta^2(c_{s_L s_L}\lambda^\alpha+c_{1s_L}\lambda_1^\alpha+c_{2 s_L}\lambda_2^\alpha+\rho^\alpha)\delta^2(c_{s_R s_R}\lambda^\alpha+c_{3 s_R}\lambda_3^\alpha+c_{4 s_R}\lambda_4^\alpha+\rho^\alpha)
\end{align}
We find,
\begin{align}
    &\tilde{\lambda}^{\Dot{\alpha}}=\frac{-\sum_{i=1}^{2}c_{s_L i}\tilde{\lambda}_i^{\Dot{\alpha}}-\sum_{i=3}^{4}c_{s_R i}\tilde{\lambda}_i^{\Dot{\alpha}}}{(c_{s_L s_L}-c_{s_R s_R})},~~~\tilde{\rho}^{\Dot{\alpha}}=\frac{c_{s_R s_R}\sum_{i=1}^{2}c_{s_L i}\tilde{\lambda}_i^{\Dot{\alpha}}+c_{s_L s_L}\sum_{i=3}^{4}c_{s_R i}\tilde{\lambda}_i^{\Dot{\alpha}}}{(c_{s_L s_L}-c_{s_R s_R})},\notag\\
    &\lambda^\alpha=\frac{-\sum_{i=1}^{2}c_{is_L}\lambda_i^\alpha+\sum_{i=3}^{4}c_{i s_R}\lambda_i^\alpha}{(c_{s_L s_L}-c_{s_R s_R})},~~~\rho^\alpha=\frac{c_{s_R s_R}\sum_{i=1}^{2}c_{is_L}\lambda_i^\alpha-c_{s_L s_L}\sum_{i=3}^{4}c_{is_R}\lambda_i^\alpha}{(c_{s_L s_L}-c_{s_R s_R})}.
\end{align}
This also results in the Jacobian factor,
\begin{align}
    \Gamma_1=\frac{1}{(c_{s_L s_L}-c_{s_R s_R})^4}.
\end{align}
We can now combine the remaining delta function constraints to the form,
\begin{align}
    \delta^{4\times 2}(C_{4 ij,I}\tilde{\Lambda}_{4j}^{I\Dot{\alpha}})\delta^{4\times 2}(\tilde{C}_{4 ij,I}\Lambda_{4j}^{I\alpha}),
\end{align}
where the 4 point Grassmannian matrices corresponding to factorization kinematics are given by,
\begin{align}
    &C_4=\begin{pmatrix}
c_{11}-\dfrac{c_{1s_L}c_{s_L1}}{c_{s_Ls_L}-c_{s_Rs_R}}
&
1
&
c_{12}-\dfrac{c_{1s_L}c_{s_L2}}{c_{s_Ls_L}-c_{s_Rs_R}}
&
0
&
-\dfrac{c_{1s_L}c_{s_R3}}{c_{s_Ls_L}-c_{s_Rs_R}}
&
0
&
-\dfrac{c_{1s_L}c_{s_R4}}{c_{s_Ls_L}-c_{s_Rs_R}}
&
0
\\[6pt]
c_{21}-\dfrac{c_{2s_L}c_{s_L1}}{c_{s_Ls_L}-c_{s_Rs_R}}
&
0
&
c_{22}-\dfrac{c_{2s_L}c_{s_L2}}{c_{s_Ls_L}-c_{s_Rs_R}}
&
1
&
-\dfrac{c_{2s_L}c_{s_R3}}{c_{s_Ls_L}-c_{s_Rs_R}}
&
0
&
-\dfrac{c_{2s_L}c_{s_R4}}{c_{s_Ls_L}-c_{s_Rs_R}}
&
0
\\[6pt]
\dfrac{c_{3s_R}c_{s_L1}}{c_{s_Ls_L}-c_{s_Rs_R}}
&
0
&
\dfrac{c_{3s_R}c_{s_L2}}{c_{s_Ls_L}-c_{s_Rs_R}}
&
0
&
c_{33}+\dfrac{c_{3s_R}c_{s_R3}}{c_{s_Ls_L}-c_{s_Rs_R}}
&
1
&
c_{34}+\dfrac{c_{3s_R}c_{s_R4}}{c_{s_Ls_L}-c_{s_Rs_R}}
&
0
\\[6pt]
\dfrac{c_{4s_R}c_{s_L1}}{c_{s_Ls_L}-c_{s_Rs_R}}
&
0
&
\dfrac{c_{4s_R}c_{s_L2}}{c_{s_Ls_L}-c_{s_Rs_R}}
&
0
&
c_{43}+\dfrac{c_{4s_R}c_{s_R3}}{c_{s_Ls_L}-c_{s_Rs_R}}
&
0
&
c_{44}+\dfrac{c_{4s_R}c_{s_R4}}{c_{s_Ls_L}-c_{s_Rs_R}}
&
1
\end{pmatrix},
\end{align}
and,
\begin{align}
    \tilde{C}_4=
\begin{pmatrix}
c_{11}-\dfrac{c_{1s_L}c_{s_L1}}{c_{s_Ls_L}-c_{s_Rs_R}}
&
1
&
c_{21}-\dfrac{c_{2s_L}c_{s_L1}}{c_{s_Ls_L}-c_{s_Rs_R}}
&
0
&
\dfrac{c_{3s_R}c_{s_L1}}{c_{s_Ls_L}-c_{s_Rs_R}}
&
0
&
\dfrac{c_{4s_R}c_{s_L1}}{c_{s_Ls_L}-c_{s_Rs_R}}
&
0
\\[6pt]
c_{12}-\dfrac{c_{1s_L}c_{s_L2}}{c_{s_Ls_L}-c_{s_Rs_R}}
&
0
&
c_{22}-\dfrac{c_{2s_L}c_{s_L2}}{c_{s_Ls_L}-c_{s_Rs_R}}
&
1
&
\dfrac{c_{3s_R}c_{s_L2}}{c_{s_Ls_L}-c_{s_Rs_R}}
&
0
&
\dfrac{c_{4s_R}c_{s_L2}}{c_{s_Ls_L}-c_{s_Rs_R}}
&
0
\\[6pt]
-\dfrac{c_{1s_L}c_{s_R3}}{c_{s_Ls_L}-c_{s_Rs_R}}
&
0
&
-\dfrac{c_{2s_L}c_{s_R3}}{c_{s_Ls_L}-c_{s_Rs_R}}
&
0
&
c_{33}+\dfrac{c_{3s_R}c_{s_R3}}{c_{s_Ls_L}-c_{s_Rs_R}}
&
1
&
c_{43}+\dfrac{c_{4s_R}c_{s_R3}}{c_{s_Ls_L}-c_{s_Rs_R}}
&
0
\\[6pt]
-\dfrac{c_{1s_L}c_{s_R4}}{c_{s_Ls_L}-c_{s_Rs_R}}
&
0
&
-\dfrac{c_{2s_L}c_{s_R4}}{c_{s_Ls_L}-c_{s_Rs_R}}
&
0
&
c_{34}+\dfrac{c_{3s_R}c_{s_R4}}{c_{s_Ls_L}-c_{s_Rs_R}}
&
0
&
c_{44}+\dfrac{c_{4s_R}c_{s_R4}}{c_{s_Ls_L}-c_{s_Rs_R}}
&
1
\end{pmatrix}.
\end{align}
We now introduce the matrices,
\begin{align}\label{Bmatrix}
    B_4=\begin{pmatrix}
b_{11} & 1 & b_{12} & 0 & b_{13} & 0 & b_{14} & 0\\
b_{21} & 0 & b_{22} & 1 & b_{23} & 0 & b_{24} & 0\\
b_{31} & 0 & b_{32} & 0 & b_{33} & 1 & b_{34} & 0\\
b_{41} & 0 & b_{42} & 0 & b_{43} & 0 & b_{44} & 1
\end{pmatrix},
\end{align}
and,
\begin{align}\label{Btildematrix}
    \tilde{B}_4=\begin{pmatrix}
b_{11} & 1 & b_{21} & 0 & b_{31} & 0 & b_{41} & 0\\
b_{12} & 0 & b_{22} & 1 & b_{32} & 0 & b_{42} & 0\\
b_{13} & 0 & b_{23} & 0 & b_{33} & 1 & b_{43} & 0\\
b_{14} & 0 & b_{24} & 0 & b_{34} & 0 & b_{44} & 1
\end{pmatrix}.
\end{align}
We set them equal to the matrices $C_4$ and $\tilde{C}_4$ by introducing the delta functions (the below delta functions are only over the columns that are not gauge-fixed to form the identity matrix).
\begin{align}
    \delta^{4\times 4}(C_4-B_4).
\end{align}
The aim of introducing the $B$ matrix is simply to re-write the product of the three point functions as an integral of the form $\int \prod_{i=1,2,3,4}d b_{ij}$ as is appropriate for a four point function. By inserting this delta function, one can now solve for $c_{11},c_{12},c_{21},c_{22},c_{33},c_{34},c_{43},c_{44}$ without picking up any Jacobian factors. Using the constraints coming from the (first row, fifth column), (first row, seventh column), (third row, first column) and (third row, third column) of equating $C_4$ and $B_4$ results in,
\begin{align}
    &c_{sR_3}=\frac{-b_{13}(c_{s_L s_L}-c_{s_R s_R})}{c_{1s_L}},c_{s_R 4}=\frac{-b_{14}(c_{s_L s_L}-c_{s_R s_R})}{c_{1 s_L}}\notag\\
    &c_{s_L 1}=\frac{b_{31}(c_{s_L s_L}-c_{s_R s_R})}{c_{3 s_R}},c_{s_L 2}=\frac{b_{32}(c_{s_L s_L}-c_{s_R s_R})}{c_{3s_R}}.
\end{align}
This results in a non-trivial Jacobian factor,
\begin{align}
    \Gamma_2=\frac{(c_{s_L s_L}-c_{s_R s_R})^4}{c_{1s_L}^2 c_{3s_R}^2}.
\end{align}
Further, we can use the constraints from the (second row, fifth column) and (fourth row, first column) from equating $C_4$ and $B_4$ which results in,
\begin{align}
    c_{2s_L}=\frac{b_{23}c_{1s_L}}{b_{13}},c_{4s_R}=\frac{b_{41}c_{3s_R}}{b_{31}}.
\end{align}
This brings about the Jacobian factor,
\begin{align}
    \Gamma_3=\frac{c_{1s_L}c_{3s_R}}{b_{13}b_{31}}.
\end{align}
We then end up with two delta functions,
\begin{align}
    \delta(b_{13}b_{24}-b_{14}b_{23})\delta(b_{31}b_{42}-b_{41}b_{32}),
\end{align}
with the factor,
\begin{align}
    \Gamma_4=b_{31}b_{13}.
\end{align}
The net Jacobian factor after all these integrals is,
\begin{align}
    \Gamma=\Gamma_1\Gamma_2\Gamma_3\Gamma_4=\frac{1}{c_{1s_L}c_{3s_R}}
\end{align}
Thus, we obtain the following integral representation.
\begin{align}
    &\int \prod_{i,j=1,2,3,4}db_{ij}\delta^{4\times 2}(\tilde{B}_{4i,jI}\Lambda_{4j}^{I\alpha})\delta^{4\times 2}(B_{4i,jI}\tilde{\Lambda}_{4j}^{I\Dot{\alpha}})\delta(b_{13}b_{24}-b_{14}b_{23})\delta(b_{31}b_{42}-b_{41}b_{32})\notag\\&\times\int \frac{dc_{1s_L}dc_{3s_R}dc_{s_L s_L}dc_{s_R s_R}}{\text{Vol}(GL(2))c_{1s_L}c_{3s_R}}A_{3L}\times A_{3R}|_{C_4=B_4}.
\end{align}
One can now peform the Faddeev-Popov method to choose $c_{1s_L}=1,c_{3s_R}=1,c_{s_L s_L}=1,c_{s_R s_R}=-1$. Another key point to note is that for the $B_4$ and $\tilde{B}_4$ matrices we respectively have,
\begin{align}
    &\tilde{S}\sim(1_1 1_2 2_1 2_2)=b_{14}b_{23}- b_{13}b_{24},\notag\\
    &S\sim (\tilde{1}_1\tilde{1}_2\tilde{2}_1\tilde{2}_2)=b_{41}b_{32}-b_{31}b_{42}.
\end{align}
Thus, we can write down the integral as,
\begin{align}
     &\int \prod_{i,j=1,2,3,4}db_{ij}\delta^{4\times 2}(\tilde{B}_{4i,jI}\Lambda_{4j}^{I\alpha})\delta^{4\times 2}(B_{4i,jI}\tilde{\Lambda}_{4j}^{I\Dot{\alpha}})\delta(S)\delta(\tilde{S})A_{3L}\times A_{3R}.
\end{align}
which is a beautiful and simple result. Its covariantization is,
\begin{align}\label{3x3eqn1}
    \int \frac{d^{4\times 2}C}{\text{Vol}(GL(2))}\int\frac{d^{4\times 2}\tilde{C}}{\text{Vol}(GL(2))}\delta^{4\times 4}(C\cdot\Omega\cdot\tilde{C}^T)\delta^{4\times 2}(\tilde{C}_{ij,I}\Lambda_j^{I\alpha})\delta^{4\times 2}(C_{ij,I}\tilde{\Lambda}_j^{I\Dot{\alpha}})\delta(S)\delta(\tilde{S})A_{3L}\times A_{3R}.
\end{align}
This is very analogous to the factorization formula in CFT$_3$/AdS$_4$ that was derived in \cite{Arundine:2026fbr} with the difference being that we have two, rather than one minor being localized in the CFT$_4$/AdS$_5$ context.
We will now compute the LHS of \eqref{SHDisc1} and equate it to the \eqref{3x3eqn1}, to obtain the factorization formula for $A_4$.
\subsection{The discontinuity of the four point function}
Next, we deal with the discontinuity of the four point function which is the left hand side of \eqref{SHDisc1}. Let us write down the (gauge-fixed) Grassmannian representation of the four point function,
\begin{align}
    \psi_4=\int \prod_{i,j=1,2,3,4}db_{ij}\delta^{4\times 2}(B_{ij,I}\tilde{\Lambda}_j^{I\Dot{\alpha}})\delta^{4\times 2}(\tilde{B}_{ij,I}\Lambda_j^{I\alpha})A_4(B,\tilde{B}),
\end{align}
where $B$ and $\tilde{B}$ are given in \eqref{Bmatrix} and \eqref{Btildematrix}.
We make the assumption that $A_4(B,\tilde{B})$ only has simple poles in $\tilde{S}\sim b_{13}b_{24}-b_{14}b_{23}$ and $S\sim b_{31}b_{42}-b_{41}b_{32}$. Thus we write,
\begin{align}
    A_4(B,\tilde{B})=\frac{R_4(B,\tilde{B})}{S \tilde{S}}+\text{Less singular terms},
\end{align}
where $R_4(B,\tilde{B})$ is regular at $S=\tilde{S}=0$. To figure out how the discontinuity with respect to $s^2$ acts on this quantity, we need to determine how $S$ and $\tilde{S}$ depend on $s^2$. To proceed, let us split the external data as is appropriate for the $s-$channel. Define the $2\times 2$ matrices,
\begin{align}
    &\lambda_L=\begin{pmatrix}
        \lambda_1^\alpha\\
        \lambda_2^\alpha
    \end{pmatrix},\rho_L=\begin{pmatrix}
        \rho_1^\alpha\\
        \rho_2^\alpha
    \end{pmatrix},\tilde{\lambda}_L=\begin{pmatrix}
        \tilde{\lambda}_1^{\Dot{\alpha}}\\
        \tilde{\lambda}_2^{\Dot{\alpha}}
    \end{pmatrix},\tilde{\rho}_L=\begin{pmatrix}
        \tilde{\rho}_1^{\Dot{\alpha}}\\
        \tilde{\rho}_2^{\Dot{\alpha}}.
    \end{pmatrix}
\end{align}
\begin{align}
    &\lambda_R=\begin{pmatrix}
        \lambda_3^\alpha\\
        \lambda_4^\alpha
    \end{pmatrix},\rho_R=\begin{pmatrix}
        \rho_3^\alpha\\
        \rho_4^\alpha
    \end{pmatrix},\tilde{\lambda}_R=\begin{pmatrix}
        \tilde{\lambda}_3^{\Dot{\alpha}}\\
        \tilde{\lambda}_4^{\Dot{\alpha}}
    \end{pmatrix},\tilde{\rho}_R=\begin{pmatrix}
        \tilde{\rho}_3^{\Dot{\alpha}}\\
        \tilde{\rho}_4^{\Dot{\alpha}}.
    \end{pmatrix}
\end{align}
In the gauge we are working in,
\begin{align}
    \psi_4&=\int \prod_{i,j=1,2,3,4}db_{ij}\delta^{4\times 2}(B_{ij,I}\tilde{\Lambda}_j^{I\Dot{\alpha}})\delta^{4\times 2}(\tilde{B}_{ij,I}\Lambda_j^{I\alpha})A_4(B,\tilde{B})\notag\\
    &=\int \prod_{i,j=1,2,3,4}db_{ij}\delta^{4\times 2}(\rho_i+b_{ji}\lambda_j)\delta^{4\times 2}(\tilde{\rho}_i+b_{ij}\tilde{\lambda}_j)A_4(b_{ij}).
\end{align}
We also define,
\begin{align}
\begin{pmatrix}
    b_{11}&b_{12}&b_{13}&b_{14}\\
    b_{21}&b_{22}&b_{23}&b_{24}\\
    b_{31}&b_{32}&b_{33}&b_{34}\\
    b_{41}&b_{42}&b_{43}&b_{44}
\end{pmatrix}=\begin{pmatrix}
    A_{2\times 2}&B_{2\times 2}\\
    c_{2\times 2}&D_{2\times 2}.
\end{pmatrix}
\end{align}
The four point function constraints can then be written as,
\begin{align}
    &\begin{pmatrix}
        \rho_L\\
        \rho_R
    \end{pmatrix}=-\begin{pmatrix}
        A^T&c^T\\
        B^T&D^T
    \end{pmatrix}\begin{pmatrix}
        \lambda_L\\
        \lambda_R
    \end{pmatrix},\notag\\
    & \begin{pmatrix}
        \tilde{\rho}_L\\
        \tilde{\rho}_R
    \end{pmatrix}=-\begin{pmatrix}
        A&B\\
        c&D
    \end{pmatrix}\begin{pmatrix}
        \tilde{\lambda}_L\\
        \tilde{\lambda}_R
    \end{pmatrix}.
\end{align}
Therefore,
\begin{align}
    &\rho_L=-A^T\lambda_L-c^T \lambda_R\label{eq1}\\
    &\rho_R=-B^T \lambda_L-D^T \lambda_R\label{eq2}\\
    &\tilde{\rho}_L=-A \tilde{\lambda}_L-B\tilde{\lambda}_R\label{eq3}\\
    &\tilde{\rho}_R=-c\tilde{\lambda}_L-D\tilde{\lambda}_R\label{eq4}.
\end{align}
From \eqref{eq1}, we find,
\begin{align}
    &A^T\lambda_L=-\rho_L-c^T\lambda_R\implies A^T=-\rho_L(\lambda_L)^{-1}-c^T\lambda_R(\lambda_L)^{-1}\notag\\
    &\implies A=-(\lambda_L)^{-T}\big(\rho_L^T+\lambda_R^T c\big).
\end{align}
For \eqref{eq4}, we find,
\begin{align}
    &D\tilde{\lambda}_R=-\tilde{\rho}_R-c\tilde{\lambda}_L\implies D=-(\tilde{\rho}_R+c\tilde{\lambda}_L)(\tilde{\lambda}_R)^{-1}.
\end{align}
From \eqref{eq3} we obtain,
\begin{align}
    &B\tilde{\lambda}_R=-(\tilde{\rho}_L+A\tilde{\lambda}_L)\implies B=-(\tilde{\rho}_L+A\tilde{\lambda}_L)(\tilde{\lambda}_R)^{-1}\notag\\
    &\implies B=(-\tilde{\rho}_L+(\lambda_L)^{-T}(\rho_L^T+\lambda_R^T c)\tilde{\lambda}_L)(\tilde{\lambda}_R)^{-1}.
\end{align}
Substituting the solutions for $A,D$ and $B$ in \eqref{eq2} yields,
\begin{align}
    \rho_R&=-(\tilde{\lambda}_R)^{-T}\bigg(-\tilde{\rho}_L^T\lambda_L+\tilde{\lambda}_L^T(\rho_L+c^T\lambda_R)\bigg)+(\tilde{\lambda}_R)^{-T}(\tilde{\rho}_R^T+\tilde{\lambda}_L^T c^T)\lambda_R\notag\\
    &=(\tilde{\lambda}_R)^{-T}\bigg(\tilde{\rho}_L^T\lambda_L-\tilde{\lambda}_L^T\rho_L-\tilde{\lambda}_L^T c^T\lambda_R+\tilde{\rho}_R^{T}\lambda_R+\tilde{\lambda}_L^T c^T \lambda_R\bigg)\notag\\
    &=(\tilde{\lambda}_R)^{-T}\bigg(\tilde{\rho}_L^T\lambda_L-\tilde{\lambda}_L^T\rho_L+\tilde{\rho}_R^{T}\lambda_R\bigg)\implies (\tilde{\lambda}_R)^{T}\rho_R=\tilde{\rho}_R^{T}\lambda_R+\tilde{\rho}_L^T\lambda_L-\tilde{\lambda}_L^T\rho_L\notag\\
    &\implies \tilde{\rho}_L^T\lambda_L+\tilde{\rho}_R^T\lambda_R-(\tilde{\lambda}_L^T\rho_L+\tilde{\lambda}_R^T\rho_R)=0,
\end{align}
which is the statement of momentum conservation. We now define the $s-$channel exchanged momentum,
\begin{align}
    p=p_{\alpha\Dot{\alpha}}=-p_{1\alpha\Dot{\alpha}}-p_{2\alpha\Dot{\alpha}}=-\lambda_L^T\tilde{\rho}_L+\rho_L^T\tilde{\lambda}_L.
\end{align}
Define,
\begin{align}
    c_0=(\lambda_R)^{-T}p\tilde{\lambda}_L^{-1}.
\end{align}
We can then show that,
\begin{align}
   p=\lambda_R^T c_0 \tilde{\lambda}_L\implies (\lambda_L)^{-T}\lambda_R^T c_0\tilde{\lambda}_L=(\lambda_L)^{-T}(-\lambda_L^T\tilde{\rho}_L+\rho_L^T\tilde{\lambda}_L)=-\tilde{\rho}_L+(\lambda_L)^{-T}\rho_L^T\tilde{\lambda}_L.
\end{align}
Substituting this into the solution for $B$ results in,
\begin{align}
    &B=\big((\lambda_L)^{-T}\lambda_R^Tc_0\tilde{\lambda}_L+(\lambda_L)^{-T}\lambda_R^T c\tilde{\lambda}_L\big)(\tilde{\lambda}_R)^{-1}\notag\\
    &\implies B=(\lambda_L)^{-T}\lambda_R^T(c_0+c)\tilde{\lambda}_L\tilde{\lambda}_R^{-1}.
\end{align}
We now compute $S\sim (1_1 1_2 2_1 2_2)$ and $\tilde{S}\sim (\tilde{1}_1\tilde{1}_2\tilde{2}_1\tilde{2}_2)$ in terms of these matrices. It is clear that,
\begin{align}
    &S\sim (b_{41}b_{32}-b_{31}b_{42})\propto \text{det}(c),\notag\\
    &\tilde{S}\sim (b_{14}b_{23}-b_{13}b_{24})\propto \text{Det}(B)=\frac{\text{Det}(\lambda_R)\text{Det}(\tilde{\lambda}_L)}{\text{Det}(\lambda_L)\text{Det}(\tilde{\lambda}_R)}\text{Det}(c+c_0).
\end{align}
Therefore, after solving for $A,B,D$ in terms of the $c$ matrix we find the following expression for the $s-$channel exchange four point function,
\begin{align}
    \psi_4=(2\pi)^4\delta^4(p_1+p_2+p_3+p_4)\mathcal{J}\int \frac{d^4 c}{\text{Det}(c)\text{Det}(c+c_0)}R_4(A(c),B(c),c,D(c)),
\end{align}
where we recall that $R_4$ is regular in $S$ and $\tilde{S}$. $\mathcal{J}$ is a Jacobian factor that is independent of the $c$ matrix. Let us now suppress $R_4$ and focus on,
\begin{align}
    \int \frac{d^4 c}{\text{Det}(c)\text{Det}(c+c_0)}=\int \frac{d^4 c}{c^2(c+c_0)^2}.
\end{align}
This integral requires an $i\epsilon$ prescription to be well defined. Using the Feynman $i\epsilon$ prescription we have,
\begin{align}
    \frac{1}{(c^2+i\epsilon)((c+c_0)^2+i\epsilon)}
\end{align}
It is easy to show that,
\begin{align}
    \text{Det}(c_0)=\frac{\text{Det}(p)}{\text{Det}(\lambda_R)\text{Det}(\tilde{\lambda}_L)}=-\frac{s^2}{\text{Det}(\lambda_R)\text{Det}(\tilde{\lambda}_L)}.
\end{align}
We can gauge-fix the external kinematics such that $\text{Det}(\lambda_R)=1$ and $\text{Det}(\tilde{\lambda}_L)=1$. Thus,
\begin{align}
    c_0^2=-\text{Det}(c_0)=s^2.
\end{align}
We thus have,
\begin{align}\label{int1}
    \int \frac{d^4 c}{(c^2+i\epsilon)((c-p)^2+i\epsilon)},
\end{align}
as the local structure of the integral where $p_{\alpha\Dot{\alpha}}=-p_{1\alpha\Dot{\alpha}}-p_{2\alpha\Dot{\alpha}}$. Then, we need to take a discontinuity with respect to $s^2$ which is the magnitude square of $p_{\alpha\Dot{\alpha}}$. \eqref{int1} can be interpreted as a time-ordered two point correlator (due to the two propagator like terms). The discontinuity of the time ordered correlator is the Wightman correlator (symmetrized Wightman correlator to be precise) which is schematically,
\begin{align}
    \int d^4 c~\delta(c^2)\delta((c-p)^2).
\end{align}
Thus, the discontinuity operation essentially tells us that,
\begin{align}
    \frac{1}{S\tilde{S}}\to \delta(S)\delta(\tilde{S}).
\end{align}
Undoing the gauge fixing and using this essential result we obtain,
\begin{align}\label{3x3eqn2}
   \text{Disc}_{s^2}\psi_4=\int \frac{d^{4\times 2}C}{\text{Vol}(GL(2))}\int\frac{d^{4\times 2}\tilde{C}}{\text{Vol}(GL(2))}\delta^{4\times 4}(C\cdot\Omega\cdot\tilde{C}^T)\delta^{4\times 2}(\tilde{C}_{ij,I}\Lambda_j^{I\alpha})\delta^{4\times 2}(C_{ij,I}\tilde{\Lambda}_j^{I\Dot{\alpha}})\delta(S)\delta(\tilde{S})R_4,
\end{align}
where,
\begin{align}
    R_4=\text{Res}_{S=0}\text{Res}_{\tilde{S}=0}A_4.
\end{align}
Equating \eqref{3x3eqn2} to \eqref{3x3eqn1} and allowing for the possibility of multiple exchanges we find,
\begin{align}
    \text{Res}_{S=0}\text{Res}_{\tilde{S}=0}A_4=\sum_{s}a_s A_{3L}^{(s)}\times A_{3R}^{(s)},
\end{align}
which is our factorization formula to bootstrap four point functions \eqref{A4inA3A3}.

\section{The half-Fourier transform of the Grassmannian}\label{app:HFT}
In this appendix, we perform the details of the half-Fourier transform computation from spinor-helicity to twistor space that we used to derive \eqref{twistorGrassmannian}. Our starting point is the product of the delta functions,
\begin{align}
    \delta^{n\times 2}(C\cdot \Omega\cdot \tilde{\Lambda})\delta^{n\times 2}(\tilde{C}\cdot\Omega\cdot\Lambda)=\delta^{n\times 2}(C_{ij,I}\tilde{\Lambda}_j^{I\Dot{\alpha}})\delta^{n\times 2}(\tilde{C}_{ij,I}\Lambda_j^{I\alpha}).
\end{align}
We work in the chart,
\begin{align}
C_{ij,I} \equiv  C =
\left(
\begin{array}{cc|cc|c|cc}
1_1 & 1_2 & 2_1 & 2_2 & \cdots & n_1 & n_2 \\
\hline
c_{11} & 1 & c_{21} & 0 & \cdots & c_{1n} & 0 \\
c_{12} & 0 & c_{22} & 1 & \cdots & c_{n2} & 0 \\
\vdots & \vdots & \vdots & \vdots & \ddots & \vdots & \vdots \\
c_{1n} & 0 & c_{2n} & 0 & \cdots & c_{nn} & 1
\end{array}
\right),\tilde C_{ij,I} \equiv \tilde C =
\left(
\begin{array}{cc|cc|c|cc}
\tilde{1}_1 & \tilde{1}_2 & \tilde{2}_1 & \tilde{2}_2 & \cdots & \tilde{n}_1 & \tilde{n}_2 \\
\hline
c_{11} & 1 & c_{12} & 0 & \cdots & c_{1n} & 0 \\
c_{21} & 0 & c_{22} & 1 & \cdots & c_{2n} & 0 \\
\vdots & \vdots & \vdots & \vdots & \ddots & \vdots & \vdots \\
c_{n1} & 0 & c_{n2} & 0 & \cdots & c_{nn} & 1
\end{array}
\right).
\end{align}
The delta function product then takes the form,
\begin{align}
    \delta^{n\times 2}(C_{ij,I}\tilde{\Lambda}_j^{I\Dot{\alpha}})\delta^{n\times 2}(\tilde{C}_{ij,I}\Lambda_j^{I\alpha})=\delta^{n\times 2}(\tilde{\rho}_{i}^{\Dot{\alpha}}+c_{ij}\tilde{\lambda}_j^{\Dot{\alpha}})\delta^{n\times 2}(\rho_i^{\alpha}+c_{ji}\lambda_j^\alpha).
\end{align}
where we wrote $\tilde{\lambda}^{I\Dot{\alpha}}=(\tilde{\lambda}^{\Dot{\alpha}},\tilde{\rho}^{\Dot{\alpha}})$ and $\lambda^{I\alpha}=(\lambda^\alpha,\rho^\alpha)$. We can now perform the half-Fourier transform (we ignore factors of $2\pi$),
\begin{align}
    \prod_{i=1}^{n}\big(\int d^2\tilde{\rho}_i d^2\tilde{\lambda}_i e^{i(\mu_{i\Dot{\alpha}}\tilde{\rho}_i^{\Dot{\alpha}}-\nu_{i\Dot{\alpha}}\tilde{\lambda}_i^{\Dot{\alpha}})}\big)\delta^{n\times 2}(\tilde{\rho}_{i}^{\Dot{\alpha}}+c_{ij}\tilde{\lambda}_j^{\Dot{\alpha}}),
\end{align}
where we have defined $\mu_i^{I\Dot{\alpha}}=(\mu_i^{\Dot{\alpha}},\nu_i^{\Dot{\alpha}})$. Performing the $\tilde{\rho_i}$ integrals is trivial and results in,
\begin{align}
    \prod_{i=1}^{n}\int d^2\tilde{\lambda}_i e^{-i(\mu_{i\Dot{\alpha}}c_{ij}\tilde{\lambda}_{j}^{\Dot{\alpha}}+\nu_{i\Dot{\alpha}}\tilde{\lambda}_i^{\Dot{\alpha}})}=\delta^{n\times 2}(\nu_{i\Dot{\alpha}}+c_{ji}\mu_{i\Dot{\alpha}}).
\end{align}
Combining with the other delta function this yields,
\begin{align}
    \delta^{n\times 2}(\rho_i^\alpha+c_{ji}\lambda_j^\alpha)\delta^{n\times 2}(\nu_{i\Dot{\alpha}}+c_{ji}\mu_{i\Dot{\alpha}})=\delta^{n\times 2}(\tilde{C}\cdot \Omega\cdot \Lambda)\delta^{n\times 2}(\tilde{C}\cdot \Omega\cdot M)=\delta^{n\times 4}(\tilde{C}\cdot\Omega\cdot Z),
\end{align}
where we restored the gauge redundancy and defined the twistor variable $Z$ as in the main-text. 
\section{Connection to Ambitwistors}\label{app:AmbiTwistors}
In this appendix, we convert our manifest $GL(2,\mathbb{R})$ twistors to the ambi-twistor formalism which employs a pair $(Z^A,W_A)$ to describe each operator. Consider the general $n-$point function in twistor space \eqref{twistorGrassmannian},
\begin{align}
    \psi_n(Z^{IA})=\int D\tilde{C}~\delta^{n\times 4}(\tilde{C}_{ij,I}Z_j^{IA})\tilde{A}_n(\tilde{C}).
\end{align}
We write down each twistor as,
\begin{align}
    Z_1^{IA}=\begin{pmatrix}
        Z_1^A\\
        Y_1^A
    \end{pmatrix},
\end{align}
and so on. We then Fourier transform $Y_i^A$ to go to ambi-twistor space. 
\begin{align}\label{YtoW}
    f(Z_1,W_1,\cdots,Z_n,W_n)=\int d^4 Y_1\cdots d^4 Y_n e^{iW_1\cdot Y_1+\cdots+iW_n\cdot Y_n}f(Z_1,Y_1,\cdots,Z_n,Y_n).
\end{align}
Let us gauge-fix the $\tilde{C}$ matrix and then perform this twistor Fourier transform.
\begin{align}
\tilde C_{ij,I} \equiv \tilde C =
\left(
\begin{array}{cc|cc|c|cc}
\tilde{1}_1 & \tilde{1}_2 & \tilde{2}_1 & \tilde{2}_2 & \cdots & \tilde{n}_1 & \tilde{n}_2 \\
\hline
c_{11} & 1 & c_{12} & 0 & \cdots & c_{1n} & 0 \\
c_{21} & 0 & c_{22} & 1 & \cdots & c_{2n} & 0 \\
\vdots & \vdots & \vdots & \vdots & \ddots & \vdots & \vdots \\
c_{n1}& 0 & c_{n2} & 0 & \cdots & c_{nn} & 1
\end{array}
\right).
\end{align}
This yields,
\begin{align}
    \delta^{n\times 4}(C_{ij,I}Z_j^{IA})=
        \delta^{n\times 4}(Y_i^{A}+c_{ji}Z_j^{A}).
\end{align}
The Grassmannian then takes the form,
\begin{align}
    \psi_n(Z^A,Y^A)=\int d^{n\times n}c_{ij}\delta^{n\times 4}(Y_i^{A}+c_{ji}Z_j^{A})\tilde{A}_n(\tilde{C}).
\end{align}
This form makes it trivial to perform the Fourier transform \eqref{YtoW}. The result is,
\begin{align}
    \psi_n(Z^A,W_A)=\int d^{n\times n}c_{ij}e^{-i c_{ji}W_{i}\cdot Z_j}\tilde{A}_n(\tilde{C}).
\end{align}
A few comments are now in order. For scalar two point functions, $\tilde{A}_2$ is a constant and thus the integral becomes trivial resulting in,
\begin{align}
    \psi_2^{\varphi,\varphi}=\delta(W_1\cdot Z_1)\delta(W_2\cdot Z_2)\delta(W_1\cdot Z_2)\delta(W_2\cdot Z_1).
\end{align}
For scalar three point functions, $\tilde{A}_3(\tilde{C})=\frac{1}{\mathcal{K}}$ which is the three point invariant. In the gauge we are working in, it does not depend on the diagonal elements so a factor of $\delta(W_i\cdot Z_i)$ automatically factorizes out. In general we find,
\begin{align}
    \psi_n^{\varphi,\cdots\varphi}(Z,W)=\prod_{i=1}^{n}\delta(W_i\cdot Z_i)f_n(Z,W).
\end{align}
The constraint $Z_i\cdot W_i$ is precisely the defining relation of ambi-twistor space \cite{Adamo:2016rtr,CarrilloGonzalez:2026eum}. For spinning correlators, the ambi-twistor notation is less useful than for scalars. The reason is that $\tilde{A}_n(\tilde{C})$ carries the $SL(2)$ little group indices of the external currents whereas the fundamental variables $Z$ and $W$ do not. Thus, this representation at first glance, obscures the manifest little group covariance. Conformal invariance on the other hand is clearly manifest just like the manifest $GL(2)$ twistor formalism. This is because $W_i\cdot Z_j$ are the natural $SL(4)$ invariants formed from a fundamental and anti-fundamental representation. It would be interesting to make the connection between this formalism and that of \cite{CarrilloGonzalez:2026eum}.

\section{A Direct derivation of the supersymmetric Grassmannian}\label{app:SHGrassmann}
In this appendix, we provide a sketch of the direct derivation of the supersymmetric extension of the symplectic bi-Grassmannian \eqref{SHSUSYGrassmannian}. Our starting point are the $\mathcal{N}=1$ supersymmetry generators,
\begin{align}
    &Q_\alpha=\frac{\partial}{\partial\theta^\alpha}+\frac{\tilde{\theta}^{\Dot{\alpha}}p_{\alpha\Dot{\alpha}}}{2},\notag\\
    &\tilde{Q}_{\Dot{\alpha}}=\frac{\partial}{\partial\tilde{\theta}^{\Dot{\alpha}}}+\frac{\theta^{\alpha}p_{\alpha\Dot{\alpha}}}{2}.
\end{align}
They obey the anti-commutation relation,
\begin{align}
    \{Q_\alpha,\tilde{Q}_{\Dot{\alpha}}\}=p_{\alpha\Dot{\alpha}}.
\end{align}
We now define the dimensionless Grassmann variables $\xi^I$ and $\chi^I$ implicitly via,
\begin{align}
    \theta^\alpha=\frac{\lambda^{I\alpha}\xi_I}{\text{Det}(\lambda)},\tilde{\theta}^{\Dot{\alpha}}=\frac{\tilde{\lambda}^{I\Dot{\alpha}}\chi_I}{\text{det}(\tilde{\lambda})}.
\end{align}
We then find,
\begin{align}
    \frac{\partial}{\partial\theta^\alpha}=\frac{\partial\xi^J}{\partial\theta^\alpha}\frac{\partial}{\partial \xi^J}=\lambda^J_\alpha\frac{\partial}{\partial\xi^J},~~\frac{\partial}{\partial\tilde{\theta}^{\Dot{\alpha}}}=\tilde{\lambda}^I_{\Dot{\alpha}}\frac{\partial}{\partial \chi^J}.
\end{align}
Using these formulae we obtain,
\begin{align}
    &Q_{\alpha}=\lambda_{\alpha}^J\bigg(\frac{\partial}{\partial \xi^J}-\frac{\chi_J}{2}\bigg),\notag\\
    &\tilde{Q}_{\Dot{\alpha}}=\tilde{\lambda}_{\Dot{\alpha}}^J\bigg(\frac{\partial}{\partial \chi^J}+\frac{\xi_J}{2}\bigg).
\end{align}
We now perform a half-Fourier transform,
\begin{align}
    f(\xi,\rho)=\int d^2 \chi e^{\rho_I \chi^I}f(\xi,\chi)\iff f(\xi,\chi)=\int \frac{d^2\rho}{(2\pi)^2} e^{-\rho_I \chi^I}f(\xi,\rho).
\end{align}
This yields,
\begin{align}
    &Q_{\alpha}=\lambda^J_{\alpha}\bigg(\frac{\partial}{\partial \xi^J}+\frac{1}{2}\frac{\partial}{\partial \rho^J}\bigg),\notag\\
    &\tilde{Q}_{\Dot{\alpha}}=\tilde{\lambda}^J_{\Dot{\alpha}}\bigg(\rho_J+\frac{\xi_J}{2}\bigg).
\end{align}
We define the Grassmann twistor variable,
\begin{align}
    \eta_J=\rho_J+\frac{\xi_J}{2}\implies \frac{\partial}{\partial \eta^J}=\frac{1}{2}\frac{\partial}{\partial \rho^J}+\frac{\partial}{\partial \xi^J}.
\end{align}
The supersymmetry generators then take the form,
\begin{align}
    Q_\alpha=\lambda_{\alpha}^J\frac{\partial}{\partial\eta^J},~~\tilde{Q}_{\Dot{\alpha}}=\tilde{\lambda}_{\Dot{\alpha}}^I\eta_I.
\end{align}
Acting on an $n-$point function these quantities take the form,
\begin{align}
    &\sum_{i=1}^{n}Q_{i\alpha}=(\lambda_{1\alpha}^{J}\cdots\lambda_{n\alpha}^J)\begin{pmatrix}
        \frac{\partial}{\partial\eta_1^J}\\
        \vdots\\
        \frac{\partial}{\partial\eta_n^J}
    \end{pmatrix}=\Lambda_{i\alpha}^{I}\frac{\partial}{\partial\eta_i^J}\notag\\
    &\sum_{i=1}^{n}\tilde{Q}_{i\Dot{\alpha}}=(\tilde{\lambda}_{1\Dot{\alpha}}^{I}\cdots\tilde{\lambda}_{n\Dot{\alpha}}^I)\begin{pmatrix}
       \eta_{1I}\\
       \vdots\\
       \eta_{nI}
    \end{pmatrix}=\tilde{\Lambda}_{i\Dot{\alpha}}^{I}\eta_{iI}
    .
\end{align}
Using these generators, one can prove that the supersymmetric symplectic bi-Grassmannian \eqref{SHSUSYGrassmannian} is indeed invariant under the action of the super-conformal group. We start with the conditions,
\begin{align}
    C\cdot \tilde{\Lambda}=0,\tilde{C}\cdot\Lambda=0,C\cdot\Omega\cdot\tilde{C}^T=0,\tilde{C}\cdot\eta=0,
\end{align}
where we absorbed $\Omega$ into $\Lambda,\tilde{\Lambda}$ and $\eta$ since its role is just to raise the $SL(2)$ index. Then $C\cdot\tilde{\Lambda}=0\implies \tilde{\Lambda}=\Omega\cdot\ \tilde{C}^T\cdot P_{\tilde{\Lambda}}$ where $P_{\tilde{\Lambda}}$ is a $n\times 2$ matrix. Then, 
\begin{align}
    \tilde{Q}_{\Dot{\alpha}}=\tilde{\Lambda}^T\cdot\eta=-P_{\tilde{\Lambda}}^T\tilde{C}\cdot\Omega\cdot\eta\sim P_{\tilde{\Lambda}}^T\tilde{C}\cdot\eta,
\end{align}
where we absorbed $\Omega$ into $\eta$ by raising the $SL(2)$ index using $\Omega^{ij,IJ}\eta_{jJ}=\eta_{i}^{I}$. Therefore, acting on the supersymmetric delta function we find,
\begin{align}
    \tilde{C}\cdot\eta\delta(\tilde{C}\cdot\eta)=0,
\end{align}
by virtue of the Grassmann property, thus proving invariance under $\tilde{Q}$. As for invariance under $Q_{\alpha}$, consider,
\begin{align}
    Q_{\alpha}=\Lambda^T\cdot \frac{\partial}{\partial \eta}.
\end{align}
Under the support of the delta functions we can write $\Lambda=\Omega\cdot C^T P_{\Lambda}$. Thus,
\begin{align}
    Q_{\alpha}=-P_{\Lambda}^T C\cdot \Omega\frac{\partial}{\partial\eta}.
\end{align}
Then,
\begin{align}
    \frac{\partial}{\partial \eta_i^I}\delta(\tilde{C}_{ij,I}\eta_{j}^I)=\frac{\partial(\tilde{C}_{jk,J}\eta_k^J)}{\partial \eta_i^I}\frac{\partial}{\partial (\tilde{C}_{jk,J}\eta_k^J)}\delta(\tilde{C}_{ij,I}\eta_{j}^I)=\tilde{C}_{ji,I}\frac{\partial}{\partial (\tilde{C}_{jk,J}\eta_k^J)}\delta(\tilde{C}_{ij,I}\eta_{j}^I).
\end{align}
Therefore,
\begin{align}
    Q_{\alpha}\delta(\tilde{C}\cdot\eta)=-P_{\Lambda}^TC.\Omega\cdot\tilde{C}^T\frac{\partial}{\partial (\tilde{C}\cdot\eta)}\delta(\tilde{C}\cdot\eta)=0,
\end{align}
since $C\cdot\Omega\cdot\tilde{C}^T=0$ by virtue of the symplectic orthogonality constraint. The proof for the $SL(4)$ conformal invariance is the same as in \cite{Bala:2026trw}. As for invariance under special super-conformal transformations, it is also guaranteed since $[K,Q]=\tilde{S}$ and $[K,\tilde{Q}]=S$. Therefore, \eqref{SHSUSYGrassmannian} is invariant under the action of the entire super-conformal group. Of course, the super-twistor space representation \eqref{supertwistorGrassmannian} made this much more manifest.

\section{Higher supersymmetry}\label{app:ExtendedSUSY}
In this appendix, we speculate about the extension of our formalism to theories with $\mathcal{N}>1$ supersymmetry.
First, what is the $R-$symmetry group in Klein signature? Consider the supersymmetry relation,
\begin{align}
    \{Q_{\alpha}^a,\tilde{Q}_{\Dot{\alpha}b}\}=p_{\alpha\Dot{\alpha}}\delta^{a}_{b}.
\end{align}
These indices represent (anti-)fundamental indices of $SL(\mathcal{N})$. This is in contrast to Lorentzian signature where the $R-$symmetry group is $SU(\mathcal{N})$. This is because of the reality property of spinors in Klein space (they are real). There is a natural extension of the supersymmetric symplectic Grassmannian namely,
\begin{align}
    &\mathbf{\Psi}_n(\Lambda,\tilde{\Lambda},\eta)\notag\\&=\int \frac{d^{n\times 2n}C}{\text{Vol}(GL(n))}\int \frac{d^{n\times 2n}\tilde{C}}{\text{Vol}(GL(n))}\delta^{n\times n}(C\cdot\Omega\cdot\tilde{C}^T)\delta^{n\times 2}(C\cdot\Omega\cdot\tilde{\Lambda})\delta^{n\times 2}(\tilde{C}\cdot\Omega\cdot\Lambda)\delta^{n\times \mathcal{N}}(\tilde{C}\cdot\Omega\cdot\eta)\mathbf{A}_n(C,\tilde{C}).
\end{align}
The projectiveness of the integrals implies that,
\begin{align}
    \mathbf{A}_n(GC,\tilde{G}\tilde{C})=\frac{1}{\text{Det}(G)^{n-2}\text{Det}(\tilde{G})^{n+\mathcal{N}-2}}\mathbf{A}_n(C,\tilde{C}).
\end{align}
For the correlator with $C$ integrated out, we have (in super-twistor space for example),
\begin{align}
    \mathbf{\Psi}_n(\mathcal{Z})=\int \frac{d^{n\times 2n}\tilde{C}}{\text{Vol}(GL(n))}\delta^{n\times 4|n\times \mathcal{N}}(\tilde{C}\cdot\Omega\cdot \mathcal{Z})\tilde{A}_n(\tilde{C}),
\end{align}
where the super-twistor now takes the form,
\begin{align}
    \mathcal{Z}^{I\mathcal{A}}=(Z^{IA},\eta^{Ia}).
\end{align}
We require for this integral to be well defined,
\begin{align}
    \mathbf{\tilde{A}}_n(\tilde{G}\tilde{C})=\frac{1}{\text{Det}(\tilde{G})^{2n+\mathcal{N}-4}}\mathbf{\tilde{A}}_n(\tilde{C}).
\end{align}
As for the supersymmetric Penrose transform \eqref{superPenrose}, a natural generalization to extended supersymmetry is,
\begin{align}\label{superPenroseExtendedSUSY}
    \mathbf{J}_{s,\alpha_1\cdots \alpha_m,\Dot{\alpha}_1\cdots\Dot{\alpha}_n}(x,\theta,\tilde{\theta})=\int \frac{d^4 \lambda}{\text{Vol}(GL(2))}\lambda_{I_1\alpha_1}\cdots\lambda_{I_m\alpha_m}\frac{\partial}{\partial\mu^{J_1}_{\Dot{\alpha}_1}}\cdots\frac{\partial}{\partial \mu^{J_n}_{\Dot{\alpha}_n}}\mathbf{J}_{s}^{I_1\cdots I_m J_1\cdots J_n}(\mathcal{Z})|_{\mathcal{X}},
\end{align}
where the incidence relations are now,
\begin{align}
    \mathcal{X}=\{\mu^{I\Dot{\alpha}}=x^{\alpha\Dot{\alpha}}\lambda^I_\alpha+\frac{i}{2}\theta^{\alpha a}\tilde{\theta}^{\Dot{\alpha}}_{a}\lambda_\alpha^I,~~\eta^{Ia}= \theta^{\alpha,a}\lambda_\alpha^I\}.
\end{align}
We leave a more detailed study of extended supersymmetry to the future.

\bibliographystyle{JHEP}
\bibliography{biblio}
\end{document}